\documentclass[twocolumn,times,twocolappendix]{aastex631}
\usepackage{amsmath,bm}
\usepackage{xcolor}
\usepackage[english]{babel}
\usepackage[latin1]{inputenc} 
\usepackage{enumitem}
\usepackage{float}
\usepackage[title]{appendix}
\usepackage{natbib}
\usepackage{makecell}

\newcommand{\mpchi}{\,h^{-1}{\rm {Mpc}}}

\newcommand{\msun}{M_{\sun}}
\newcommand{\msunh}{h^{-1}M_{\sun}}

\newcommand{\hi}{H{~\sc i}}
\newcommand{\hj}{\textrm{H \textsc{i}}}

\shorttitle{\textsc{NeutralUniverseMachine}}
\shortauthors{Guo et al.}

\begin{document}
	
	\title{\textsc{NeutralUniverseMachine}: An Empirical Model for the Evolution of \hi\ and H$_2$ Gas in the Universe}
	
	\author[0000-0003-4936-8247]{Hong Guo}
	\affiliation{Shanghai Astronomical Observatory, Chinese Academy of Sciences, Shanghai 200030, China; guohong@shao.ac.cn}
	
	\author[0000-0002-6593-8820]{Jing Wang}
	\affiliation{Kavli Institute for Astronomy and Astrophysics, Peking University, Beijing 100871, China}
	
	\author[0000-0002-5434-4904]{Michael G. Jones}
	\affiliation{Steward Observatory, University of Arizona, 933 N Cherry Ave., Tucson, AZ 85721, USA}
	
	\author[0000-0002-2517-6446]{Peter Behroozi}
	\affiliation{Steward Observatory, University of Arizona, 933 N Cherry Ave., Tucson, AZ 85721, USA}
	\affiliation{Division of Science, National Astronomical Observatory of Japan, 2-21-1 Osawa, Mitaka, Tokyo 181-8588, Japan}
	
	\begin{abstract}
		Accurately modeling the cold gas content in the universe is challenging for current theoretical models. We propose a new empirical model \dataset[\textsc{NeutralUniverseMachine}]{https://halos.as.arizona.edu/UniverseMachine/DR1/Gas_Masses_NeutralUniverseMachine/} for the evolution of \hi\ and H$_2$ gas along with dark matter halos based on the \dataset[\textsc{UniverseMachine}]{http://halos.as.arizona.edu/UniverseMachine/DR1/} catalog. It is able to accurately describe the observed \hi\ and H$_2$ mass functions, molecular-to-atomic ratio, \hi-halo mass relation, \hi/H$_2$-stellar mass relations at $z\sim0$, as well as the evolution of cosmic gas densities $\rho_{\hj}$ and $\rho_{\rm H_2}$ in $0<z<6$. The predictions from our model include: (i) There is weak evolution of \hi\ mass function in $0<z<3$, but the evolution of H$_2$ mass function is much stronger at the massive end. (ii) The average \hi\ and H$_2$ masses at a given stellar mass decrease by around 1~dex since $z=3$ for the star-forming galaxies, but the evolution for the quenched galaxies is much weaker. (iii) Star-forming galaxies have varying \hi\ depletion time $\tau_\hj$ from 0.1~Gyr to 10~Gyr, and the dependence of $\tau_\hj$ on stellar mass and redshift is much stronger than those of H$_2$ depletion time. The quenched galaxies have much longer gas depletion time and weaker redshift evolution. (iv) The cosmic baryon density associated with galaxies is dominated by stars for $z<1.2$ and mainly contributed by \hi\ gas at higher redshifts. (v) The \hi\ bias gradually increases with the redshift from 0.69 to 2.33 in $0<z<3$ and is consistent with recent \hi\ intensity mapping experiments.
	\end{abstract}

	\section{Introduction}
	Cold neutral gas, both in the atomic and molecular phases, is a key component of the cosmic baryon cycle \citep[see e.g.][for reviews]{Tumlinson2017,Peroux2020}. As fuel for star formation, cold gas is fully involved in the complex baryon physics of gas accretion, heating, and cooling, and stellar and active galactic nucleus (AGN) feedback in the galactic ecosystem \citep[e.g.,][]{Tacconi2020,Saintonge2022}. The distribution and evolution of cold gas thus provide essential information about the formation histories of galaxies, as well as their coevolution with the host dark matter halos.
	
	The distribution of atomic gas in the local universe has been well mapped with the \hi\ 21~cm  surveys, such as the \hi\ Parkes All-Sky Survey \citep[HIPASS;][]{Barnes2001,Meyer2004}, the Arecibo Fast Legacy ALFA Survey \citep[ALFALFA;][]{Giovanelli2005,Haynes2011}, the HI Nearby
	Galaxy Survey \citep[THINGS;][]{Walter2008}, the GALEX Arecibo SDSS Survey \citep[GASS;][]{Catinella2010} and the Apertif survey \citep{Adams2022}. At higher redshifts, the detection of individual galaxies becomes increasingly difficult because of faint signals and radio frequency interference. The \hi\ spectra stacking and intensity mapping techniques have been applied to infer the cosmic \hi\ density at $z<1$ \citep[e.g.,][]{Lah2007,Delhaize2013,Masui2013,Rhee2013,Rhee2018,Kanekar2016,Bera2019}. At even higher redshifts of $z>1.5$, the cosmic \hi\ density can be observed through the damped Ly$\alpha$ (DLA) systems \citep{Wolfe2005} arising from background quasar absorption lines \citep[e.g.,][]{Peroux2003,Prochaska2005,Prochaska2009,Noterdaeme2009,Noterdaeme2012,Zafar2013,Crighton2015,Neeleman2016,Bird2017}. 
	
	The measurements of molecular hydrogen (H$_2$) are commonly probed with surveys of CO rotational lines over a large redshift range \citep[see e.g.,][for a review]{Tacconi2020}, such as FCRAO Extragalactic CO Survey \citep{Young1995}, HERACLES \citep[$z\sim0$,][]{Leroy2009}, xCOLD GASS \citep[$z\sim0$,][]{Saintonge2017}, ASPECS \citep[$1<z<3.5$,][]{Decarli2016,Decarli2019}, COLDz \citep{Riechers2019} PHIBSS2 \citep[$0.48<z<5.25$,][]{Lenkic2020}, and NOEMA HDF-N Survey \citep{Boogaard2023}. The molecular gas mass can also be estimated with the far-IR dust continuum using reasonable molecular gas-to-dust mass ratios \citep[see e.g.,][]{Carilli2013,Berta2016,Scoville2017,Magnelli2020}. 
	
	Together with galaxy stellar properties, e.g., stellar mass ($M_\ast$) and star formation rate (SFR), obtained from optical surveys such as SDSS \citep{York2000} and CANDELS \citep{Grogin2011,Koekemoer2011}, well-defined scaling relations between gas mass and other galaxy properties have been established with statistical and representative samples \citep[see summaries in e.g.,][]{Tacconi2018,Saintonge2022}. Although the \hi and H$_2$ masses typically increase with $M_\ast$, the ratios of $f_{\hj}\equiv M_{\hj}/M_\ast$ and $f_{\rm H_2}\equiv M_{\rm H_2}/M_\ast$ are anticorrelated with $M_\ast$, i.e., galaxies become progressively more gas-poor as their stellar mass grows. There are also positive tight correlations between gas fractions and the specific SFR (${\rm sSFR}\equiv{\rm SFR/M_\ast}$), with the $f_{\rm H_2}$--sSFR relation at $z\sim0$ showing a small intrinsic scatter of 0.12~dex \citep{Saintonge2017}. In terms of molecular gas depletion time, $\tau_{\rm H_2}\equiv M_{\rm H_2}/{\rm SFR}$, it is typically around $\sim1$~Gyr in the local universe and decreases slightly towards higher redshifts \citep{Tacconi2020}.
	
	Combining atomic and molecular gas with stars, we can quantify the total baryon content associated with galaxies \citep[e.g.,][]{Walter2020}. The exact contributions from different components provide valuable constraints to galaxy formation and evolution models. However, it is still a long-standing question how the different baryonic components of galaxies evolve with the mass assembly histories of dark matter halos. Various theoretical models have been proposed to understand the cosmic evolution of the cold gas reservoir, using hydrodynamical simulations \citep[e.g.,][]{Lagos2015,Crain2017,Diemer2019,Dave2020} and semi-analytical models \citep[e.g.,][]{Lagos2011,Fu2013,Popping2014,Kim2017,Xie2017,Baugh2019,Chauhan2020,Spinelli2020}. By modeling the star formation and feedback mechanisms in simulations, these models show reasonable agreement in reproducing some of the observed cold gas properties, such as the cold gas mass functions. However, different galaxy formation models have varying predictions of gas accretion and consumption, causing large discrepancies when comparing other gas scaling relations, e.g., \hi-halo mass relation \citep{Guo2020,Li2022a,Li2022,Dev2023,Rhee2023} and \hi-stellar mass relation \citep{Catinella2018,Janowiecki2020,Guo2021}.
	
	\cite{Popping2015} proposed a semi-empirical approach based on the subhalo abundance matching model of \cite{Behroozi2013} that provides realistic star formation histories for each subhalo in the simulations by fitting to the observed stellar mass functions, cosmic SFRs and sSFRs. They estimated the cold gas masses with the gas density--SFR relation of \cite{Bigiel2008} and pressure-regulated \hi-to-H$_2$ transition model of \cite{Blitz2006} by assuming exponential distributions of gas disks. After calibrating with the observed \hi\ and H$_2$ masses, it shows reasonable agreement with the local \hi\ and H$_2$ mass functions for gas-rich galaxies. Although it still has difficulty reproducing the distributions of gas-poor galaxies and cosmic evolution of cold gas densities, such an approach shows the advantage of empirical models in characterizing the realistic evolution of gas content in halos, without introducing complex galaxy formation recipes.
	
	Other simpler empirical models have been suggested to match some of the observed gas properties. For example, conditional \hi\ mass distributions based on galaxy luminosity and color \citep{Paul2018,Dutta2021},  stellar mass and morphology \citep{Calette2021,Calette2021a}, or even a more complex multi-parameter \hi mass estimator \citep{Li2022a}, show better agreement with fitted observations and provide reasonable predictions for other gas properties. Currently, all of these models are limited to $z\sim0$, where a wealth of direct \hi\ observations are available to constrain the models. In principle, it is possible to construct empirical models to coherently trace the evolution of cold gas within the dark matter halos. 
	
	In this paper, we propose a new empirical model framework, \textsc{NeutralUniverseMachine}, that can self-consistently predict the evolution of both \hi\ and H$_2$ in a broad redshift range of $0<z<6$. Our work is based on the empirical model of \textsc{UniverseMachine} \citep{Behroozi2019} that successfully matches many of the observed galaxy stellar properties, such as stellar mass functions, SFRs, quenched fractions, and correlation functions over $0<z<10$. The key parameters that we adopt from the \textsc{UniverseMachine} model are $M_\ast$ and SFR, which are essential in modeling the cold gas content. Our \hi\ and H$_2$ models based on \textsc{UniverseMachine} would then provide a comprehensive description of the baryon content of galaxies.
	
	The organization of this paper is as follows. In Section~\ref{sec:data}, we introduce our modeling method, as well as the observational data used in the constraints. We show the results in Section~\ref{sec:results}. The discussion and conclusions are presented in Section~\ref{sec:discussion} and Section~\ref{sec:conclusion}, respectively. Throughout the paper, all masses are expressed in units of $\msun$. We assume a flat $\Lambda$CDM cosmology of $\Omega_{\rm m}=0.307$, $h=0.678$, $\Omega_{\rm b}=0.048$ and $\sigma_8=0.823$, consistent with the Planck15 results \citep{Planck2016}.
	
	\section{Data and Method}\label{sec:data}
	\subsection{UNIVERSEMACHINE Catalog}
	The \textsc{UniverseMachine} model parametrizes the probability distribution of galaxy SFRs in halos as a function of $v_{\rm M_{peak}}$, $z$ and $\Delta v_{\rm max}$, where $v_{\rm M_{peak}}$ is the maximum halo circular velocity ($v_{\rm max}$) at the redshift of the peak halo mass ($M_{\rm peak}$) and $\Delta v_{\rm max}$ is the relative change in $v_{\rm max}$ over the past halo dynamical time. $v_{\rm M_{peak}}$ and $\Delta v_{\rm max}$ are used as proxies for the halo mass and mass accretion rate, respectively. An SFR will be assigned to each halo (or subhalo) according to the probability distribution. Meanwhile, halos of larger $\Delta v_{\rm max}$ are assigned higher SFRs, allowing for random scatters. The galaxy stellar masses can be obtained from integrating SFRs along the halo merger trees. In this way, the galaxy mass growth history is self-consistently calculated from the halo assembly and star formation histories. The 44 model-free parameters are determined from fitting to the observations of stellar mass functions, cosmic SFRs and sSFRs, quenched fractions, UV luminosity functions, UV--stellar mass relations and correlation functions. We refer the readers to \cite{Behroozi2019} for more details.
	
	In this paper, we use the public catalogs of \dataset[\textsc{UniverseMachine} DR1]{http://halos.as.arizona.edu/UniverseMachine/DR1/}, which are run on the Bolshoi-Planck $N$-body simulation \citep{Klypin2016} with a box size of $250\mpchi$ on a side and a dark matter particle mass resolution of $2.3\times10^8\msun$. The cosmological parameters are the same as our adopted cosmology. Dark-matter halos (and subhalos) are identified using the \textsc{Rockstar} halo finder \citep{Behroozi2013a}, while halo merger trees are constructed through the \textsc{Consistent Trees} algorithm \citep{Behroozi2013b}.
	
	We will use the following quantities in the \textsc{UniverseMachine} catalogs, $M_{\rm vir}$ (halo or subhalo virial mass, defined as in \citealt{Bryan1998}), $M_{\rm peak}$ (the peak halo or subhalo virial mass over its entire merger history), $v_{\rm M_{peak}}$ (as defined above), $z_{\rm form}$ (halo formation time, defined as the redshift when the most massive progenitor reaches $0.5M_{\rm peak}$), $R_{\rm vir}$ (halo virial radius, also defined in \citealt{Bryan1998}), $M_\ast$, and SFR. To match the observational measurements, we adopted the galaxy properties of $M_\ast$ and SFR with random and systematic uncertainties included (see the discussions in Section 3.5 of \citealt{Behroozi2019}).  
	
	It is suggested in \cite{Behroozi2019} that the inclusion of orphan galaxies (i.e., tidally stripped subhalos with masses below the simulation resolution) will improve the matching with observed clustering measurements and produce the correct satellite evolution.
	We also include the orphan galaxies from the \textsc{UniverseMachine} catalogs in our cold-gas model and trace their merger trees to derive the corresponding measurements of $z_{\rm form}$. Our modeling results would also be less affected by the simulation resolution of Bolshoi-Planck after the inclusion of orphans.  
	
	\subsection{\textsc{NeutralUniverseMachine} Model}
	As noted in \cite{Saintonge2022}, the common practice in the \hi\ community is to calculate the \hi\ mass using observed 21~cm fluxes, without accounting for the contribution of helium and heavier elements. However, the H$_2$ mass measurements typically included an upward correction of 1.36 to obtain all molecular gas masses. To avoid confusion and be consistent with the literature, we still use $M_{\hj}$ to represent \hi\ mass only and use $M_{\rm H_2}$ to indicate the total mass of the molecular gas in the following sections.
	
	\subsubsection{H{~\scriptsize I} Model} 
	The \hi\ mass of each halo or subhalo is determined as a function of its own $M_{\rm vir}$, $z_{\rm form}$, SFR, and $z$, as follows.
	\begin{eqnarray}
		M_{\hj}&=&\frac{\kappa M_{\rm vir}}{\mu^{-\alpha}+\mu^{\beta}} \left(\frac{1+z}{1+z_{\rm form}}\right)^{\gamma}  \left(\frac{\rm SFR}{\rm SFR_{MS,obs}}\right)^{\lambda}\label{eq:mhi}\\ 
		\mu &=&M_{\rm vir}/M_{\rm crit}\\
		\log\kappa&=&\kappa_0+\kappa_1z+\kappa_2 z^2\\
		\log M_{\rm crit}&=&M_0+M_1z+M_2z^2,
	\end{eqnarray}
	where SFR is in units of $\msun/{\rm yr}$ and ${\rm SFR_{MS,obs}}$ is the best-fitting median (observed) SFR of star-forming galaxies defined in \cite{Behroozi2019},
	\begin{eqnarray}
		{\rm SFR_{MS,obs}}&=&{\rm SFR_{MS}}+0.041-0.044z/(1+z)+\nonumber\\
		&&0.314\exp\left(-\frac{(z-2)^2}{2}\right)\label{eq:sfrobs}\\
		{\rm SFR_{MS}}&=&\epsilon\left[(v^{\alpha^\prime}+v^{\beta^\prime})^{-1}+\gamma^\prime\exp\left(-\frac{(\log v)^2}{2\delta^2}\right)\right]\\
		v&=&v_{\rm M_{peak}}/V\\
		\log V&=&2.151-1.658z/(1+z)\nonumber\\
		&&+1.680\ln(1+z)-0.233z \label{eq:logv}\\
		\log\epsilon&=&0.109-3.441z/(1+z)+\nonumber\\
		&&5.079\ln(1+z)-0.781z\\
		\alpha^\prime&=&-5.598-20.731z/(1+z)+\nonumber\\
		&&13.455\ln(1+z)-1.321z\\
		\beta^\prime&=&-1.911+0.395z/(1+z)+0.747z\\
		\log\gamma^\prime&=&-1.699+4.206z/(1+z)-0.809z\\
		\delta&=&0.055
	\end{eqnarray}	
	Equation~(\ref{eq:sfrobs}) corrects for the systematic offset between observed and intrinsic SFRs as in \cite{Behroozi2019}. The functional formal of Equation~(\ref{eq:mhi}) has three parts, the double power law relation of $M_{\rm vir}$, the scaling with the halo formation time $z_{\rm form}$ and the dependence on SFR. In summary, we have 10 free model parameters for the \hi\ gas in halos, $\kappa_0$, $\kappa_1$, $\kappa_2$, $M_0$, $M_1$, $M_2$, $\alpha$, $\beta$, $\gamma$, and $\lambda$.
	
	\begin{figure*}
		\centering
		\includegraphics[width=1\textwidth]{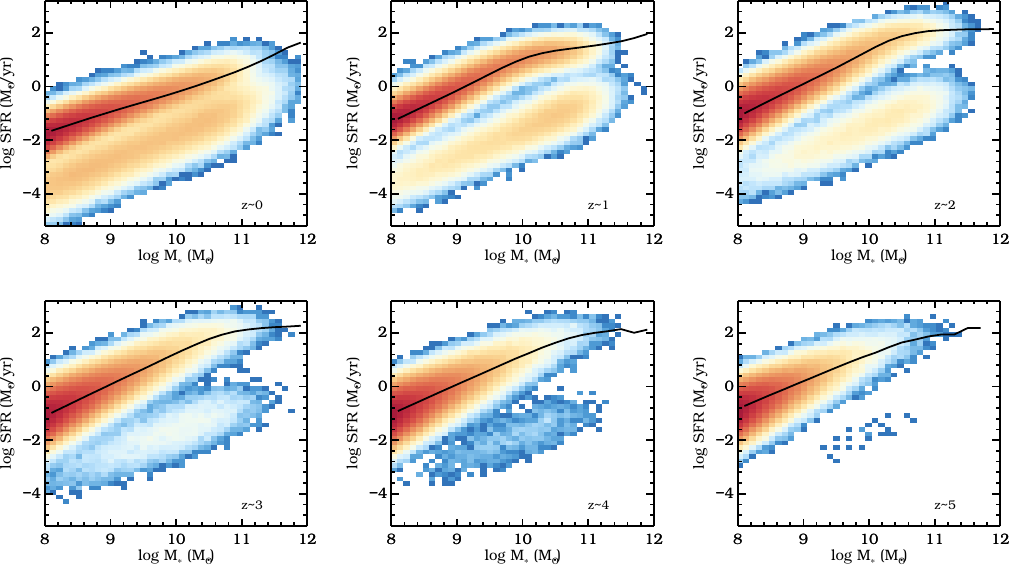}
		\caption{Distributions of galaxies as a function of $M_\ast$ and SFR from $z=0$ to $z=5$ in the \textsc{UniverseMachine} catalog. The color is coded by the logarithmic number densities of galaxies, with redder colors for higher densities. The solid black line in each panel indicates the positions of ${\rm SFR_{MS,obs}}$.} 
		\label{fig:sfr}
	\end{figure*}
	In Figure~\ref{fig:sfr}, we show the distributions of galaxies as a function of $M_\ast$ and SFR from $z=0$ to $z=5$ in the \textsc{UniverseMachine} catalog. The black solid line in each panel indicates the positions of ${\rm SFR_{MS,obs}}$. The feature of bimodal distributions in SFR becomes more prominent at lower redshifts. The SFR distribution in \textsc{UniverseMachine} is simply parameterized as the sum of two lognormal distributions that mimic the bimodal distribution. 
	
	The double power law of $M_{\rm vir}$ is motivated by the observed \hi-halo mass relation of \cite{Guo2020}, where \hi\ mass increases in general with the halo mass and the slope becomes steeper at the lower mass end, which is also confirmed by a recent study of \cite{Dev2023}. Physically, a galaxy's \hi\ disk is typically much more extended than its stellar disk \citep{Wang2016} and thus more affected by the halo environment for low-mass galaxies with low binding energy, as found in both central \citep{Guo2021} and satellite galaxies \citep{Brown2017,Stevens2019}. The redshift evolution of normalization parameter $\kappa$ is to reproduce the change of cosmic \hi\ density at higher redshifts \citep{Rhee2018,Walter2020}. We assume a similar functional form for the evolution of the characteristic mass $M_{\rm crit}$. The parameters $\alpha$ and $\beta$ are the slopes of the low and high mass halos, respectively. We assume no evolution in these slopes, consistent with the trend found in hydrodynamical simulations and semi-analytical models \citep[e.g.,][]{Villaescusa2018,Spinelli2020,Chauhan2020}. 
	
	The dependence of $M_{\hj}$ on both the halo mass and the halo formation time is found by \cite{Guo2017} using the spatial two-point correlation function measurements of the \hi-selected galaxies in ALFALFA. They found that the clustering of \hi-rich galaxies is significantly lower than the predictions from simple abundance matching models, strongly favoring the additional dependence on the halo formation time \citep[see also][]{Li2022}. It is also confirmed by \cite{Stiskalek2021} using a different abundance matching method. Late-forming halos tend to host more \hi\ gas, which is probably caused by the more recent accretion of cold gas or the new gas reservoir brought in by recent wet mergers. We include the dependence of formation time as $M_{\hj}\propto[(1+z)/(1+z_{\rm form})]^\gamma$, based on the results of \cite{Guo2017} and \cite{Li2022}. In this way, the dependence becomes weaker at higher redshifts when the halos are in the rapid growth phase with smaller differences between $z$ and $z_{\rm form}$ \citep{Zhao2009}.
	
	As shown in \cite{Guo2021}, the logarithmic growth/reduction of a galaxy's \hi\ mass is proportional to its logarithmic offset from the star formation main sequence (SFMS), $\log({\rm SFR/SFR_{MS}})$. For galaxies in the SFMS, we can define the \hi\ main sequence as,
	\begin{equation}
		M_{\hj,{\rm MS}}=\frac{\kappa M_{\rm vir}}{\mu^{-\alpha}+\mu^{\beta}} \left(\frac{1+z}{1+z_{\rm form}}\right)^{\gamma}.
	\end{equation}
	Then the scaling of \hi\ mass with SFR is simply $\log(M_{\hj}/M_{\hj,{\rm MS}})=\lambda\log({\rm SFR/SFR_{MS,obs}})$, where $\lambda$ is found to be $\sim0.4$ at $z\sim0$ \citep{Guo2021}. 
	
	\subsubsection{H$_2$ Model}
	\cite{Tacconi2020} proposed a fitting formula for $M_{\rm H_2}$, as a function of $M_\ast$, ${\rm SFR/SFR_{MS}}$ and $\log(1+z)$, which shows good agreement with a compilation of $M_{\rm H_2}$ measurements in the literature. We follow their functional form for the molecular gas as,
	\begin{eqnarray}
		M_{\rm H_2}&=& \zeta M_\ast^\nu\left(\frac{\rm SFR}{\rm SFR_{MS,obs}}\right)^{\eta} \label{eq:mh2}\\
		\log\zeta&=&\zeta_0+\zeta_1\ln(1+z)+\zeta_2\left[\ln(1+z)\right]^2, \label{eq:mh2z}
	\end{eqnarray}
	The five free parameters are $\zeta_0$, $\zeta_1$, $\zeta_2$, $\nu$, and $\eta$. As in the case of \hi, for galaxies on the SFMS, their H$_2$ main sequence is described by $M_{\rm H_2,MS}=\zeta M_\ast^\nu$. The star formation law is then $\log(M_{\rm H_2}/M_{\rm H_2,MS})=\eta\log({\rm SFR/SFR_{MS,obs}})$. Although the effect of the environment on the molecular gas content is still debated \citep[e.g.,][]{Noble2017,Darvish2018,Tadaki2019}, it is generally found to be small. Interestingly, since the molecular gas is distributed mostly in the stellar disks, it would be less affected by the halo environment than in the case of \hi. Thus, our H$_2$ main sequence is defined on the basis of stellar mass rather than halo mass.
	
	\subsection{Observational Measurements}\label{sec:obs_g21}
	In order to fully constrain the parameters of the cold gas model, we have included recent observations of the \hi\ mass function, \hi -halo mass relation, \hi -stellar mass relations, H$_2$ mass function, H$_2$-stellar mass relation, and molecular-to-atomic mass ratio for $z\sim0$, as summarized in Table~\ref{tab:obs}. For higher redshift observations, we used the collections of cosmic \hi\ density and molecular gas density measurements in \cite{Walter2020}, as well as the \hi-stellar mass relation of \cite{Chowdhury2022} at $z\sim1.1$. 
	The details are described as follows.
	\begin{table}
		\caption{Observational Constraints}
		\label{tab:obs}
		\centering
		\begin{tabular}{llr}
			\hline
			Measurements & References & Redshifts \\
			\hline
			\hi\ Mass Function & This work (Appendix~\ref{sec:himf}) & $z\sim0$ \\
			\hi-Halo Mass Relation & \cite{Guo2020} & $z\sim0$ \\
			\hi-Stellar Mass Relation & \cite{Guo2021} & $z\sim0$ \\
			H$_2$ Mass Function & \cite{Fletcher2021} & $z\sim0$ \\
			H$_2$-Stellar Mass Relation & \cite{Saintonge2017} & $z\sim0$ \\
			H$_2$-to-\hi\ mass ratio & \cite{Catinella2018} & $z\sim0$ \\
			\hline
			\hi-Stellar Mass Relation & \cite{Chowdhury2022} & $z\sim1.1$\\
			cosmic \hi\ density & \cite{Walter2020} & $0<z<5$\\
			cosmic H$_2$ density & \cite{Walter2020} & $0<z<6$\\
			\hline
		\end{tabular}
	\end{table}
	
	\subsubsection{Local Universe}\label{sec:z0}
	
	The \hi\ mass function (HIMF), $\phi(M_{\hj})$, in the local universe has been measured from the final catalog of ALFALFA \citep{Haynes2018} covering $\sim6900~{\rm deg}^2$ over the redshift range of $0<z<0.05$ in \cite{Jones2018}. It describes the average number densities of galaxies in given \hi\ mass bins. By including an effective volume weight of each galaxy in the catalog, the sample selections of ALFALFA and the influence of local large-scale structures can be robustly taken into account in the final measurements of $\phi(M_{\hj})$ \citep{Martin2010}. The \hi\ observations of ALFALFA depend on both the \hi\ flux and the line profile width. After selecting \hi\ targets above the 50\% completeness limit  \cite{Jones2018} measured the HIMF using the two-dimensional stepwise maximum likelihood (2DSWML) method \citep[see e.g.,][]{Zwaan2005,Martin2010}. They have approximated the completeness limit as a sharp cut-off at the location of 50\% completeness \citep[Eq.~5 of][]{Haynes2011}. However, given the large number of sources now available in the ALFALFA catalog, the missing sources between 50\% and 100\% completeness could potentially lead to an overestimation of the effective volumes for low-$M_\hj$ sources and thus underestimate low-mass bins of the HIMF. We have therefore decided to remeasure the HIMF following the same steps of \cite{Jones2018}, but only consider sources above the 90\% completeness limit \citep[Eq.~4 of][]{Haynes2011}. It avoids the need to heavily weight the \hi\ sources with low completeness. We verify that changing from 90\% completeness to even higher thresholds does not have any visible effect on HIMF. We estimate the errors of $\phi(M_{\hj})$ using the fractional errors from 64 mock observations of the ALFALFA final sample, as will be detailed in Section~\ref{subsec:mock}. We list the corrected HIMF measurements in Appendix~\ref{sec:himf}. 
	
	The \hi-halo mass relation, $\langle M_{\hj,{\rm tot}}|M_{\rm vir}\rangle$, describes the total \hi\ mass (i.e. including the \hi\ gas of all member galaxies) in halos of different masses. By selecting halos in the overlapping regions between ALFALFA and the galaxy groups constructed from SDSS DR7 \citep{Lim2017}, \cite{Guo2020} stacked the total \hi\ spectra for halos in different $M_{\rm vir}$ bins, using the aperture size determined from the halo virial radius with a minimum diameter of $8\arcmin$, which is about twice the ALFALFA beam width. The average \hi\ mass is then obtained by dividing the total \hi\ mass in each $M_{\rm vir}$ bin by the corresponding number of galaxy groups. The uncertainties in $\langle M_{\hj,{\rm tot}}\rangle$ were determined from bootstrapping the stacking samples.
	
	The \hi-stellar mass relation, $\langle M_{\hj}(M_\ast)\rangle$, provides additional constraints to the correlation between the stellar and gas components of galaxies.  By applying the \hi\ spectra stacking technique, \cite{Guo2021} accurately measured $\langle M_{\hj}(M_\ast)\rangle$ for star-forming and quenched \emph{central} galaxies (i.e., divided by cuts in SFR) using the same galaxy sample as in \cite{Guo2020}. This is especially important for quenched galaxies, most of which are usually without individual 21~cm detections \citep{Catinella2018}. The use of $\langle M_{\hj}(M_\ast)\rangle$ for star-forming and quenched central galaxies will also constrain the power-law index $\eta$ of SFR in Equation~(\ref{eq:mhi}). 
	
	The H$_2$ mass function (H$_2$MF), $\phi(M_{\rm H_2})$, at $z\sim0$ is robustly measured by \cite{Fletcher2021} using a large set of H$_2$ measurements (532 galaxies with a detection rate of 63\%) in xCOLD GASS \citep{Saintonge2017}. We use their $\phi(M_{\rm H_2})$ values for $8<\log(M_{\rm H_2}/\msun)<10.16$, where the measurements are more complete. When integrating over the best-fitting Schechter function, they found a cosmic H$_2$ density of $\rho_{\rm H_2}=1.04\times10^7\msun/{\rm Mpc^{3}}$.
	
	We also utilize the H$_2$-stellar mass relation, in terms of the molecular gas fraction $f_{\rm H_2}(M_\ast)$ measured by \cite{Saintonge2017} using xCOLD GASS. We adopt their stacking measurements of $\log\langle f_{\rm H_2}(M_\ast)\rangle$, which properly include the contribution from galaxies without detection.  We adopt their measurements for the main-sequence galaxies, defined as $|\log({\rm SFR}/{\rm SFR_{MS}})|<0.4$. For fair comparisons, we would apply the same SFMS definition of \cite{Saintonge2016} (their Eq.~5) in our model when fitting to $f_{\rm H_2}(M_\ast)$ of the main sequence galaxies.  
	
	To connect the H$_2$ model with that of \hi, it is important to have constraints from the molecular-to-atomic mass ratio (i.e. $R_{\rm mol}\equiv M_{\rm H_2}/M_{\hj}$). We adopt the weighted average relation between $\langle\log R_{\rm mol}\rangle$ and $M_\ast$ from \cite{Catinella2018} (their Table~3). We note that the H$_2$ measurements in \cite{Catinella2018} were obtained from xCOLD GASS, but without the correction for helium and heavier elements. Therefore, we weighted their $R_{\rm mol}$ measurements by a factor of 1.36 to be consistent with our definition.   
	
	The above observational measurements at $z\sim0$ come from the ALFALFA, xGASS, and xCOLD GASS surveys. Both ALFALFA and xGASS data were observed with the Arecibo telescope \citep{Catinella2010}. The use of these self-consistent measurements will then provide tight constraints on our model. We also note that the \hi\ measurements adopted here do not include the self-absorption correction, which is quite uncertain and still under debate \citep{Jones2018}.
	
	\subsubsection{High Redshifts}
	There are fewer robust gas scaling relations available at higher redshifts, due to the lack of large statistical samples, especially for the \hi\ observations. To anchor the correct \hi\ mass distribution, we adopt the latest \hi\ stacking measurements of \cite{Chowdhury2022} using the Giant Metrewave Radio Telescope (GMRT) for star-forming galaxies in two different stellar mass bins at $0.74<z<1.45$ ($\langle z\rangle\sim1.1$).
	
	It is much easier to estimate the cosmic \hi\ and H$_2$ densities than to obtain robust scaling relations. Since our model only includes 6 parameters ($\kappa_1$, $\kappa_2$, $M_1$, $M_2$, $\zeta_1$, $\zeta_2$) to describe the redshift dependence, they can be well constrained with the cosmic gas density measurements at various redshifts. We adopt the collections of various cosmic \hi\ density ($\rho_\hj$) and H$_2$ density ($\rho_{\rm H_2}$) measurements in the literature from \cite{Walter2020} in the redshift range of $0<z<6$. However, we note that their \hi\ densities (their Table~2) have included the helium contribution. To match our definition, we have corrected for the different cosmologies and removed the helium contribution using the original measurements in the literature. 
	
	As summarized in \cite{Walter2020}, the cosmic \hi\ densities can be derived by measuring the 21~cm emission in the local universe, stacking \hi\ spectra and 21~cm intensity mapping at intermediate redshifts, and using DLA systems at $z>1.5$. The uncertainties of these methods become increasingly larger at higher redshifts, especially for DLAs.  DLA measurements are generally limited to systems above an \hi\ column density threshold of $\log N({\hj})>20.3\,{\rm cm}^{-2}$ \citep[e.g.,][]{Peroux2003,OMeara2007}. The ignorance of lower $N({\hj})$ systems will cause an underestimate of \hi\ density by 10--20\% \citep{Zafar2013,Berg2019,Peroux2020}.
	
	On the other hand, the DLAs probe \hi\ gas in and around galaxies. In fact, there is an increased contribution of \hi\ gas outside the galaxies (i.e., the intergalactic medium) toward high redshifts \citep{Peroux2020}. Using the hydrodynamical simulation of IllustrisTNG \citep{Nelson2019}, \cite{Villaescusa2018} found that about 80\% and 90\% of the contributions to cosmic \hi\ densities at $z=5$ are still from galaxies and within halos, respectively. It means that our model will also likely miss 10--20\% of \hi\ signals at high redshifts, counterbalancing the underestimation of \hi\ densities from DLAs. Thus, DLA measurements at high redshifts will still provide reasonable constraints to the \hi\ gas within halos. 
	
	The uncertainties associated with the cosmic molecular gas density inferred from CO line luminosity lie mainly in the systematics of the CO to H$_2$ conversion factor (commonly denoted as $\alpha_{\rm CO}$) \citep[see discussions in][]{Tacconi2020}. $M_{\rm H_2}$ measurements in xCOLD GASS were derived with a conversion factor that depends on both the metallicity and the offset from the SFMS \citep{Accurso2017}. The median value of $\alpha_{\rm CO}$ is around $3.3\msun {\rm (K\,km\,s^{-1}\,pc^2)}^{-1}$ in the xCOLD GASS sample. But other CO measurements, as listed in Table~3 of \cite{Walter2020} adopted a constant $\alpha_{\rm CO}=3.6\msun {\rm (K\,km\,s^{-1}\,pc^2)}^{-1}$. It will cause a small level of inconsistency among the CO measurements, but the large uncertainties for $\rho_{\rm H_2}$ at high redshifts make it less an issue. Using the dust continuum to infer $\rho_{\rm H_2}$ also requires the assumptions of a metallicity-dependent gas-to-dust mass ratio, dust temperature, and emissivity, causing the systematic uncertainties of $\rho_{\rm H_2}$ in the literature. 
	
	There are still some \hi\ and H$_2$ gas scaling relation measurements available at high redshift in the literature. However, their sample selections would vary from one to another. We will use these measurements as consistency check of our model predictions in the following sections, rather than apply the different selection cuts to constrain our model.
	
	\subsection{Model Fitting}
	To fit the observational \hi\ and H$_2$ measurements using our 15-parameter \textsc{NeutralUniverseMachine} model, we apply the Bayesian inference tool of \textsc{MultiNest} \citep{Feroz2009} to explore the parameter space. For each set of model parameters in the Monte Carlo Markov Chain (MCMC), we can generate the corresponding \hi\ and H$_2$ masses for each halo (central galaxy) and subhalo (satellite galaxy) following Equations~(\ref{eq:mhi}) and~(\ref{eq:mh2}). To fit the observational data of $\phi(M_{\hj})$, $\langle M_{\hj,{\rm tot}}|M_{\rm vir}\rangle$, $\langle M_{\hj}(M_\ast)\rangle$, $\phi(M_{\rm H_2})$, $f_{\rm H_2}(M_\ast)$ and $R_{\rm mol}$, we adopt the same binning schemes as in the references (Table~\ref{tab:obs}) to avoid the binning effect. To be consistent with the observations using \hi\ spectra stacking, we calculate the average \hi\ mass in $\langle M_{\hj,{\rm tot}}|M_{\rm vir}\rangle$ and $\langle M_{\hj}(M_\ast)\rangle$ as $\Sigma_i M_{{\hj}, i}/N$, where $N$ is the total number of halos (or galaxies) in each bin. In order to fit the observed H$_2$MF, we have added a Gaussian random scatter of $0.2$~dex to $M_{\rm H_2}$ for each galaxy as suggested in \cite{Saintonge2022}, mimicking measurement errors.   
	
	To match the observations of $\langle M_{\hj}(M_\ast)\rangle$ for star-forming and quenched galaxies in \cite{Guo2021}, we adopt the same SFR cut of $\log({\rm SFR_{cut}}/{\rm yr}^{-1}\msun)=0.65\log(M_\ast/\msun)-7.25$ to separate the two populations. To fit the $\langle M_{\hj}(M_\ast)\rangle$ relation at $z\sim1.1$ for star-forming galaxies from \cite{Chowdhury2022}, we define our star-forming population using $\log({\rm SFR/SFR_{MS,obs}})>-1$. As shown in Figure~\ref{fig:sfr}, it provides a reasonable cut to separate star-forming and quenched galaxies.
	
	Other observables are straightforward to measure in the \textsc{UniverseMachine} catalogs. There are 149 redshift outputs between $z=0$ and $z=6$ in the \dataset[\textsc{UniverseMachine} DR1]{http://halos.as.arizona.edu/UniverseMachine/DR1/} of the Bolshoi-Planck simulation. We calculate $\rho_{\hj}$ and $\rho_{\rm H_2}$ for all redshift outputs in each run of MCMC, and interpolate within the outputs to derive gas densities at the observed redshifts.
	
	The $z\sim0$ measurements are self-consistent with each other and the uncertainties are also small. But the high-redshift gas density measurements suffer from large errors and systematic uncertainties, as well as inconsistency between different sets of observations. It is more practical and reliable to first constrain the nine redshift-independent model parameters using $z\sim0$ measurements. The remaining 6 parameters ($\kappa_1$, $\kappa_2$, $M_1$, $M_2$, $\zeta_{1}$, $\zeta_{2}$) can then be better constrained with the cosmic gas densities at higher redshifts. Taking into account the large uncertainties in the \hi\ and H$_2$ densities, we emphasize that our best-fitting models at these high redshifts are based on the available data sets and may be improved with future observations. 
	
	The likelihood of each MCMC run is proportional to $\exp(-\chi^2/2)$, where the total $\chi^2$ is determined as the sum of $\chi^2$ for each set of measurements in Table~\ref{tab:obs}. For different sets of measurements, the covariance between the data points is hard to quantify, and we simply use the error of each data point to calculate $\chi^2$. Minimizing the total $\chi^2$ seems to bias the fit toward data sets with more data points (e.g., the HIMF). Intuitively, this corresponds to weighting the data sets by information content, which is the optimal way to use the available data. In cases where one data set is not fit well, this could be due to some combination of: (1) a fitting function that is less flexible than needed, (2) unmodeled systematics, instrumental effects or analysis assumptions that cause tension between data sets, and (3) statistical fluctuations. In the first two cases, the statistically valid solution is to add parameters (to the fitting function and to the nuisance parameter set, respectively) to better account for the data, as in \cite{Behroozi2019}. In the latter case, the statistically valid solution is to weight the points by the total $\chi^2$.
	
	\section{Results}\label{sec:results}
	
	\subsection{Best-fitting Model}
\begin{figure*}
		\centering
		\includegraphics[width=\textwidth]{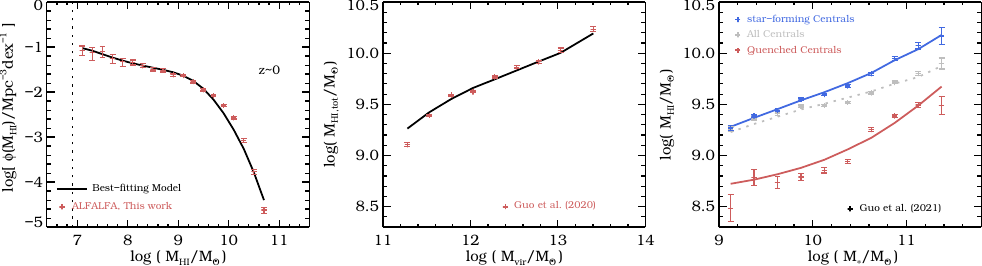}
		\caption{Comparisons between the observed measurements and the best-fitting model predictions, for HIMF (top left panel), \hi-halo mass relation (top right panel), \hi-stellar mass relations of central galaxies (bottom left panel) and projected 2PCF (bottom right panel) at $z\sim0$. The observational data are shown as points with errors, while the best-fitting models are represented by solid lines. The estimated \hi\ mass limit of our halo-based model is shown as the vertical dotted line in the left panel (see text for details). The measurements of \hi-halo mass relation with confusion correction in \cite{Guo2020} are shown as the open blue circles in the top right panel. The \hi-stellar mass relations of central galaxies are shown for both star-forming (blue points and blue line) and quenched galaxies (red points and red line). We also show for comparison the $\langle M_{\hj}(M_\ast)\rangle$ measurements for all central galaxies (gray points), which are not used in the model fittings. The corresponding prediction of the best-fitting model is shown as the gray dotted line. } 
		\label{fig:hifit}
	\end{figure*}

\begin{figure*}
		\centering
		\includegraphics[width=1\textwidth]{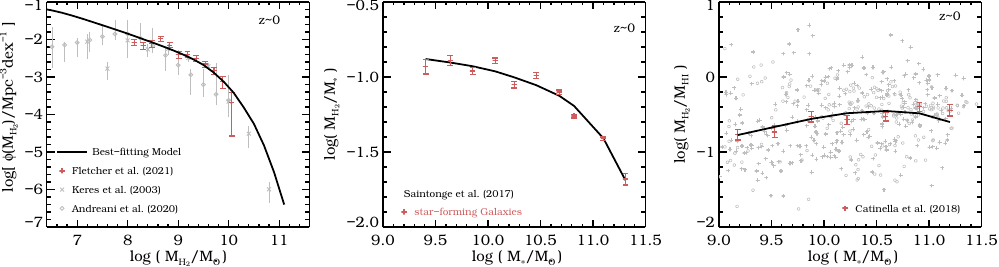}
		\caption{Comparisons between the best-fitting models and the observed measurements, for H$_2$MF $\phi(M_{\rm H_2})$ (left panel), H$_2$ mass fraction $f_{\rm H_2}(M_\ast)$ (middle panel) and H$_2$-to-\hi\ mass ratio $R_{\rm mol}$ (right panel) at $z\sim0$. The best-fitting model is shown as the solid line in each panel. In addition to the observational constraints of H$_2$MF in \cite{Fletcher2021}, we also display in the left panel the measurements of \cite{Keres2003} (crosses) and \cite{Andreani2020} (diamonds). The $R_{\rm mol}$ measurements of galaxies in the xCOLD GASS sample are shown as the gray symbols in the right panel. The nondetections in the sample are displayed as the open circles, using the upper limits of the mass measurements.} 
		\label{fig:h2fit}
	\end{figure*}

\begin{figure*}
		\centering
		\includegraphics[width=0.9\textwidth]{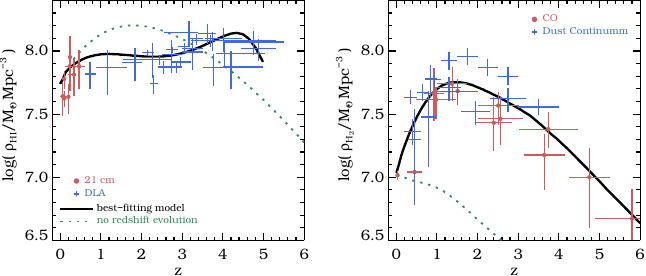}
		\caption{Comparison between observed cosmic gas densities and the best-fitting models for \hi\ (left) and H$_2$ (right). The measurements from \hi\ 21~cm emission and intensity mapping are shown as red symbols in the left panel, while those from DLA are shown as blue symbols. Similarly, the H$_2$ densities of the CO line luminosity and dust continuum are shown as red and blue symbols in the right panel, respectively. The best-fitting models are shown as solid lines, while the dotted lines indicate the model predictions without redshift evolution (i.e., $\kappa_1$, $\kappa_2$, $M_1$, $M_2$, $\zeta_1$, and $\zeta_2$ set as 0). }
		\label{fig:rhohih2}
	\end{figure*}

	Our best-fitting model is determined from the MCMC run with the maximum likelihood, which is capable of reproducing all the observations in Table~\ref{tab:obs}. The observational data, as well as the best-fitting models, are shown for $\phi(M_{\hj})$ (Fig.~\ref{fig:hifit}, left panel), $\langle M_{\hj,{\rm tot}}|M_{\rm vir}\rangle$ (Fig.~\ref{fig:hifit}, middle panel), $\langle M_{\hj}(M_\ast)\rangle$ (Fig.~\ref{fig:hifit}, right panel), $\phi(M_{\rm H_2})$ (Fig.~\ref{fig:h2fit}, left panel), $f_{\rm H_2}(M_\ast)$ (Fig.~\ref{fig:h2fit}, middle panel), $R_{\rm mol}$ (Fig.~\ref{fig:h2fit}, right panel), $\rho_{\hj}(z)$ (Fig.~\ref{fig:rhohih2}, left panel), $\rho_{\rm H_2}(z)$ (Fig.~\ref{fig:rhohih2}, right panel), and $\langle M_{\hj}(M_\ast)\rangle$ at $z\sim1.1$ (Fig.~\ref{fig:mhiz}, right panel). The observational measurements are shown as the points with errors, whereas our model predictions are displayed as solid lines.
	
	The best-fitting model parameters are,
	\begin{eqnarray}
		&&\log\kappa=-0.972^{+0.012}_{-0.062}-0.180^{+0.046}_{-0.126}z+0.053^{+0.067}_{-0.026}z^2\nonumber\\
		\\
		&&\log M_{\rm crit}=10.832^{+0.045}_{-0.016}+0.835^{+0.068}_{-0.253}z-0.246^{+0.134}_{-0.044}z^2\nonumber\\
		\\
		&&\alpha=1.346^{+0.162}_{-0.071}, \ \ \beta=0.604^{+0.011}_{-0.017}\\
		&&\gamma=2.233^{+0.037}_{-0.137},\ \ \lambda=0.433^{+0.010}_{-0.007} \\
		&&\nu=0.921^{+0.009}_{-0.016}, \quad \eta=0.896^{+0.036}_{-0.021}\\
		&&\log\zeta=-0.384^{+0.166}_{-0.094}+1.420^{+0.084}_{-0.071}\ln(1+z)\nonumber\\
		&&\qquad\quad\ -0.425^{+0.059}_{-0.069}\left[\ln(1+z)\right]^2,
	\end{eqnarray}
	where $M_{\rm crit}$ is in units of $\msun$. In Appendix~\ref{sec:para}, we display the density distributions of the model parameters in the MCMC chains.
	
	Our best-fitting model shows good agreement with the HIMF measurements $\phi(M_{\hj})$ in the left panel of Figure~\ref{fig:hifit}. The HIMF is typically fitted with a Schechter function in the literature \citep{Zwaan2005,Martin2010,Jones2018}. But it is naturally explained in our halo model by integrating the halo mass function weighted by the \hi\ mass. We note that our halo-based \hi\ model would be affected by the dark matter particle mass resolution of the Bolshoi-Planck simulation ($2.3\times10^8\msun$). The halo mass function is only accurate for halos with $M_{\rm vir}>10^{10}\msun$ \citep{Klypin2016}, which corresponds to an average \hi\ mass limit of $10^{6.9}\msun$ for our best-fitting model. We indicate this mass limit as the vertical dotted line in the left panel of Figure~\ref{fig:hifit}. Galaxies with $M_\hj$ below this limit should be used with caution.
	
	Our predicted \hi-halo mass relation $\langle M_{\hj,{\rm tot}}|M_{\rm vir}\rangle$ is shown as the solid line in the middle panel. It seems to slightly overestimate the total \hi\ mass in halos of $M_{\rm vir}\sim10^{11.25}\msun$. As will be shown in Section~\ref{sec:otherHI}, it is likely caused by both the sample selection of SDSS galaxies and the halo mass uncertainties in the SDSS group catalog. As shown in Fig.~10 of \cite{Lim2017}, the estimates of the halo mass from the group finder are biased high for the halo of $M_{\rm vir}<10^{11}\msun$. We note that the minimal halo mass in the group catalog is $10^{11.14}\msun$, which is potentially overestimated for those low-mass halos.
	
	In the right panel of Figure~\ref{fig:hifit}, the \hi-stellar mass relations are shown for both star-forming (blue symbols) and quenched \emph{central} galaxies (red symbols). As expected, the best-fitting power index $\lambda$ is consistent with the observation ($\sim0.4$). The fitting to quenched galaxies is somewhat worse at the low-mass end, mainly due to the slight differences between the SFR distributions of SDSS and \textsc{UniverseMachine} model. For comparison, we also display the $\langle M_{\hj}(M_\ast)\rangle$ measurements for all central galaxies (gray symbols), which are not included in observational constraints but are still well reproduced by our model (dotted line), because it is simply the summation of $M_{\hj}$ in the star-forming and quenched populations weighted by the corresponding numbers of galaxies.

	
	In the left panel of Figure~\ref{fig:h2fit}, we show the best-fitting model for the H$_2$MF $\phi(M_{\rm H_2})$. For comparisons, we also display the H$_2$MF measurements at $z\sim0$ from \cite{Keres2003} (crosses) and \cite{Andreani2020} (diamonds). While the H$_2$MF measurements of \cite{Keres2003} were estimated from the CO luminosity function based on the FCRAO Extragalactic CO Survey, the measurements of \cite{Andreani2020} were derived from the bivariate K-band luminosity-H$_2$ mass function of the Herschel Reference Survey.
	The different measurements are consistent with each other within errors. But the H$_2$MF is most robustly measured with the xCOLD GASS sample \citep{Fletcher2021}, where sample selection bias is carefully taken into account \citep{Saintonge2022}. Our model agrees with the measurements of \cite{Fletcher2021}, but predicts a somewhat higher $\phi(M_{\rm H_2})$ for $M_{\rm H_2}<10^8\msun$ compared to \cite{Andreani2020}. Since the H$_2$MF measurements at these low-mass ranges are quite uncertain, the discrepancies can be verified with future surveys. 
	
	Reasonable agreement between data and the model is seen for the H$_2$-stellar mass relation, as well as the average H$_2$-to-\hi\ mass ratio $R_{\rm mol}$ in the middle and right panels of Figure~\ref{fig:h2fit}. The H$_2$ fraction $f_{\rm H_2}$ of star-forming galaxies is decreasing with $M_\ast$ from $13\%$ to $2\%$ in the mass range probed by xCOLD GASS. Since the \textsc{UniverseMachine} model accurately fits the observed galaxy stellar mass function with $M_\ast>10^7\msun$ and the average H$_2$ fraction is approaching a constant of $13\%$ for low-mass galaxies, the mass resolution effect of Bolshoi-Planck simulation is then negligible for $M_{\rm H_2}>10^{6.1}\msun$, which is well below the current detection limit of H$_2$ gas. 
	
	The average H$_2$-to-\hi\ mass ratio $R_{\rm mol}$ has only a weak dependence on $M_\ast$. The $R_{\rm mol}$ measurements of galaxies in the xCOLD GASS sample are shown as gray symbols. The non-detections in the xCOLD GASS survey are shown as the open circles using the upper limits provided in the catalog. The small level of inconsistency between $R_{\rm mol}$ of our best-fitting model and those of \cite{Catinella2018} is likely due to the treatment of non-detections in xCOLD GASS, since only the upper limits for non-detections are used in the estimation of $R_{\rm mol}$. The scatter around the mean $R_{\rm mol}$ is still quite large and varies by two orders of magnitude.  Our model also predicts a similarly large scatter of $\sim0.5$~dex for $R_{\rm mol}$, consistent with the results of \cite{Catinella2018}.
	
	The predictions of the best-fitting model for $\rho_{\hj}(z)$ and $\rho_{\rm H_2}(z)$ are shown in Figure~\ref{fig:rhohih2} as solid lines. The measurements from \hi\ 21~cm emission and intensity mapping are shown as red symbols in the left panel of Figure~\ref{fig:rhohih2}, while those from DLAs are shown as blue symbols. Similarly, the H$_2$ densities from the CO line luminosity and the dust continuum are shown as red and blue symbols in the right panel, respectively. To see the effects of the redshift evolution of the model parameters in Equations~(\ref{eq:mhi}) and~(\ref{eq:mh2}), we also show the model predictions without redshift evolution as green dotted lines (that is, $\kappa_1$, $\kappa_2$, $M_1$, $M_2$, $\zeta_1$ and $\zeta_2$ set as 0). It shows the importance of including the redshift evolution, otherwise $\rho_{\hj}$ would quickly increase to the peak at around $z\sim1.8$ and significantly decrease towards higher redshifts, following the evolution of cosmic SFR densities \citep{Behroozi2019}. 
	
	Our best-fitting model shows that $\rho_{\hj}$ has a fast increase by $\sim0.25$~dex from $z=0$ to $z=1$, but it remains flat in $1<z<3$. The increase and decrease of $\rho_{\hj}$ in $3<z<5$ is caused by our functional form of $\kappa$ and $M_{\rm crit}$ to fit the available data. More complex functional forms as in \cite{Behroozi2019} might be necessary to explore the behavior at these high redshifts, but they cannot be well constrained by the current data. For even higher redshifts of $z>5$, our current model is not accurate enough, because there is an increasing fraction of \hi\ gas outside the halo boundary \citep{Villaescusa2018}. Our halo-based model may not encompass the correct amount of \hi\ gas at these redshifts. As will be shown in the following section, the simulation resolution of Bolshoi-Planck is also not able to resolve the ambient \hi\ gas in low-mass halos below $10^{10}\msun$ at these redshifts. Without the redshift evolution of $\zeta$ in Eq.~(\ref{eq:mh2}), $\rho_{\rm H_2}$ is decreasing with redshift (green dotted line), following the trend of stellar mass accretion. The molecular fraction of the cosmic neutral gas density is highest around $z\sim1.5$ and quickly decreases for both lower and higher redshifts. 
	
	We emphasize that the current measurements of $\rho_\hj$ and $\rho_{\rm H_2}$ from DLAs at high redshifts (especially $z>3$) have large uncertainties. Our best-fitting model is based on all these available measurements and will be improved with future cold gas surveys.  Moreover, other tracers of \hi\ densities (e.g., the [C\,{\scriptsize II}] emission) have been suggested to provide additional constraints to the \hi\ gas within galaxies \citep{Heintz2021,Heintz2022}, which will be useful to further understand the neutral gas distribution in and around galaxies.
	
	\subsection{Mock Observation}\label{subsec:mock}
	Since the ALFALFA sample only covers a small volume in the local universe, it is important to investigate the effect of cosmic variance on the available measurements \citep[see e.g.,][]{Chauhan2019,Chen2019}. Both HIMF and \hi\ clustering would be affected by the cosmic variance effect \citep{Li2022a}. The \hi\ clustering measurements would be underestimated on large scales due to the integral constraint effect \citep[see discussions in Section 3.3 of][]{Guo2017}. This is more severe for galaxies with lower $M_\hj$, as they are observed in smaller volumes due to lower \hi\ fluxes. 
	
	To fully account for this effect, we construct realistic mock catalogs as in observations to measure the HIMF and \hi\ clustering by adopting the best-fitting model parameters. We first transform the Cartesian coordinates in the \textsc{UniverseMachine} catalogs into celestial coordinates and then apply the same geometry as in ALFALFA. The galaxy redshift is determined by including the distortion from the LOS velocity of each subhalo. The mock line width $W_{50}$ of the \hi\ profile for each galaxy is estimated to be $W_{50}=2V_{\rm max}\sin(i)$, where $V_{\rm max}$ is the maximum circular velocity of each halo/subhalo and $i$ is a random inclination angle. It is found in \cite{Chauhan2019} (their Fig.~5) that $W_{50}$ has a strong correlation with $V_{\rm max}$. But our results are not very sensitive to the details of the modeling $W_{50}$. We further apply the 90\% completeness limit of ALFALFA for the \hi\ flux at a given $W_{50}$ as in Eq.~(5) of \cite{Haynes2011}. We divide the Bolshoi-Planck simulation into 64 sub-boxes. By applying the periodic boundary condition, we then construct 64 mock catalogs by placing an observer at the center of each sub-box.  
	
\begin{figure*}
		\centering
		\includegraphics[width=0.9\textwidth]{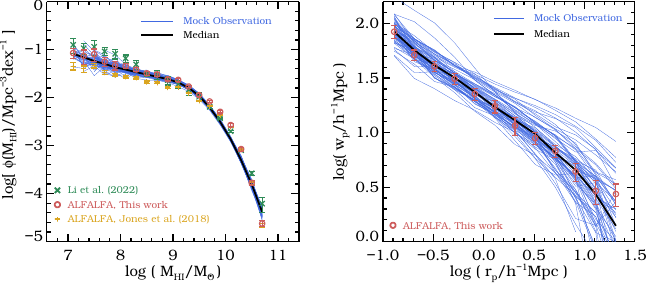}
		\caption{Left: Measurements of HIMF from 64 mock observations (blue lines). The median of the mocks is shown as the black line, whereas our corrected ALFALFA measurements are shown as the red open circles. For comparison, we also show the original ALFALFA measurements of \cite{Jones2018} (yellow pluses) using the 50\% completeness cut and the HIMF estimates (green crosses) of \cite{Li2022a} applying the \hi\ mass estimator on the SDSS optical galaxies. Right: Projected \hi\ clustering measurements $w_{\rm p}(r_{\rm p})$ for the mock observations (blue lines) and their median values (black line). The corresponding measurements for the ALFALFA sample with galaxies above the 90\% completeness limit are shown as the red open circles.} 
		\label{fig:mockhimf}
	\end{figure*}

	To measure the HIMF in each mock, we apply the 2DSWML algorithm as in observation to derive the effective volumes for mock galaxies \citep{Jones2018}. The HIMF $\phi(M_\hj)$ is then simply calculated as the sum of inverse volumes of galaxies in each $M_\hj$ bin. The resulting $\phi(M_\hj)$ for the 64 mocks are shown as the blue lines in the left panel of Figure~\ref{fig:mockhimf}. The median HIMF is shown as the black thick line, which agrees with our corrected ALFALFA HIMF (shown as red open circles) as expected. The scatters among the different mock HIMFs are relatively small, demonstrating that the cosmic variance effect is significantly suppressed using the 2DSWML method. For comparison, we also display the original ALFALFA HIMF of \cite{Jones2018} using the 50\% completeness cut as the yellow pluses. Although the two ALFALFA measurements agree with each other for $M_\hj>10^{9.5}\msun$, the low-mass end HIMF is enhanced by 50\% using the 90\% completeness cut. But the \cite{Jones2018} measurements are still within the range of mock variations. Using the comprehensive \hi\ mass estimator of \cite{Li2012}, \cite{Li2022a} also estimated the HIMF using the optical SDSS galaxies and correct for the cosmic variance effect using the method of \cite{Chen2019}. Their HIMF (green crosses) agrees with our measurements for $M_\hj>10^{8.4}\msun$, but slightly higher for smaller $M_\hj$. This difference is likely due to the fact that the underdense distribution of galaxies in the local universe would cause the galaxy stellar and \hi\ mass functions to be underestimated at the low mass end, as noted in \cite{Chen2019} and \cite{Li2022a}.
	
	In the right panel of Figure~\ref{fig:mockhimf}, we show the \hi\ clustering measurements of the 64 mocks (blue lines), as well as the median measurement (black line). The observed \hi\ clustering measurements of ALFALFA are shown as red open circles. Based on the method of \cite{Guo2017}, \cite{Li2022} measured the projected two-point correlation function (2PCF), $w_{\rm p}(r_{\rm p})$, for the \hi-selected galaxies in ALFALFA, where $r_{\rm p}$ is the projected separation of galaxy pairs and is measured in logarithmic bins of $\Delta\log r_{\rm p}=0.2$ covering $0.1\mpchi$--$25.1\mpchi$. As detailed in \cite{Li2022}, $w_{\rm p}(r_{\rm p})$ was obtained by integrating the 3D 2PCF $\xi(r_{\rm p},r_{\rm \pi})$ along the line of sight (LOS) to a distance of $r_{\rm\pi,max}=20\mpchi$ with $r_{\rm\pi}$ being the pair separation along LOS, i.e.,
	\begin{equation}
		w_{\rm p}(r_{\rm p})=2\int_0^{r_{\rm\pi,max}}\xi(r_{\rm p},r_{\rm \pi})d r_{\rm\pi},
	\end{equation}
	where $\xi(r_{\rm p},r_{\rm \pi})$ is measured using the Landy-Szalay estimator \citep{Landy1993}. To measure the clustering of \hi\ gas and correct for the selection bias of the ALFALFA galaxies, each galaxy pair (between galaxies $i$ and $j$) should be weighted by $M_{{\hj},i}M_{{\hj},j}/V_{ij}$, where $V_{ij}={\rm min}(V_{{\rm eff},i},V_{{\rm eff},j})$, with $V_{{\rm eff},i}$ and $V_{{\rm eff},j}$ being the effective volumes accessible to the galaxy pair. In this paper, we update their measurements by using the ALFALFA galaxies above the 90\% completeness limit. The effect of completeness cuts on the \hi\ clustering measurements is relatively minor. 
	
	The median $w_{\rm p}(r_{\rm p})$ measurements of the mock observations agree well with the observed ALFALFA \hi\ clustering. It demonstrates that the \hi\ clustering can be recovered with the information of the HIMF and \hi-halo mass relation. However, the cosmic variance effect is still very significant for clustering measurements even with effective volume weights, necessitating future \hi\ surveys of much larger volumes.	The best-fitting value of $\gamma$ ($2.233^{+0.037}_{-0.137}$) strongly favors the dependence of \hi\ clustering on the halo formation time, confirms the findings of \cite{Guo2017} and \cite{Li2022}. The $M_{\hj}$--$z_{\rm form}$ relation in our model is in agreement with the result of \cite{Guo2017} (their Fig.~12), where the $z_{\rm form}$ dependence is constrained with the clustering measurements in their extended subhalo abundance matching model.
	
	\subsection{Comparison with Literature}
	By fitting the observational constraints listed in Table~\ref{tab:obs}, our model is able to explain a set of important gas scaling relations at various redshifts. However, comparison to other measurements not used in the model fitting will provide an independent opportunity to further verify our model.
	
	\subsubsection{H{~\scriptsize I} Measurements for Satellites}\label{sec:otherHI}
	\begin{figure*}
		\centering
		\includegraphics[width=0.9\textwidth]{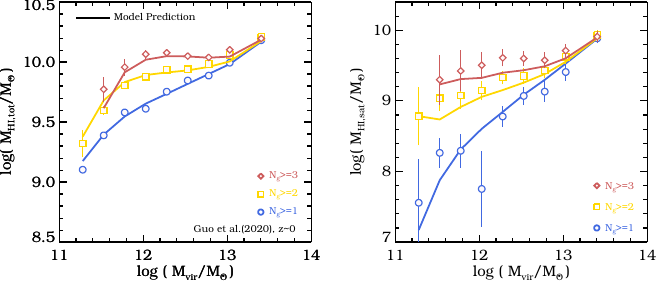}
		\caption{Left: Comparison between the observed $M_{\hj, {\rm tot}}$--$M_{\rm vir}$ relation and the model predictions for halos of different richness values ($N_{\rm g}$). The predictions of the model are shown as solid lines, while the observational measurements of \cite{Guo2020} are displayed as symbols of different colors. Right: Corresponding measurements for the total \hi\ masses of all satellite galaxies ($M_{\hj, {\rm sat}}$) in each halo. } 
		\label{fig:hihm_nsat}
	\end{figure*}
	
	\begin{figure}
		\centering
		\includegraphics[width=0.45\textwidth]{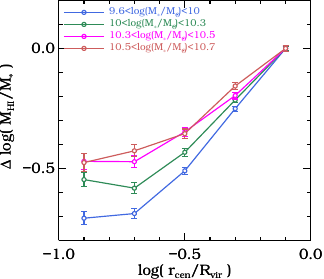}
		\caption{Differences in \hi\ fraction with respective to the values at virial radius, $\Delta\log(M_\hj/M_\ast)$, as a function of halo-centric distance $r_{\rm cen}$ scaled by the halo virial radius ($R_{\rm vir}$). We investigate the \hi\ fraction in the same four stellar mass bins as in \cite{Zhang2013}, represented by lines of different colors. } 
		\label{fig:hisatdrvir}
	\end{figure}
	\cite{Guo2020} found that the total $M_{\hj,{\rm tot}}$ in halos depends on both the halo mass and halo richness, with halos of higher richness possessing larger amount of \hi\ gas. They explain this behavior as the dependence of \hi\ mass on the halo assembly history. In the left panel of Figure~\ref{fig:hihm_nsat}, we compare our model predictions (solid lines) for halos of different richness values ($N_{\rm g}$) with measurements of \cite{Guo2020} (symbols of different colors). We follow \cite{Guo2020} by applying a stellar mass threshold of $M_\ast>10^9\msun$ to the \textsc{UniverseMachine} catalog, to mimic the optical flux limit of SDSS observations. The halo richness is then determined by calculating the number of galaxies (including the orphans) with $M_\ast>10^9\msun$ for each halo. We compare our model predictions to the observed measurements of \cite{Guo2020} after applying the confusion correction. Our model is in remarkably good agreement with observation. The slight overestimation of $M_{\hj,{\rm tot}}$ at $\log(M_{\rm vir}/\msun)\sim11.25$ in Figure~\ref{fig:hifit} is improved, because those halos with $N_{\rm g}=0$ (i.e., all member galaxies have masses smaller than $10^9\msun$) also form later than halos hosting more massive galaxies and thus have higher \hi\ masses. The trend of $M_{\hj,{\rm tot}}$ with halo richness is well reproduced. It was shown in \cite{Wechsler2006} that the halos with later formation time have higher richness. Our result further confirms that the dependence of \hi\ mass on halo richness originates from the dependence on halo formation time. 
	
	In our model constraints of \hi\ content at $z\sim0$, we include the \hi-halo mass relation for the total \hi\ gas and the \hi-stellar mass relation for central galaxies. We do not explicitly constrain the \hi\ content for satellite galaxies. It is helpful to investigate how well the model works for the satellite galaxies. In the right panel of Figure~\ref{fig:hihm_nsat}, we show the measurements of the total \hi\ mass contributed by the satellite galaxies (symbols of different colors for halos of different $N_{\rm g}$), measured as the subtraction between $M_{\hj,{\rm tot}}$ and $M_\hj$ of central galaxies in \cite{Guo2020}. Our model predictions are shown as the corresponding solid lines. Measurements and model predictions agree with each other for $N_{\rm g}\ge1$. The satellite \hi\ masses for halos of higher $N_{\rm g}$ are slightly underestimated in our model, but the agreement is still reasonable considering the large errors.
	
	\cite{Zhang2013} quantified the effect of \hi\ depletion for cluster galaxies using groups and clusters identified in SDSS. They found a smaller \hi\ fraction $M_\hj/M_\ast$ for satellite galaxies with smaller cluster-centric distances (their Fig.~5) \citep[see also,][]{Wang2020,Wang2021}. This effect is stronger for lower-mass galaxies. The changes of $M_\hj/M_\ast$ with cluster-centric distance are around 0.37~dex and 0.68~dex for $M_\ast\sim10^{10.6}\msun$ and $M_\ast\sim10^{9.8}\msun$, respectively. In our model, the gas depletion effect is naturally included along with the decrease of subhalo mass after infall. While the cluster-centric distances in \cite{Zhang2013} suffer from the redshift-space distortion effect, we can compare the relative change of $M_\hj/M_\ast$ for satellite galaxies from outer parts to inner parts. 
	
	In Figure~\ref{fig:hisatdrvir}, we show the differences in \hi\ fraction with respective to the values at the virial radius as a function of halo-centric distance ($r_{\rm cen}$) scaled by the virial radius ($R_{\rm vir}$). We adopt the same four stellar mass bins as in \cite{Zhang2013}. We find similar levels of decrease in $M_\hj/M_\ast$ between our model and \cite{Zhang2013} for all stellar mass bins, further confirming that the \hi\ depletion of satellite galaxies is reasonably reproduced in our empirical model. The distributions of \hi\ gas for satellite galaxies in halos of different masses can also be investigated using the conditional HIMF as in \cite{Li2022a}. We will compare the predictions of the conditional HIMF in our future work.
	
	\subsubsection{HIMF in Groups}
	\begin{figure}
		\centering
		\includegraphics[width=0.45\textwidth]{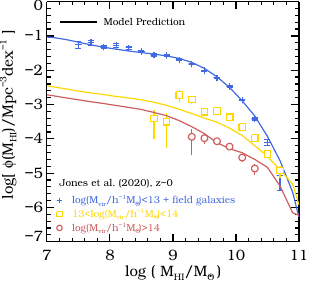}
		\caption{Comparison between the observed HIMFs for galaxies in groups of different halo masses from \cite{Jones2020} (symbols) (updated using ALFALFA galaxies above the 90\% completeness limit) and the model predictions (solid lines). } 
		\label{fig:himfgroup}
	\end{figure}
	\cite{Jones2020} measured the HIMF for galaxies residing in groups by making use of the overlapping regions between the ALFALFA survey and SDSS group catalogs. In this paper, we have updated their measurements by using the ALFALFA galaxies above 90\% completeness limit. However, the calculation of HIMF in \cite{Jones2020} is relative to the volume covered by the groups. For fair comparisons with the HIMF shown in Figure~\ref{fig:hifit}, we normalized their HIMF measurements using the volume of the entire galaxy sample. The galaxies in \cite{Jones2020} were separated into field galaxies and group galaxies of different halo masses. But the definition of field galaxies is somewhat ambiguous and mainly refers to isolated galaxies in low-mass halos, which might be affected by the flux limit of the SDSS sample. Therefore, we combine their HIMF measurements of field galaxies with those in halos of $M_{\rm vir}<10^{13}\msunh$, and compare them to the HIMF measurements for galaxies in more massive halos, shown as the colored symbols in Figure~\ref{fig:himfgroup}. Our model predictions are shown as solid lines with the corresponding colors. 
	
	We find reasonable agreement between observation and model for different halo mass bins. But our model predictions of HIMF in massive halos of $13<\log(M_{\rm vir}/\msunh)<14$ are slightly underestimated. Since more than $95\%$ of the \hi-selected galaxies in these halos are satellites. The misidentification of central and satellite galaxies in the SDSS group catalog, as well as the errors in the halo mass estimates, could easily cause the level of discrepancy seen in the figure \citep{Campbell2015}. We will explore more accurate models for satellite galaxies in our future work. But the overall trend of HIMF with the halo mass is reasonably reproduced in our model. \cite{Jones2020} pointed out that the quick drop-off in the measured HIMFs at the low mass end in massive halos is caused by the small volume of the ALFALFA survey. The effect is more severe for larger halos. As confirmed by our model prediction, the low-mass slopes of the HIMF measurements in different halos are quite similar. But HIMFs in massive halos have much shallower slopes at the massive end. 
	
	\subsubsection{H{~\scriptsize I} and H$_2$ Measurements at High Redshifts}\label{subsec:mh2z}
	\begin{figure*}
		\centering
		\includegraphics[width=0.9\textwidth]{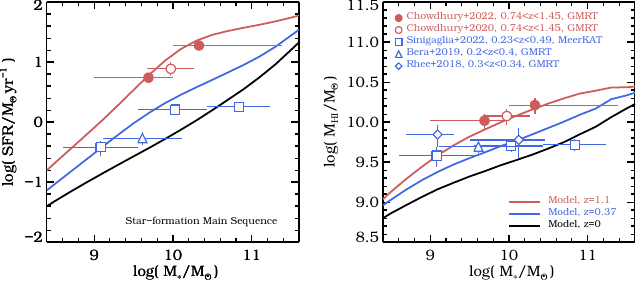}
		\caption{Comparison between observed data and best-fitting models for the SFR--$M_\ast$ (left) and \hi-$M_\ast$ relations (right) at $z\sim0.37$ and $z\sim1.1$. The fitted \hi-stack measurements of \cite{Chowdhury2022} are shown as filled cycles, while the other measurements of \cite{Chowdhury2020}, \cite{Rhee2018}, \cite{Bera2019} and \cite{Sinigaglia2022} are shown as open symbols. We display the best-fitting model at $z=0$ as the black solid lines for comparison.} 
		\label{fig:mhiz}
	\end{figure*}
	In the right panel of Figure~\ref{fig:mhiz}, we show our best-fitting model of $M_{\hj}$-$M_\ast$ relation at $z\sim0.37$ and $z\sim1.1$ for star-forming galaxies as the blue and red solid lines, respectively. The fitted \hi-stacking measurements of \cite{Chowdhury2022} are shown as filled circles. For comparison, we also display the \hi-stacking measurements of \cite{Chowdhury2020} at $z\sim1.1$ and those of \cite{Rhee2018}, \cite{Bera2019}, and \cite{Sinigaglia2022} at $z\sim0.37$ as open symbols. Except for the measurements of \cite{Sinigaglia2022} using the MeerKAT radio telescope, all other measurements were observed with GMRT. For comparison, the model predictions at $z=0$ are shown as the black solid line. Although our model predicts a consistent slope of the \hi-stellar mass relation at $0<z<1$, the observed $M_\hj$ measurements at $z\sim0.37$ seem to have a weak dependence on $M_\ast$, with $M_\hj\sim10^{9.7}\msun$.
	
	We note that all these measurements were made for the star-forming galaxies, as only these galaxies can have reliable measurements of \hi\ gas even with the stacking method. We show in the left panel of Figure~\ref{fig:mhiz} the available SFR measurements from the corresponding references using the same symbols, and the positions of the SFMS in the best-fitting models are shown as solid lines. In \cite{Sinigaglia2022} it is claimed that the slope of the \hi-stellar mass relation is becoming flat at the massive end at $z\sim0.37$, compared to the measurements of $z=0$. However, \cite{Chowdhury2022a} found no evolution in slope for the \hi-stellar mass relation by comparing measurements between $z=0$ and $z\sim1.1$ \citep[see also][]{Bera2023}. We emphasize that $M_\hj$ depends on both $M_\ast$ and SFR. It is still important to check their SFR distributions when comparing the measurements with the best-fitting models. Part of the decrease in $M_\hj$ at $M_\ast\sim10^{11}\msun$ seen for \cite{Sinigaglia2022} can be attributed to the lower SFR values relative to SFMS. High redshift measurements of \hi-stellar mass relation and HIMF \citep[e.g.,][]{Bera2022} are still scarce at the moment, but we expect the upcoming \hi\ surveys to provide more insight.
	
	\begin{figure}
		\centering
		\includegraphics[width=0.45\textwidth]{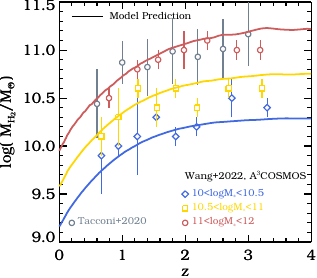}
		\caption{Comparison between observed H$_2$ gas masses of \cite{Wang2022} using the A$^3$COSMOS catalog (symbols) and our model predictions (solid lines) for main sequence galaxies. The measurements in different stellar mass bins are shown as symbols and lines of different colors. For comparison, the average H$_2$ masses derived from the compilation of measurements in the literature \citep{Tacconi2020} in the mass range of $11<\log(M_\ast/M_\sun)<12$ are also shown as the gray circles.} 
		\label{fig:mh2z_wang22}
	\end{figure}
	Using a mass-complete sample of main sequence galaxies in the archival Atacama Large Millimeter/submillimeter Array (ALMA) observations of COSMOS field \cite[A$^3$COSMOS;][]{Liu2019}, \cite{Wang2022} measured the redshift evolution of stacked molecular gas mass in different stellar mass bins. In Figure~\ref{fig:mh2z_wang22}, we show the comparison between their measurements (symbols) and our model predictions (solid lines) for main sequence galaxies ($|\log({\rm SFR}/{\rm SFR_{MS,obs}})|<0.5$). Measurements in different stellar mass bins are shown as different colors. 
	
	We find good agreement between their measurements and our model predictions in $0.5<z<3.5$. For comparison, we also derive the average H$_2$ masses for the most massive main-sequence galaxies ($11<\log(M_\ast/M_\sun)<12$) using the compilation of H$_2$ measurements from \cite{Tacconi2020}, as measurements for these massive galaxies are more complete and representative. They are shown as gray circles in Figure~\ref{fig:mh2z_wang22}, also in agreement with our model predictions. 
	
	It is interesting that at a given stellar mass bin, $M_{\rm H_2}$ is generally increasing with redshift, but the trend is becoming flatter at $z>2$. Combining with the fact that there are significantly less massive galaxies at higher redshifts \citep[e.g., Figure~2 of][]{Behroozi2019}, the cosmic molecular gas density $\rho_{\rm H_2}$ would then reach the peak at round $z\sim2$. It would be more intriguing to trace the evolution of gas densities along with the galaxy stellar mass growth, as will be shown in the following section.

	\subsection{Model Predictions}
	With the best-fitting model that is able to explain most \hi\ and H$_2$ observations, we can make valuable predictions about the evolution of the properties of cold gas for the upcoming surveys.
	
	\subsubsection{Cosmic Baryon Budget}
	\begin{figure}
		\centering
		\includegraphics[width=0.45\textwidth]{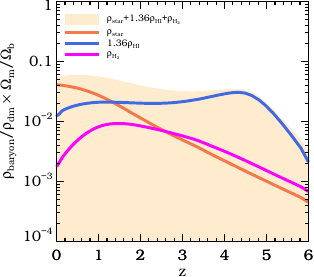}
		\caption{Fractional contributions of different baryon components to the total cosmic baryon density, $\rho_{\rm dm}\Omega_{\rm b}/\Omega_{\rm m}$. Contributions from stars, \hi\ and H$_2$ gas are shown as the orange, blue, and pink lines, respectively. The overall contribution of $\rho_{\rm baryon}=\rho_{\rm star}+1.36\rho_{\hj}+\rho_{\rm H_2}$ is displayed as the shaded area.} 
		\label{fig:baryonz}
	\end{figure}
	\begin{figure*}
		\centering
		\includegraphics[width=1\textwidth]{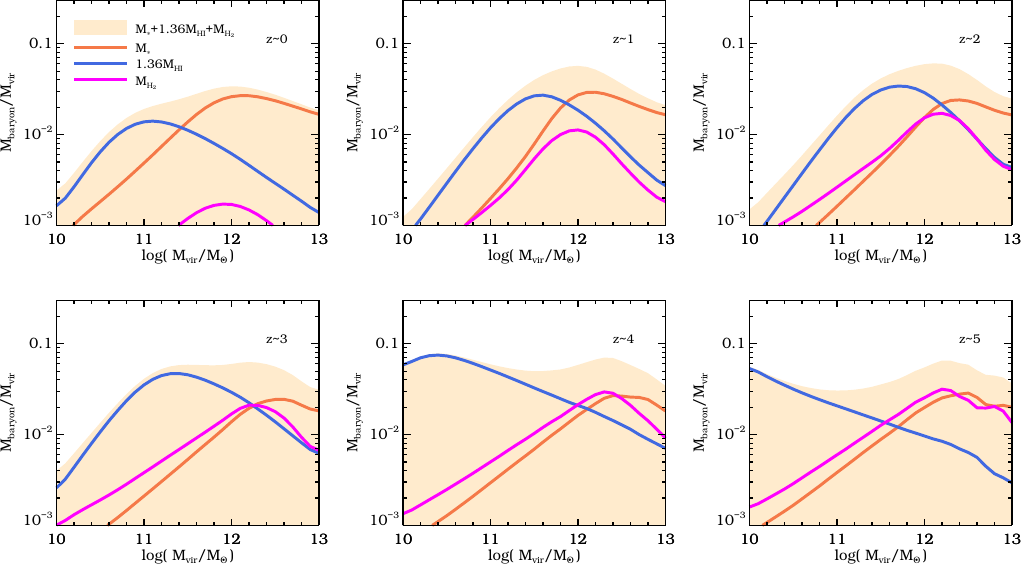}
		\caption{Similar to Figure~\ref{fig:baryonz}, but for the ratios between the masses of different baryon components and the halo mass $M_{\rm vir}$ at six typical redshifts.} 
		\label{fig:baryon}
	\end{figure*}
	Since our model includes the stellar, \hi\ and H$_2$ gas components, we are able to quantify the evolution of cosmic baryon density associated with galaxies, i.e., $\rho_{\rm baryon}=\rho_{\rm star}+1.36\rho_{\hj}+\rho_{\rm H_2}$. We have multiplied the \hi\ density by a factor of 1.36 to account for the contribution of helium and heavier elements. We can normalize the cosmic densities with $\rho_{\rm dm}\Omega_{\rm b}/\Omega_{\rm m}$ ($\rho_{\rm dm}$ is the dark matter density) to quantify the fractional contributions of different components to the total baryon density, as shown in Figure~\ref{fig:baryonz}. Different components are represented as lines of different colors, while the overall contribution of $\rho_{\rm baryon}$ is displayed as the shaded area. 
	
	It is clear that the baryon densities associated with galaxies only account for about 5\% of the total baryon budget at $z=0$ and this fraction is slowly decreasing at higher redshifts. Since our halo-based \hi\ model has already included the contribution of \hi\ gas in the circumgalactic medium (CGM), the majority baryons are in the form of hot gas distributed in the intergalactic medium (IGM) and CGM \citep[see more dicussions in][]{Tumlinson2017}.
	
	We note that $\rho_{\rm baryon}$ is dominated by \hi\ gas at $z>1.2$ and by stars thereafter. The contribution of $\rho_{\rm H_2}$ to $\rho_{\rm baryon}$ is generally less than $23\%$ and it is even dominating over $\rho_{\rm star}$ at $z>2.6$. At low redshifts of $z<1.2$, the decrease of \hi\ and H$_2$ gas is roughly comparable to stellar mass growth, which indicates the persistent conversion of cold gas into stars in a quasi-steady state \citep{Lilly2013}. 
	
	To better understand the roles of different components, we show in Figure~\ref{fig:baryon} the ratios between the masses of different components and $M_{\rm vir}$, as well as the combined contribution of stars and cold gas, $M_{\rm baryon}=M_\ast+1.36M_{\hj}+M_{\rm H_2}$, similarly as in Figure~\ref{fig:baryonz}. In the calculation of mass for each component, we have included the contribution from all galaxies in the halos. The ratios of $M_\ast/M_{\rm vir}$ and $M_{\rm H_2}/M_{\rm vir}$ peak around $M_{\rm vir}\sim10^{12}\msun$ and slightly shift toward more massive halos at higher redshifts. The two ratios also have similar shapes at all redshifts, arising from the tight correlation between $M_{\rm H_2}$ and $M_\ast$ in Eq.~\ref{eq:mh2}.
	
	However, the peak of $M_{\hj}/M_{\rm vir}$ occurs around $M_{\rm vir}\sim10^{11.4}\msun$ for $z<3$ and quickly moves to halos of lower mass at higher redshifts. It highlights the importance of using high-resolution simulations when modeling the \hi\ gas at high redshifts. With the particle mass resolution of $2.3\times10^8\msun$, the Bolshoi-Planck simulation is unable to correctly sample halos of $M_{\rm vir}<10^{10}\msun$ (i.e., with less than 50~particles). Our current model using Bolshoi-Planck simulation is less accurate for $z\sim5$, where the contribution of \hi\ gas to halos of $M_{\rm vir}<10^{10}\msun$ is significant.
	
	At higher redshifts, particularly above $z=2$, we only have constraints on the cosmic \hi\ and H$_2$ densities, as opposed to \hi\ or H$_2$ as a function of galaxy mass.  Therefore, it is useful to examine whether the relationship between the gas mass and the halo mass in Fig.~\ref{fig:baryon} continues to be reasonable at $z>2$.  At $z=3$--$5$, the H$_2$ mass scales roughly proportionally to the stellar mass, which is as expected for galaxies on a star-forming main sequence with sSFR roughly independent of stellar mass.  The behavior of \hi\ is more interesting, with the peak of $M_\hj/M_{\rm vir}$ dramatically shifting to lower halo masses, e.g., at $z=4$--$5$.  We tested many alternate forms for the redshift scaling of \hi\ mass with the halo mass, but all forms that matched the observational data required a similarly rapid shifting of the peak of $M_\hj/M_{\rm vir}$ to lower masses.  Fig.~\ref{fig:baryon} provides a simple explanation for this behavior: even with the rapid change in the shape of the $M_\hj$--$M_{\rm vir}$ relation, the total baryon fraction sits relatively close to the cosmic baryon fraction of 0.16. Therefore, to match the high observed cosmic density of \hi\ in DLAs, any less rapid shift in the peak of $M_\hj/M_{\rm vir}$ would result in halos where the total mass in \hi, H$_2$, and stars exceeded the cosmic baryon fraction. Put another way, the shape of the $M_\hj$--$M_{\rm vir}$ relation at $z>2$ is constrained above by the total cosmic baryon fraction and below by the high observed number density of DLAs at high redshifts.  Physically, we can interpret this as cooling becoming more and more efficient in high-redshift halos of all masses, such that a larger fraction of the available baryons in the halos are cooling to \hi\ and H$_2$. High star formation rates in massive halos (e.g., $M_{\rm vir}\sim10^{12}\msun$) mean that most of the gas in those halos is in the form of H$_2$ rather than \hi, and lower star formation rates in less massive halos mean that most of the gas in the lower mass halos will be in \hi\ instead of H$_2$.
	
	\subsubsection{Evolution of Cold Gas Mass Functions and Scaling Relations}
	
\begin{figure*}
		\centering
		\includegraphics[width=0.9\textwidth]{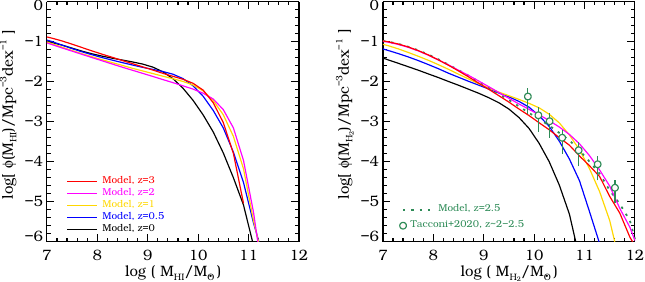}
		\caption{Best-fitting model predictions for the \hi\ (left panel) and H$_2$ mass functions (right panel) at $0<z<3$, shown as lines of different colors. For comparison, The collected measurements of H$_2$MF at $2<z<2.5$ from \cite{Tacconi2020} are shown as open circles in the right panel. Our model prediction of H$_2$MF at $z\sim2.5$ is shown as the dotted line in the right panel.} 
		\label{fig:hihighz}
	\end{figure*}

	In Figure~\ref{fig:hihighz}, we show the predictions of the best fit model for the evolution of \hi\ (left panel) and H$_2$ mass functions (right panel) from $z=3$ to $z=0$, shown as lines of different colors. We have included the predictions of the model at $z=0.5$ that could be compared with observations in the upcoming \hi\ surveys probing similar redshifts. The evolution of HIMF is relatively weak, and the trends at low- and high-mass ends vary with redshift. But generally we have a higher chance of detecting \hi-rich galaxies in the future deep \hi\ surveys than those in the local universe \citep[see e.g.,][]{Xi2021,Ponomareva2023}.
	
	There is also weak evolution of H$_2$MF for $M_{\rm H_2}<10^9\msun$, but the evolution is much stronger at the massive end. There would be many more galaxies with a large H$_2$ reservoir at high redshifts. For comparison, we show the collected measurements of H$_2$MF at $2<z<2.5$ from \cite{Tacconi2020}, which agrees with our model predictions at $z\sim2.5$ (shown as the dotted line).
	
	\begin{figure*}
		\centering
		\includegraphics[width=0.9\textwidth]{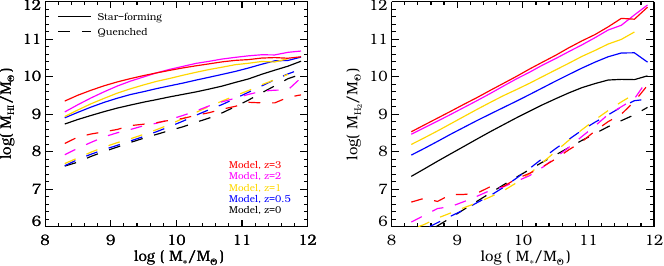}
		\caption{Best-fitting model predictions for the $M_{\hj}$--$M_\ast$ (left panel) and $M_{\rm H_2}$--$M_\ast$ relations (right panel) at $0<z<3$, shown as lines of different colors. The measurements of star-forming and quenched galaxies are displayed as solid and dashed lines, respectively.} 
		\label{fig:sfqg}
	\end{figure*}
	\begin{figure*}
		\centering
		\includegraphics[width=0.9\textwidth]{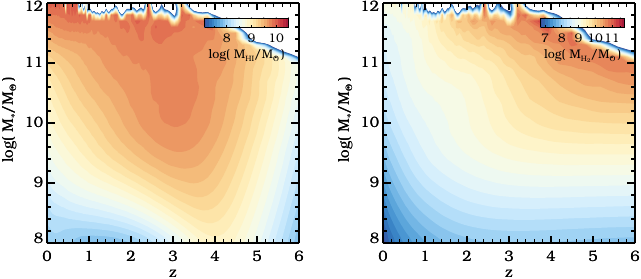}
		\caption{Average \hi\ (left panel) and H$_2$ masses (right panel) ($\langle\log M_\hj\rangle$ and $\langle\log M_{\rm H_2}\rangle$) as a function of $M_\ast$ and $z$. The values of $\langle\log M_\hj\rangle$ and $\langle\log M_{\rm H_2}\rangle$ are color coded as labeled.} 
		\label{fig:starhi}
	\end{figure*}
	In Figure~\ref{fig:sfqg}, we show the scaling relations of $M_{\hj}$--$M_\ast$ (left panel) and $M_{\rm H_2}$--$M_\ast$ (right panel) for the star-forming (solid lines) and quenched galaxies (dashed lines) at various redshifts as in Figure~\ref{fig:hihighz}. We show in Figure~\ref{fig:sfqg} the model predictions of $\log\langle M_{\hj}\rangle$ and $\log\langle M_{\rm H_2}\rangle$. Star-forming and quenched galaxies are separated by the demarcation line of $\log({\rm SFR}/{\rm SFR_{MS,obs}})=-1$. The measurements at $z=0$ are slightly different from the results in the bottom left panel of Figure~\ref{fig:hifit}, due to the different selection cuts of star-forming and quenched galaxies. 
	
	The \hi\ mass is generally increasing with $M_\ast$ and redshift, but the slope of the $M_{\hj}$--$M_\ast$ relation for massive galaxies becomes shallower at higher redshifts. Since in our model the slopes $\alpha$ and $\beta$ do not vary with redshift, the shallower slope is caused by the slope changes of the stellar-halo mass relation \citep{Behroozi2019}. The offsets of $M_{\hj}$ between star-forming and quenched galaxies become larger at higher redshifts. It means that the massive end of HIMF is mainly contributed by star-forming galaxies at high redshifts. It is then reasonable to target star-forming galaxies for \hi\ surveys at high redshifts. 
	
	The evolution of the $M_{\rm H_2}$--$M_\ast$ relation is similar to the case of \hi. But the dependence of $M_{\rm H_2}$ on $M_\ast$ for quenched galaxies has almost no evolution, with a slope of $M_{\rm H_2}\propto M_\ast$. The variation of $M_{\rm H_2}$ with redshift is larger for star-forming galaxies, but there is a very weak evolution of $M_{\rm H_2}$--$M_\ast$ relation for $z>2$. The change of $\rho_{\rm H_2}$ at these high redshifts is due to the evolution of the stellar mass function.
	
	For a complete description of the redshift evolution of the $M_\hj$--$M_\ast$ and $M_{\rm H_2}$--$M_\ast$ relations, we show in Figure~\ref{fig:starhi} the average \hi\ (left panel) and H$_2$ masses (right panel), $\langle\log M_{\hj}\rangle$ and $\langle\log M_{\rm H_2}\rangle$, as a function of $M_\ast$ and $z$. As shown in Figure~\ref{fig:sfqg}, the high mass end slope of the $M_\hj$--$M_\ast$ relation is significantly flatter at $z\sim3$, slightly before the peak of cosmic SFR density, likely indicating that the consumption of \hi\ gas is much higher than the cooling and cold gas accretion for these massive galaxies. The evolutionary trend for the $M_{\rm H_2}$--$M_\ast$ relation has a roughly constant slope, originating from the H$_2$ model in Eq.~(\ref{eq:mh2}).
	
	\begin{figure*}
		\centering
		\includegraphics[width=0.9\textwidth]{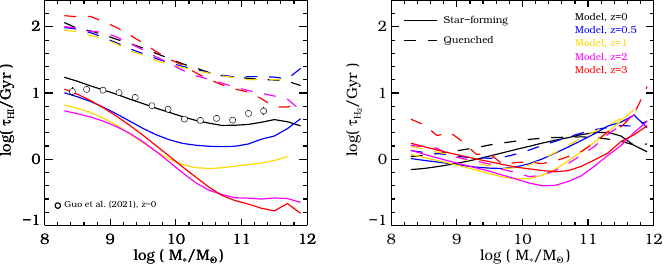}
		\caption{Similar to Figure~\ref{fig:sfqg}, but for the depletion time scales of \hi\ (left panel) and H$_2$ (right panel). For comparison, we show the measurements of depletion time scales for star-forming galaxies from \cite{Guo2021} at $z\sim0$ as the open circles. } 
		\label{fig:tau}
	\end{figure*}
	Another important parameter is the cold gas depletion time scale $\tau$. Similarly to Figure~\ref{fig:sfqg}, we show in Figure~\ref{fig:tau} the depletion time scales for \hi\ ($\log\tau_{\hj}\equiv\log\langle M_{\hj}\rangle-\langle\log{\rm SFR}\rangle$, left panel) and H$_2$ ($\log\tau_{\rm H_2}\equiv\log\langle M_{\rm H_2}\rangle-\langle\log{\rm SFR}\rangle$, right panel). We note that the \hi\ and H$_2$ masses are measured as stacked averages in each mass bin for direct comparisons with the \hi\ observations of \cite{Guo2021} at $z\sim0$ (shown as open circles). Massive galaxies typically have lower $\tau_{\hj}$ and the dependence on $M_\ast$ is stronger at higher redshifts. The \hi\ depletion time scale for star-forming galaxies varies from 0.1~Gyr to 10~Gyr, but that of the quenched galaxies is significantly longer. These quenched galaxies can hardly deplete their \hi\ reservoir (albeit small) with low SFRs. 
	
	The dependences of the H$_2$ depletion time scale $\tau_{\rm H_2}$ on $M_\ast$ and redshift are much weaker. The star-forming and quenched galaxies have similar depletion time scales around 0.3--3~Gyr and it is slightly decreasing with redshift, as also shown in \cite{Tacconi2020} (their Figure~3). Similar levels of $\tau_{\rm H_2}$ were also found in the empirical model of \cite{Padmanabhan2020}. The smallest $\tau_{\rm H_2}$ is found in galaxies of $\sim10^{10.5}\msun$ at $z=2$, indicating the fast consumption of molecular gas for the formation of stars.  We note that the lines in Figure~\ref{fig:tau} are the average values of the depletion time scales. The scatters for $\log\tau_{\hj}$ and $\log\tau_{\rm H_2}$ for individual galaxies are around 0.8~dex and 0.4~dex, respectively. 
	
	\subsubsection{Cold Gas Accretion Histories}
	\begin{figure*}
		\centering
		\includegraphics[width=0.9\textwidth]{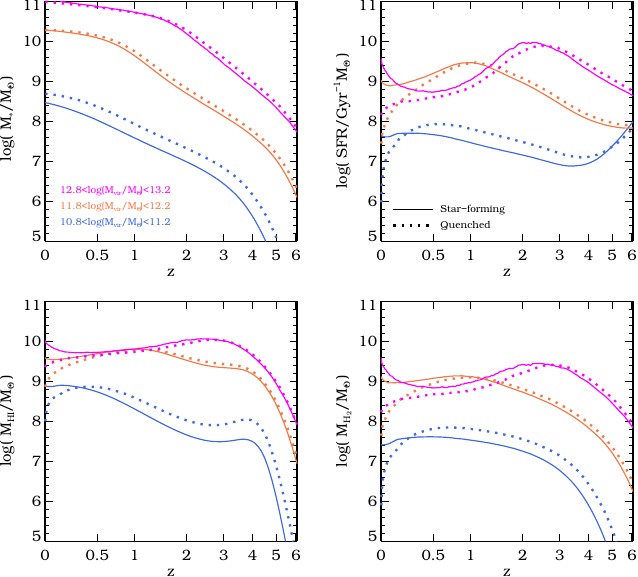}
		\caption{Best-fitting model predictions for the growth histories of central galaxies for the stellar mass (top left panel), SFR (top right panel, in units of $\msun/{\rm Gyr}$), \hi\ mass (bottom left panel) and H$_2$ mass (bottom right panel). The blue, orange and pink lines are for halos with $10.8<\log(M_{\rm vir}/\msun)<11.2$, $11.8<\log(M_{\rm vir}/\msun)<12.2$ and $12.8<\log(M_{\rm vir}/\msun)<13.2$ at $z=0$, respectively. The measurements for star-forming galaxies are shown as solid lines, while those of quenched galaxies are displayed as dotted lines.} 
		\label{fig:hievol}
	\end{figure*}
	Using the halo merger trees of the Bolshoi-Planck simulation, we are able to trace the evolution history of the cold gas content for individual halos. In Figure~\ref{fig:hievol}, we show the growth histories of central galaxies for the stellar mass (top left panel), SFR (top right panel), \hi\ mass (bottom left panel), and H$_2$ mass (bottom right panel). The blue, orange and pink lines are for halos with $10.8<\log(M_{\rm vir}/\msun)<11.2$, $11.8<\log(M_{\rm vir}/\msun)<12.2$ and $12.8<\log(M_{\rm vir}/\msun)<13.2$ at $z=0$, respectively. The measurements for star-forming galaxies are shown as solid lines, while those of quenched galaxies are displayed as dotted lines.
	
	The stellar mass growth histories for halos of different masses are quite similar. In low-mass halos of $M_{\rm vir}\sim10^{11}\msun$, the quenched galaxies have $\sim0.3$~dex higher $M_\ast$ than their star-forming counterparts. For more massive halos, there are no strong differences in the mass growth histories of the two populations. But the evolution of SFR varies significantly in halos of different masses. The SFR history peaks are shifting from low to high redshifts, with increasing $M_{\rm vir}$, as also seen in Figure~16 of \cite{Behroozi2019}. Significant differences between star-forming and quenched galaxies only happen since $z=0.5$ (i.e., within $\sim$5~Gyr of $z=0$). Along with the general decrease of SFR toward $z=0$ after reaching the peaks, the stellar mass growth becomes slower, shown as the shallower curves in the top left panel. Thus, even though the quenched galaxies have much smaller SFRs at $z=0$ than the star-forming ones, their final stellar masses are still quite similar. The higher $M_\ast$ for quenched galaxies at $z=0$ in low-mass halos is caused by their higher SFRs at $z>0.5$.
	
	When we compare the cold gas accretion histories shown in the bottom panels of Figure~\ref{fig:hievol}, it is clear that the evolution of molecular gas closely follows the star formation histories, as expected from the tight correlation in Equation~(\ref{eq:mh2}). The evolution of the \hi\ gas is also similar to that of molecular gas, but its dependence on star formation histories is much weaker, especially for massive halos. In halos of $M_{\rm vir}>10^{12}\msun$, most of the \hi\ gas has already been accreted much earlier than the redshifts of peak star formation histories, with a fast growing phase at $z>4$. The \hi\ gas will then be converted to H$_2$ and form stars. The star-forming galaxies in low-mass halos of $M_{\rm vir}\sim10^{11}\msun$ are still experiencing rapid accretion of \hi gas at low redshifts, leading to increasing SFRs toward $z=0$. 
	
	The time scale required to transform \hi\ to H$_2$ will increase the time lag between the peaks of \hi\ and H$_2$ accretion histories, which is about 0.8~Gyr for star-forming galaxies and 1~Gyr for quenched galaxies. The \hi\ gas starts to be depleted well before the star formation quenching occurs. The sharp decrease of $M_{\hj}$ for quenched galaxies at low redshifts could be caused by additional physical mechanisms, such as stellar and AGN feedback \citep{Guo2022,Ma2022}. 
	
	It is interesting that the SFRs of star-forming galaxies in massive halos of $M_{\rm vir}\sim10^{13}\msun$ will decrease from $z=2$ to $z=0.5$, but increase again since $z=0.5$. This is mainly due to the selection effect of star-forming galaxies. As most of the galaxies in these massive halos have low SFRs, selecting star-forming galaxies will tend to pick out those that have undergone recent rejuvenation effects, leading to the upturn seen in the star formation history \citep[see also, Fig.~16 of][]{Behroozi2019}. Yet, the SFRs for these galaxies are still below the SFMS and lying in the `green valley' region, consistent with the recent observations of rejuvenation events \citep{Chauke2019}. As seen in the bottom panels of Figure~\ref{fig:hievol}, the rejuvenation is accompanied by efficient accretion of cold gases in massive halos in the recent 5~Gyr.  
	
	\subsubsection{H{~\scriptsize I} Bias}
	\begin{figure*}
		\centering
		\includegraphics[width=0.9\textwidth]{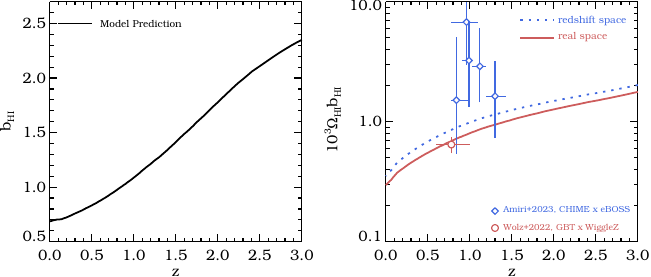}
		\caption{Best-fitting model predictions for the \hi\ bias (left panel) and the combination of $\Omega_\hj b_\hj$ (right panel) at $0<z<3$. The solid and dotted lines in the right panel are for the model predictions in the real and redshift spaces, respectively. We also show for comparison the measurements of \cite{Wolz2022} and \cite{Amiri2023} as different symbols.} 
		\label{fig:hibias}
	\end{figure*}
	Our halo-based \hi\ model can be used to predict the evolution of \hi\ bias, as well as making predictions for the future \hi\ intensity mapping surveys. The power spectrum measurements of \hi\ intensity mapping experiments constrain the product of \hi\ bias ($b_{\hj}$) and cosmic \hi\ abundance ($\Omega_{\hj}\equiv\rho_{\hj}/\rho_{\rm c}$), where $\rho_{\rm c}$ is the critical density at $z=0$. Cross-correlations between \hi\ signals and optical galaxy samples have frequently been applied in intensity mapping measurements \citep{Chang2010,Masui2013,Switzer2013,Amiri2023,Wolz2022,Cunnington2023}, because autocorrelation measurements would suffer from more severe systematic uncertainties.
	
	In our model, we can calculate the \hi\ bias as, 
	\begin{equation}
		b_{\hj}(r)=\sqrt{\frac{\xi_{\hj}(r)}{\xi_{\rm dm}(r)}},
	\end{equation}
	where $r$ is the separation between galaxy pairs. $\xi_{\hj}(r)$ is the \hi\ mass weighted real-space galaxy two-point correlation function and $\xi_{\rm dm}(r)$ is the corresponding measurement for dark matter. To calculate the large-scale linear \hi\ bias, we use the correlation functions in $3.25\mpchi<r<12.92\mpchi$ to derive the average $b_{\hj}$.  
	
	In the left panel of Figure~\ref{fig:hibias}, we show the model prediction of $b_{\hj}$ as the solid line. The \hi\ bias gradually increases from $0.69$ at $z=0$ to 2.33 at $z=3$, which is in agreement with the semi-analytical model prediction of \cite{WangZY2021}. The critical value of $b_{\hj}=1$ is reached at $z\sim0.85$. As shown in \cite{Guo2017}, the clustering amplitudes of \hi-selected galaxies will increase with $M_{\hj}$. The overall \hi\ bias then depends on the galaxy population hosting the \hi\ gas. Therefore, correctly modeling the \hi-halo mass relation is essential to obtain the accurate \hi\ bias.
	
	Our model prediction for the combined value of $\Omega_\hj b_\hj$ probed by the \hi\ intensity mapping experiments is shown in the right panel of Figure~\ref{fig:hibias} as the solid line. We also show for comparison the recent measurements of \cite{Wolz2022} and \cite{Amiri2023} as symbols of different colors. 
	
	\begin{figure*}
		\centering
		\includegraphics[width=0.9\textwidth]{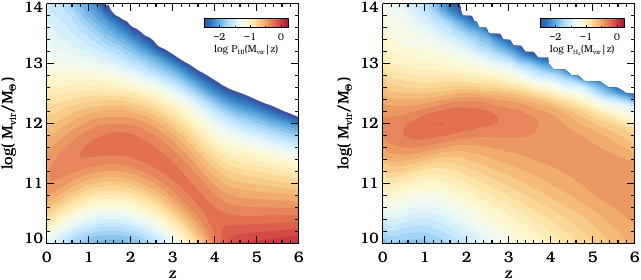}
		\caption{Host halo mass probability distributions of a random \hi\ atom (left panel) and a random H$_2$ molecule (right panel) in the universe. We show the logarithmic probabilities $\log P_\hj(M_{\rm vir}|z)$ and $\log P_{\rm H_2}(M_{\rm vir}|z)$ as contours of different colors. See the text for details. } 
		\label{fig:halohi}
	\end{figure*}
	\cite{Wolz2022} cross-correlated the \hi\ intensity mapping data from the Green Bank Telescope (GBT) with the optical galaxies from WiggleZ Dark Energy Survey \citep{Drinkwater2010} in the redshift range of $0.6<z<1$, based on the work of \cite{Masui2013}. They found that $\Omega_\hj b_\hj r_\hj=(0.58\pm0.14)\times10^{-3}$ on the effective scale $k_{\rm eff}=0.31h\,{\rm Mpc}^{-1}$, where $r_\hj$ is the galaxy-\hi\ cross-correlation coefficient. The main uncertainty lies in the estimate of $r_\hj$, which depends on both the \hi\ and the galaxy samples. We adopt their estimate of $r_\hj=0.9$ and derive the combination of $\Omega_\hj b_\hj=(0.64\pm0.16)\times10^{-3}$ (consistent with the measurements of \citealt{Switzer2013}), which shows good agreement with the solid line.
	
	\cite{Amiri2023} made the first \hi\ intensity mapping measurements with the interferometer from the Canadian Hydrogen Intensity Mapping Experiment \citep[CHIME;][]{Chime2022}. By cross-correlating the \hi\ data with luminous red galaxies (LRGs), emission line galaxies (ELGs) and quasars (QSOs) from the eBOSS sample \citep{Dawson2016}, they can constrain the effective \hi\ clustering amplitude $\mathcal{A}_\hj\equiv10^3\Omega_\hj(b_\hj+\langle f\mu^2\rangle)$, where the term $\langle f\mu^2\rangle$ is to account for the redshift-space distortion effect in observation, with $f$ the linear growth rate and $\mu$ the cosine of angle between the wavevector $\mathbf{k}$ and line of sight. They measured $\mathcal{A}_\hj$ for LRGs ($\langle z\rangle=0.84$), ELGs ($\langle z\rangle=0.96$) and QSOs ($\langle z\rangle=1.20$) that provide constraints at higher redshifts than previous measurements. For fair comparisons, we present our model prediction of $\Omega_\hj b_\hj$ for \hi\ bias measured in the redshift space as dotted line in the right panel of Figure~\ref{fig:hibias}. The redshift-space distortion effect would increase the measured $b_\hj$ by about 20\%. The measurements of \cite{Amiri2023} are slightly higher than the predictions of our model, but generally in agreement within the errors. As discussed in \cite{Amiri2023}, the higher amplitude of $\mathcal{A}_\hj$ is likely caused by the degeneracy between $\mathcal{A}_\hj$ and their model parameters that describe the Fingers-of-God effect on non-linear scales.
	
	To better understand the distributions of \hi\ and H$_2$ in the universe, we show in Figure~\ref{fig:halohi} the probability distributions of \hi\ and H$_2$ gas in halos of different masses, i.e., $P_\hj(M_{\rm vir}|z)$ (left panel) and $P_{\rm H_2}(M_{\rm vir}|z)$ (right panel). At each redshift, we calculate the probabilities as follows,
	\begin{eqnarray}
		P_{\hj}(M_{\rm vir}|z)&=& \frac{M_{\hj}dn}{d\log M_{\rm vir}} \\
		P_{\rm H_2}(M_{\rm vir}|z)&=& \frac{M_{\rm H_2}dn}{d\log M_{\rm vir}},
	\end{eqnarray}   
	where $dn/d\log M_{\rm vir}$ is the normalized halo mass function. Therefore, $P_\hj(M_{\rm vir}|z)$ and $P_{\rm H_2}(M_{\rm vir}|z)$ indicate the host halo mass distributions of any given \hi\ atom and H$_2$ molecule, respectively. As shown in Figure~\ref{fig:halohi}, the majority of \hi\ gas is distributed in halos of $10^{11}$--$10^{12}\msun$ in $0<z<3$. The probability $P_\hj(M_{\rm vir}|z=0)$ peaks at $M_{\rm vir}\sim10^{11.3}\msun$, corresponding to a halo bias of $b\sim0.7$ \citep{Guo2017}, which is in agreement with the \hi\ bias in Figure~\ref{fig:hibias}. At higher redshifts, the peak halo mass of the \hi\ gas gradually moves to lower values. Our model based on Bolshoi-Planck simulation is not accurate enough at $z>5$, as the peak halo mass drops below $10^{10}\msun$, as also indicated in Figure~\ref{fig:baryon}.
	
	The peak halo mass for the H$_2$ molecule is higher than that of \hi. It is around $M_{\rm vir}\sim10^{12}\msun$ at $0<z<4$, which is consistent with the host-halo distributions of the cosmic SFR \citep[][their Figure~13]{Behroozi2019}. We would expect that the H$_2$ bias is higher than the \hi bias, which can be verified in future H$_2$ surveys. The resolution of Bolshoi-Planck simulation is also good enough to model the H$_2$ gas.
	
	\section{Discussion}\label{sec:discussion}
	Besides the halo properties of $M_{\rm vir}$ and $z_{\rm form}$ that can be easily measured in the $N$-body simulations, our cold gas model requires additional information of $M_\ast$ and ${\rm SFR}$. All these galaxy and halo properties can be found in the \textsc{UniverseMachine} catalog, making it a perfect starting point to apply the gas model. However, the \textsc{NeutralUniverseMachine} model can also be applied to other theoretical models that include relevant information, e.g., semi-analytical models and hydrodynamical simulations. The parameter ${\rm SFR_{MS,obs}}$ in our model can be replaced with the corresponding SFMS in different models. The exact functional form of ${\rm SFR_{MS,obs}}$ as in Eq.~\ref{eq:sfrobs} is not essential for cold gas, since only the offset from SFMS is used to scale the \hi\ and H$_2$ masses. 
	
	Although our empirical model lacks detailed baryon physics, it has the advantage of correctly describing the observed cold gas properties, and the model constraints can be further improved with future observations. Since cold gas serves as the fuel for galactic star formation, it is very sensitive to the complicated physical processes involved in galaxy formation and evolution. Matching the current observational measurements of \hi\ and H$_2$ gas is still challenging for theoretical models.
	
	The first step is to match the observed HIMF and H$_2$MF. As shown in \cite{Dave2020}, current hydrodynamical simulations can reasonably fit the $z=0$ HIMF for $M_\hj>10^{10}\msun$, but the discrepancies are greater for smaller \hi\ masses \citep[e.g.,][]{Crain2017,Diemer2019}. The situation is similar when comparing the semi-analytical models that are based on the $N$-body simulations and incorporate simple recipes to describe the baryon physics in the galaxy formation and evolution \citep[e.g.,][]{Fu2013,Popping2014,Kim2017,Baugh2019,Chauhan2020,Spinelli2020}. The comparison with observed H$_2$MF is relatively better \citep[see e.g.,][]{Lagos2015,Diemer2019,Dave2020}, since the molecular gas is determined mainly by stellar mass and SFR. 
	
	The \hi-halo mass and \hi-stellar mass relations provide more stringent tests of theoretical models. The HIMF and H$_2$MF set the probability distributions of cold gas in the universe, while the scaling relations with stellar and halo masses are more sensitive to the evolution of the baryon cycle. The \hi\ content in massive halos of $M_{\rm vir}>10^{12}\msun$ varies significantly in different hydrodynamical simulations and semi-analytical models \citep[e.g.,][]{Villaescusa2018,Baugh2019,Obuljen2019,Chauhan2020,Spinelli2020,Li2022}, due to the various levels of gas accretion, heating and cooling. Comparing the gas scaling relations for star-forming and quenched galaxies also helps distinguish the dependence of cold gas on the global SFR \citep{Ma2022}.
	
	The inclusion of halo formation time $z_{\rm form}$ is an essential component of our empirical model, which was not considered in previous theoretical models. The \hi\ clustering measurements $w_{\rm p}(r_{\rm p})$ provide further verification, as they are sensitive to the parameter $\gamma$ related to the halo formation time. Accurate measurements of \hi-halo and \hi-stellar mass relations of stacked \hi\ signals have already placed tight constraints on $\gamma$. If we fix $\gamma=0$, i.e., ignoring the dependence of $M_\hj$ on $z_{\rm form}$, the \hi\ content in massive halos would be overestimated and the shape of the \hi-stellar mass relation would also be inconsistent with observation. Furthermore, without dependence on $z_{\rm form}$, the \hi\ content in halos of different richness would be quite similar to each other, different from the results shown in Figure~\ref{fig:hihm_nsat}.
	
	Overall, the empirical model is very powerful in accurately capturing the dependence of cold gas content on the various galaxy and halo properties, which is not a simple task for the methods of hydrodynamical simulations and semi-analytical models. But it is still important to understand the physical processes that determine the cold gas content. The semi-empirical approach proposed by \cite{Popping2015} shows some success in the attempt to combine the empirical model of galaxy star formation histories with a physically motivated cold gas model. Despite the failure to describe the observed HIMF at $z\sim0$ for $M_\hj<10^{10}\msun$ (caused by the overcorrection for the number of low-mass galaxies in \citealt{Behroozi2013}) and the cosmic evolution of $\rho_{\hj}$, it is still worth the effort to explore such an approach by employing more accurate physical models in the future. 
	
	\section{Conclusions}\label{sec:conclusion}
	In this paper, we propose a new empirical model that is capable of accurately describing the various statistics for the \hi\ and H$_2$ gas content in the redshift range of $0<z<6$. The functional form of the empirical model is motivated by the various observations describing the scaling relations between cold gas mass and properties of their host galaxies and halos. Our results are summarized as follows.
	
	(i) Our empirical model can accurately describe the \hi\ and H$_2$ mass functions, molecular-to-atomic mass ratio, \hi-halo mass relation, \hi- and H$_2$-stellar mass relations, and \hi\ clustering measurements at $z\sim0$. Higher redshift measurements of the \hi-stellar mass relations, as well as the cosmic gas densities of $\rho_{\hj}$ and $\rho_{\rm H_2}$, are also well reproduced. Our best-fitting model is further verified with a few sets of \hi\ and H$_2$ measurements not used in the modeling constraints and shows good agreement with all these different observations. 
	
	(ii) There is only weak evolution in the HIMF from $z=0$ to $z=3$. However, the evolution of H$_2$MF is significantly larger at the massive end and smaller for $M_{\rm H_2}<10^9\msun$. The average $M_\hj$ and $M_{\rm H_2}$ increase by around 1~dex from $z=0$ to $z=3$ for star-forming galaxies, but there is much weaker evolution for the quenched population.
	
	(iii) The \hi\ gas depletion time $\tau_{\hj}$ generally decreases with increasing stellar mass, and varies from 0.1~Gyr to 10~Gyr for the star-forming galaxies. The H$_2$ gas depletion time $\tau_{\rm H_2}$ has a weaker dependence on the redshift and stellar mass. The quenched galaxies have much longer \hi\ gas depletion time, varying from 10~Gyr to 200~Gyr, i.e. they are not likely to fully deplete their \hi\ reservoir with the low SFRs. 
	
	(iv) From the growth histories of the galaxy stellar mass, SFR, \hi\ and H$_2$ masses in halos of different masses, we find that $M_{\rm H_2}$ closely trace the evolution of the SFR, but the correlation between $M_\hj$ and SFR is weaker. There is also an apparent time lag between the evolution trends of $M_\hj$ and SFR, with $M_\hj$ reaching the peaks earlier than SFR.
	
	(v) The cosmic baryon density associated with galaxies is dominated by stars for $z<1.2$, and mainly contributed by \hi\ gas at higher redshifts. But they only account for less than 5\% of the total baryon budget. The ratios of $M_\ast/M_{\rm vir}$ and $M_{\rm H_2}/M_{\rm vir}$ closely follow each other and reach the peaks around $M_{\rm vir}\sim10^{12}\msun$, while $M_\hj/M_{\rm vir}$ peaks around $M_{\rm vir}\sim10^{11.4}\msun$ for $z<3$ and shifts to lower-mass halos at higher redshifts.
	
	(vi) Our model can predict the evolution of \hi\ clustering in the universe. The \hi\ bias $b_\hj$ is gradually evolving from $0.69$ at $z=0$ to $2.33$ at $z=3$. The combined value $\Omega_\hj b_\hj$ increases from $0.24\times10^{-3}$ to $1.81\times10^{-3}$ in the same redshift range and shows good agreement with recent \hi\ intensity mapping measurements.
	
	\begin{acknowledgments}
		We thank the anonymous reviewer for helpful suggestions that significantly improve the presentation of this paper. 
		This work is supported by the National SKA Program of China (grant No. 2020SKA0110100), National Science Foundation of China (Nos. 11922305, 11833005, 12073002, 11721303, 12011530159) and the science research grants from the China Manned Space Project with NOs. CMS-CSST-2021-A02 and CMS-CSST-2021-B01. We thank Toby Brown, Nissim Kanekar, and  Kasper E. Heintz for helpful discussions. We acknowledge the use of the High Performance Computing Resource in the Core Facility for Advanced Research Computing at the Shanghai Astronomical Observatory.
	\end{acknowledgments}
	
	\appendix
	\section{ALFALFA H{~\scriptsize I} Mass Function with 90\% Completeness Cut}\label{sec:himf}
	
	\begin{table}
		\caption{ALFALFA \hi\ Mass Function with 90\% Completeness}
		\label{tab:himf}
		\centering
		\begin{tabular}{cccc}
			\hline
			$\log M_\hj$ & $\phi(M_\hj)$ & Error on $\phi(M_\hj)$ & $N_{\rm gal}$\\
			$\msun$ & $\rm{Mpc^{-3}dex^{-1}}$ &  $\rm{Mpc^{-3}dex^{-1}}$ & \\
			\hline
			7.1 & $8.509\times10^{-2}$ & $2.038\times10^{-2}$ & 41 \\
			7.3 & $7.433\times10^{-2}$ & $2.450\times10^{-2}$ & 52\\
			7.5 & $8.045\times10^{-2}$ & $2.298\times10^{-2}$ & 106\\
			7.7 & $5.629\times10^{-2}$ & $1.252\times10^{-2}$ & 134\\
			7.9 & $4.705\times10^{-2}$ & $0.943\times10^{-2}$ & 172\\
			8.1 & $4.581\times10^{-2}$ & $0.586\times10^{-2}$ & 245\\
			8.3 & $3.851\times10^{-2}$ & $0.395\times10^{-2}$ & 308\\
			8.5 & $3.214\times10^{-2}$ & $0.279\times10^{-2}$ & 362\\
			8.7 & $3.049\times10^{-2}$ & $0.252\times10^{-2}$ & 568\\
			8.9 & $2.412\times10^{-2}$ & $0.193\times10^{-2}$ & 827\\
			9.1 & $2.346\times10^{-2}$ & $0.153\times10^{-2}$ & 1545\\
			9.3 & $1.707\times10^{-2}$ & $0.109\times10^{-2}$ & 2174\\		
			9.5 & $1.132\times10^{-2}$ & $0.072\times10^{-2}$ & 2679\\
			9.7 & $8.335\times10^{-3}$ & $0.464\times10^{-3}$ & 3479\\
			9.9 & $5.079\times10^{-3}$ & $0.297\times10^{-3}$ & 3595\\
			10.1 & $2.687\times10^{-3}$ & $0.162\times10^{-3}$ & 2795\\
			10.3 & $8.495\times10^{-4}$ & $0.816\times10^{-4}$ & 1058\\
			10.5 & $1.687\times10^{-4}$ & $0.202\times10^{-4}$ & 219\\
			10.7 & $2.386\times10^{-5}$ & $0.357\times10^{-5}$ & 31 		\\																	
			\hline
		\end{tabular} 
		\tablecomments{The HIMF measurements are measured in logarithmic $M_\hj$ bins of 0.2 dex, with the bin centers indicated in the first column. $N_{\rm gal}$ is the number of galaxies in each $M_\hj$ bin.}
	\end{table}
	We show in Table~\ref{tab:himf} the corrected ALFALFA HIMF using galaxies above the 90\% completeness cut. The number of galaxies in each \hi\ mass bin is also listed in the last column. Even by applying the stricter cut, there are still fair amount of galaxies in the ALFALFA final sample to achieve an accurate estimation of the HIMF. To better compare with the previous ALFALFA HIMF of \cite{Jones2018}, we also fit a Schechter function to our measurements, 
	\begin{equation}
		\phi(M_\hj)=\ln(10)\phi_{\rm s}\left(\frac{M_\hj}{M_{\rm s}}\right)^{\alpha_{\rm s}+1}\exp(-\frac{M_\hj}{M_{\rm s}}).
	\end{equation}
	
	Our best-fitting parameters are $\alpha_{\rm s}=-1.30\pm0.02$, $\log(M_{\rm s}/\msun)=9.91\pm0.01$, $\phi_{\rm s}=5.93\pm0.29\times10^{-3}{\rm Mpc}^{-3}{\rm dex}^{-1}$. The corresponding parameters in \cite{Jones2018} are $\alpha_{\rm s}=-1.25\pm0.02$, $\log(M_{\rm s}/\msun)=9.94\pm0.01$, $\phi_{\rm s}=4.5\pm0.2\times10^{-3}{\rm Mpc}^{-3}{\rm dex}^{-1}$. In the corrected HIMF, the low-mass end slope $\alpha_{\rm s}$ is becoming slightly steeper and the `knee' mass is shifting to lower values.
	
	\begin{figure}
		\centering
		\includegraphics[width=0.45\textwidth]{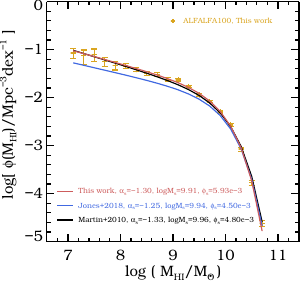}
		\caption{Comparisons of Schechter function fittings of \cite{Martin2010} (black line), \cite{Jones2018} (blue line) and our results (red line). The corrected HIMF for the ALFALFA final sample is also shown as yellow symbols. } 
		\label{fig:himf}
	\end{figure}
	In Figure~\ref{fig:himf}, we show comparisons of the Schechter function fittings between our results (red line) and that of \cite{Jones2018} (blue line), as well as the measurements of \cite{Martin2010} (black line) using the previous ALFALFA release covering 40\% of the final sample. \cite{Martin2010} also applied the 2DSWML method to derive the HIMF, but imposed a strict 100\% completeness threshold, similar to our 90\% completeness cut. 
	
	Our model fitting is in good agreement with that of \cite{Martin2010} \citep[see also,][]{Oman2022}, while the fitting from \cite{Jones2018} is systematically lower by around 50\% for $M_\hj<10^9\msun$. It emphasizes the importance of selecting complete samples or correcting for incompleteness when applying the 2DSWML method. However, we also note that there is still a large difference ($\sim30\%$) between the HIMF measurements for the ALFALFA spring and fall regions at the low-mass end, even with the 2DSWML method \citep[see e.g., Fig.~3 of][]{Jones2018}. Larger samples in future \hi\ surveys are essential to fully reduce the impact of cosmic variance.
	
	The cosmic \hi\ abundance $\Omega_\hj$ can be obtained from integrating the Schechter function \citep{Martin2010,Jones2018},
	\begin{equation}
		\Omega_\hj=\frac{1}{\rho_{\rm c}}\int M_\hj\phi(M_\hj)dM_\hj=\frac{\phi_{\rm s}}{\rho_{\rm c}}M_{\rm s}\Gamma(\alpha_{\rm s}+2),
	\end{equation}
	which gives $\Omega_\hj=(4.55\pm0.29)\times10^{-4}$ for our best-fitting parameters assuming $h=0.7$, which is about 29\% higher than the corresponding value ($3.5\times10^{-4}$) of \cite{Jones2018} before the correction of \hi\ self-absorption. 
	
	\section{Probability Distributions of Cold Gas Model Parameters}\label{sec:para}
	\begin{figure*}
		\centering
		\includegraphics[width=0.9\textwidth]{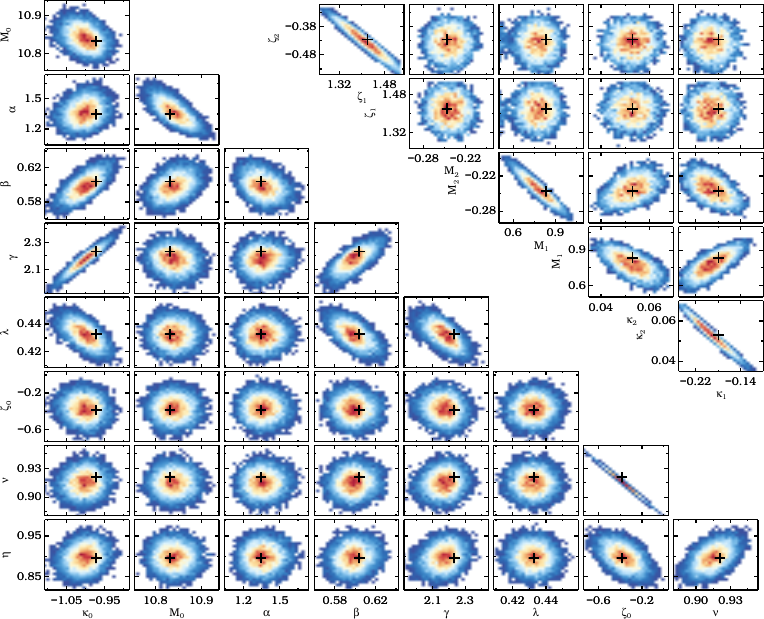}
		\caption{Pairwise Density distributions of the \textsc{NeutralUniverseMachine} model parameters of the MCMC runs. We separate the model parameters for $z=0$ (bottom left) and higher redshifts (top right), as they are not jointly fitted. The best-fitting model parameters are shown as the black plus in each panel. The distributions are color-coded by the number of galaxies in each grid, with redder colors for higher densities.} 
		\label{fig:para}
	\end{figure*}
	In Figure~\ref{fig:para}, we show the pairwise density distributions of the \textsc{NeutralUniverseMachine} model parameters from the MCMC runs. The model parameters for $z=0$ and higher redshifts are shown in the bottom left and top right, respectively. Correlations are generally weak between different parameters. But there are relatively strong correlations for the $\kappa_0$--$\gamma$, $\zeta_0$--$\nu$, $\kappa_1$--$\kappa_2$, $M_1$--$M_2$, and $\zeta_1$--$\zeta_2$ pairs. The correlation between $\kappa_0$ and $\gamma$ is caused by the small errors in the HIMF and \hi-halo mass relation. The distributions between $\zeta_0$ and $\nu$ can be well described by a tight linear relation of $\zeta_0=9.40-10.63\nu$, which is mainly constrained by small errors in the stacked H$_2$-stellar mass relations of \cite{Saintonge2017}. Similarly, the strong correlations in $\kappa_1$--$\kappa_2$, $M_1$--$M_2$, and $\zeta_1$--$\zeta_2$ pairs are caused by the stacked \hi\ measurements of \cite{Chowdhury2022} at $z\sim1.1$. Future surveys of more accurate \hi\ and H$_2$ density measurements will provide tighter constraints on the model parameters, as well as on the functional form of the redshift evolution. 
	\bibliographystyle{aasjournal}

\begin{thebibliography}{}
	\expandafter\ifx\csname natexlab\endcsname\relax\def\natexlab#1{#1}\fi
	\providecommand{\url}[1]{\href{#1}{#1}}
	\providecommand{\dodoi}[1]{doi:~\href{http://doi.org/#1}{\nolinkurl{#1}}}
	\providecommand{\doeprint}[1]{\href{http://ascl.net/#1}{\nolinkurl{http://ascl.net/#1}}}
	\providecommand{\doarXiv}[1]{\href{https://arxiv.org/abs/#1}{\nolinkurl{https://arxiv.org/abs/#1}}}
	
	\bibitem[{{Accurso} {et~al.}(2017){Accurso}, {Saintonge}, {Catinella},
		{Cortese}, {Dav{\'e}}, {Dunsheath}, {Genzel}, {Gracia-Carpio}, {Heckman},
		{Jimmy}, {Kramer}, {Li}, {Lutz}, {Schiminovich}, {Schuster}, {Sternberg},
		{Sturm}, {Tacconi}, {Tran}, \& {Wang}}]{Accurso2017}
	{Accurso}, G., {Saintonge}, A., {Catinella}, B., {et~al.} 2017, \mnras, 470,
	4750, \dodoi{10.1093/mnras/stx1556}
	
	\bibitem[{{Adams} {et~al.}(2022){Adams}, {Adebahr}, {de Blok}, {D{\'e}nes},
		{Hess}, {van der Hulst}, {Kutkin}, {Lucero}, {Morganti}, {Moss}, {Oosterloo},
		{Orr{\'u}}, {Schulz}, {van Amesfoort}, {Berger}, {Boersma}, {Bouwhuis}, {van
			den Brink}, {van Cappellen}, {Connor}, {Coolen}, {Damstra}, {van Diepen},
		{Dijkema}, {Ebbendorf}, {Grange}, {de Goei}, {Gunst}, {Holties}, {Hut},
		{Ivashina}, {J{\'o}zsa}, {van Leeuwen}, {Loose}, {Maan}, {Mancini}, {Mika},
		{Mulder}, {Norden}, {Offringa}, {Oostrum}, {Pastor-Marazuela}, {Pisano},
		{Ponomareva}, {Romein}, {Ruiter}, {Schoenmakers}, {van der Schuur}, {Sluman},
		{Smits}, {Stuurwold}, {Verstappen}, {Vilchez}, {Vohl}, {Wierenga},
		{Wijnholds}, {Woestenburg}, {Zanting}, \& {Ziemke}}]{Adams2022}
	{Adams}, E.~A.~K., {Adebahr}, B., {de Blok}, W.~J.~G., {et~al.} 2022, \aap,
	667, A38, \dodoi{10.1051/0004-6361/202244007}
	
	\bibitem[{{Amiri} {et~al.}(2023){Amiri}, {Bandura}, {Chen}, {Deng}, {Dobbs},
		{Fandino}, {Foreman}, {Halpern}, {Hill}, {Hinshaw}, {H{\"o}fer}, {Kania},
		{Landecker}, {MacEachern}, {Masui}, {Mena-Parra}, {Milutinovic},
		{Mirhosseini}, {Newburgh}, {Ordog}, {Pen}, {Pinsonneault-Marotte}, {Polzin},
		{Reda}, {Renard}, {Shaw}, {Siegel}, {Singh}, {Vanderlinde}, {Wang}, {Wiebe},
		{Wulf}, \& {CHIME Collaboration}}]{Amiri2023}
	{Amiri}, M., {Bandura}, K., {Chen}, T., {et~al.} 2023, \apj, 947, 16,
	\dodoi{10.3847/1538-4357/acb13f}
	
	\bibitem[{{Andreani} {et~al.}(2020){Andreani}, {Miyamoto}, {Kaneko}, {Boselli},
		{Tatematsu}, {Sorai}, \& {Vio}}]{Andreani2020}
	{Andreani}, P., {Miyamoto}, Y., {Kaneko}, H., {et~al.} 2020, \aap, 643, L11,
	\dodoi{10.1051/0004-6361/202038675}
	
	\bibitem[{{Barnes} {et~al.}(2001){Barnes}, {Staveley-Smith}, {de Blok},
		{Oosterloo}, {Stewart}, {Wright}, {Banks}, {Bhathal}, {Boyce}, {Calabretta},
		{Disney}, {Drinkwater}, {Ekers}, {Freeman}, {Gibson}, {Green}, {Haynes}, {te
			Lintel Hekkert}, {Henning}, {Jerjen}, {Juraszek}, {Kesteven}, {Kilborn},
		{Knezek}, {Koribalski}, {Kraan-Korteweg}, {Malin}, {Marquarding}, {Minchin},
		{Mould}, {Price}, {Putman}, {Ryder}, {Sadler}, {Schr{\"o}der}, {Stootman},
		{Webster}, {Wilson}, \& {Ye}}]{Barnes2001}
	{Barnes}, D.~G., {Staveley-Smith}, L., {de Blok}, W.~J.~G., {et~al.} 2001,
	\mnras, 322, 486, \dodoi{10.1046/j.1365-8711.2001.04102.x}
	
	\bibitem[{{Baugh} {et~al.}(2019){Baugh}, {Gonzalez-Perez}, {Lagos}, {Lacey},
		{Helly}, {Jenkins}, {Frenk}, {Benson}, {Bower}, \& {Cole}}]{Baugh2019}
	{Baugh}, C.~M., {Gonzalez-Perez}, V., {Lagos}, C. d.~P., {et~al.} 2019, \mnras,
	483, 4922, \dodoi{10.1093/mnras/sty3427}
	
	\bibitem[{{Behroozi} {et~al.}(2019){Behroozi}, {Wechsler}, {Hearin}, \&
		{Conroy}}]{Behroozi2019}
	{Behroozi}, P., {Wechsler}, R.~H., {Hearin}, A.~P., \& {Conroy}, C. 2019,
	\mnras, 488, 3143, \dodoi{10.1093/mnras/stz1182}
	
	\bibitem[{{Behroozi} {et~al.}(2013{\natexlab{a}}){Behroozi}, {Wechsler}, \&
		{Conroy}}]{Behroozi2013}
	{Behroozi}, P.~S., {Wechsler}, R.~H., \& {Conroy}, C. 2013{\natexlab{a}}, \apj,
	770, 57, \dodoi{10.1088/0004-637X/770/1/57}
	
	\bibitem[{{Behroozi} {et~al.}(2013{\natexlab{b}}){Behroozi}, {Wechsler}, \&
		{Wu}}]{Behroozi2013a}
	{Behroozi}, P.~S., {Wechsler}, R.~H., \& {Wu}, H.-Y. 2013{\natexlab{b}}, \apj,
	762, 109, \dodoi{10.1088/0004-637X/762/2/109}
	
	\bibitem[{{Behroozi} {et~al.}(2013{\natexlab{c}}){Behroozi}, {Wechsler}, {Wu},
		{Busha}, {Klypin}, \& {Primack}}]{Behroozi2013b}
	{Behroozi}, P.~S., {Wechsler}, R.~H., {Wu}, H.-Y., {et~al.} 2013{\natexlab{c}},
	\apj, 763, 18, \dodoi{10.1088/0004-637X/763/1/18}
	
	\bibitem[{{Bera} {et~al.}(2019){Bera}, {Kanekar}, {Chengalur}, \&
		{Bagla}}]{Bera2019}
	{Bera}, A., {Kanekar}, N., {Chengalur}, J.~N., \& {Bagla}, J.~S. 2019, \apjl,
	882, L7, \dodoi{10.3847/2041-8213/ab3656}
	
	\bibitem[{{Bera} {et~al.}(2022){Bera}, {Kanekar}, {Chengalur}, \&
		{Bagla}}]{Bera2022}
	---. 2022, \apjl, 940, L10, \dodoi{10.3847/2041-8213/ac9d32}
	
	\bibitem[{{Bera} {et~al.}(2023){Bera}, {Kanekar}, {Chengalur}, \&
		{Bagla}}]{Bera2023}
	---. 2023, \apjl, 950, L18, \dodoi{10.3847/2041-8213/acd0b3}
	
	\bibitem[{{Berg} {et~al.}(2019){Berg}, {Ellison}, {S{\'a}nchez-Ram{\'\i}rez},
		{L{\'o}pez}, {D'Odorico}, {Becker}, {Christensen}, {Cupani}, {Denney}, \&
		{Worseck}}]{Berg2019}
	{Berg}, T. A.~M., {Ellison}, S.~L., {S{\'a}nchez-Ram{\'\i}rez}, R., {et~al.}
	2019, \mnras, 488, 4356, \dodoi{10.1093/mnras/stz2012}
	
	\bibitem[{{Berta} {et~al.}(2016){Berta}, {Lutz}, {Genzel},
		{F{\"o}rster-Schreiber}, \& {Tacconi}}]{Berta2016}
	{Berta}, S., {Lutz}, D., {Genzel}, R., {F{\"o}rster-Schreiber}, N.~M., \&
	{Tacconi}, L.~J. 2016, \aap, 587, A73, \dodoi{10.1051/0004-6361/201527746}
	
	\bibitem[{{Bigiel} {et~al.}(2008){Bigiel}, {Leroy}, {Walter}, {Brinks}, {de
			Blok}, {Madore}, \& {Thornley}}]{Bigiel2008}
	{Bigiel}, F., {Leroy}, A., {Walter}, F., {et~al.} 2008, \aj, 136, 2846,
	\dodoi{10.1088/0004-6256/136/6/2846}
	
	\bibitem[{{Bird} {et~al.}(2017){Bird}, {Garnett}, \& {Ho}}]{Bird2017}
	{Bird}, S., {Garnett}, R., \& {Ho}, S. 2017, \mnras, 466, 2111,
	\dodoi{10.1093/mnras/stw3246}
	
	\bibitem[{{Blitz} \& {Rosolowsky}(2006)}]{Blitz2006}
	{Blitz}, L., \& {Rosolowsky}, E. 2006, \apj, 650, 933, \dodoi{10.1086/505417}
	
	\bibitem[{{Boogaard} {et~al.}(2023){Boogaard}, {Decarli}, {Walter}, {Wei{\ss}},
		{Popping}, {Neri}, {Aravena}, {Riechers}, {Ellis}, {Carilli}, {Cox}, \&
		{Pety}}]{Boogaard2023}
	{Boogaard}, L.~A., {Decarli}, R., {Walter}, F., {et~al.} 2023, \apj, 945, 111,
	\dodoi{10.3847/1538-4357/acb4f0}
	
	\bibitem[{{Brown} {et~al.}(2017){Brown}, {Catinella}, {Cortese}, {Lagos},
		{Dav{\'e}}, {Kilborn}, {Haynes}, {Giovanelli}, \&
		{Rafieferantsoa}}]{Brown2017}
	{Brown}, T., {Catinella}, B., {Cortese}, L., {et~al.} 2017, \mnras, 466, 1275,
	\dodoi{10.1093/mnras/stw2991}
	
	\bibitem[{{Bryan} \& {Norman}(1998)}]{Bryan1998}
	{Bryan}, G.~L., \& {Norman}, M.~L. 1998, \apj, 495, 80, \dodoi{10.1086/305262}
	
	\bibitem[{{Calette} {et~al.}(2021{\natexlab{a}}){Calette}, {Avila-Reese},
		{Rodr{\'\i}guez-Puebla}, {Lagos}, \& {Catinella}}]{Calette2021}
	{Calette}, A.~R., {Avila-Reese}, V., {Rodr{\'\i}guez-Puebla}, A., {Lagos}, C.
	d.~P., \& {Catinella}, B. 2021{\natexlab{a}}, \mnras, 505, 304,
	\dodoi{10.1093/mnras/stab1282}
	
	\bibitem[{{Calette} {et~al.}(2021{\natexlab{b}}){Calette},
		{Rodr{\'\i}guez-Puebla}, {Avila-Reese}, \& {Lagos}}]{Calette2021a}
	{Calette}, A.~R., {Rodr{\'\i}guez-Puebla}, A., {Avila-Reese}, V., \& {Lagos},
	C. d.~P. 2021{\natexlab{b}}, \mnras, 506, 1507,
	\dodoi{10.1093/mnras/stab1788}
	
	\bibitem[{{Campbell} {et~al.}(2015){Campbell}, {van den Bosch}, {Hearin},
		{Padmanabhan}, {Berlind}, {Mo}, {Tinker}, \& {Yang}}]{Campbell2015}
	{Campbell}, D., {van den Bosch}, F.~C., {Hearin}, A., {et~al.} 2015, \mnras,
	452, 444, \dodoi{10.1093/mnras/stv1091}
	
	\bibitem[{{Carilli} \& {Walter}(2013)}]{Carilli2013}
	{Carilli}, C.~L., \& {Walter}, F. 2013, \araa, 51, 105,
	\dodoi{10.1146/annurev-astro-082812-140953}
	
	\bibitem[{{Catinella} {et~al.}(2010){Catinella}, {Schiminovich}, {Kauffmann},
		{Fabello}, {Wang}, {Hummels}, {Lemonias}, {Moran}, {Wu}, {Giovanelli},
		{Haynes}, {Heckman}, {Basu-Zych}, {Blanton}, {Brinchmann}, {Budav{\'a}ri},
		{Gon{\c{c}}alves}, {Johnson}, {Kennicutt}, {Madore}, {Martin}, {Rich},
		{Tacconi}, {Thilker}, {Wild}, \& {Wyder}}]{Catinella2010}
	{Catinella}, B., {Schiminovich}, D., {Kauffmann}, G., {et~al.} 2010, \mnras,
	403, 683, \dodoi{10.1111/j.1365-2966.2009.16180.x}
	
	\bibitem[{{Catinella} {et~al.}(2018){Catinella}, {Saintonge}, {Janowiecki},
		{Cortese}, {Dav{\'e}}, {Lemonias}, {Cooper}, {Schiminovich}, {Hummels},
		{Fabello}, {Ger{\'e}b}, {Kilborn}, \& {Wang}}]{Catinella2018}
	{Catinella}, B., {Saintonge}, A., {Janowiecki}, S., {et~al.} 2018, \mnras, 476,
	875, \dodoi{10.1093/mnras/sty089}
	
	\bibitem[{{Chang} {et~al.}(2010){Chang}, {Pen}, {Bandura}, \&
		{Peterson}}]{Chang2010}
	{Chang}, T.-C., {Pen}, U.-L., {Bandura}, K., \& {Peterson}, J.~B. 2010, \nat,
	466, 463, \dodoi{10.1038/nature09187}
	
	\bibitem[{{Chauhan} {et~al.}(2019){Chauhan}, {Lagos}, {Obreschkow}, {Power},
		{Oman}, \& {Elahi}}]{Chauhan2019}
	{Chauhan}, G., {Lagos}, C. d.~P., {Obreschkow}, D., {et~al.} 2019, \mnras, 488,
	5898, \dodoi{10.1093/mnras/stz2069}
	
	\bibitem[{{Chauhan} {et~al.}(2020){Chauhan}, {Lagos}, {Stevens}, {Obreschkow},
		{Power}, \& {Meyer}}]{Chauhan2020}
	{Chauhan}, G., {Lagos}, C. d.~P., {Stevens}, A. R.~H., {et~al.} 2020, \mnras,
	498, 44, \dodoi{10.1093/mnras/staa2251}
	
	\bibitem[{{Chauke} {et~al.}(2019){Chauke}, {van der Wel}, {Pacifici},
		{Bezanson}, {Wu}, {Gallazzi}, {Straatman}, {Franx}, {Bari{\v{s}}i{\'c}},
		{Bell}, {van Houdt}, {Maseda}, {Muzzin}, {Sobral}, \& {Spilker}}]{Chauke2019}
	{Chauke}, P., {van der Wel}, A., {Pacifici}, C., {et~al.} 2019, \apj, 877, 48,
	\dodoi{10.3847/1538-4357/ab164d}
	
	\bibitem[{{Chen} {et~al.}(2019){Chen}, {Mo}, {Li}, {Wang}, {Yang}, {Zhou}, \&
		{Zhang}}]{Chen2019}
	{Chen}, Y., {Mo}, H.~J., {Li}, C., {et~al.} 2019, \apj, 872, 180,
	\dodoi{10.3847/1538-4357/ab0208}
	
	\bibitem[{{CHIME Collaboration} {et~al.}(2022){CHIME Collaboration}, {Amiri},
		{Bandura}, {Boskovic}, {Chen}, {Cliche}, {Deng}, {Denman}, {Dobbs},
		{Fandino}, {Foreman}, {Halpern}, {Hanna}, {Hill}, {Hinshaw}, {H{\"o}fer},
		{Kania}, {Klages}, {Landecker}, {MacEachern}, {Masui}, {Mena-Parra},
		{Milutinovic}, {Mirhosseini}, {Newburgh}, {Nitsche}, {Ordog}, {Pen},
		{Pinsonneault-Marotte}, {Polzin}, {Reda}, {Renard}, {Shaw}, {Siegel},
		{Singh}, {Smegal}, {Tretyakov}, {van Gassen}, {Vanderlinde}, {Wang}, {Wiebe},
		{Willis}, \& {Wulf}}]{Chime2022}
	{CHIME Collaboration}, {Amiri}, M., {Bandura}, K., {et~al.} 2022, \apjs, 261,
	29, \dodoi{10.3847/1538-4365/ac6fd9}
	
	\bibitem[{{Chowdhury} {et~al.}(2022{\natexlab{a}}){Chowdhury}, {Kanekar}, \&
		{Chengalur}}]{Chowdhury2022}
	{Chowdhury}, A., {Kanekar}, N., \& {Chengalur}, J.~N. 2022{\natexlab{a}},
	\apjl, 931, L34, \dodoi{10.3847/2041-8213/ac6de7}
	
	\bibitem[{{Chowdhury} {et~al.}(2022{\natexlab{b}}){Chowdhury}, {Kanekar}, \&
		{Chengalur}}]{Chowdhury2022a}
	---. 2022{\natexlab{b}}, \apjl, 941, L6, \dodoi{10.3847/2041-8213/ac9d8a}
	
	\bibitem[{{Chowdhury} {et~al.}(2020){Chowdhury}, {Kanekar}, {Chengalur},
		{Sethi}, \& {Dwarakanath}}]{Chowdhury2020}
	{Chowdhury}, A., {Kanekar}, N., {Chengalur}, J.~N., {Sethi}, S., \&
	{Dwarakanath}, K.~S. 2020, \nat, 586, 369, \dodoi{10.1038/s41586-020-2794-7}
	
	\bibitem[{{Crain} {et~al.}(2017){Crain}, {Bah{\'e}}, {Lagos}, {Rahmati},
		{Schaye}, {McCarthy}, {Marasco}, {Bower}, {Schaller}, {Theuns}, \& {van der
			Hulst}}]{Crain2017}
	{Crain}, R.~A., {Bah{\'e}}, Y.~M., {Lagos}, C. d.~P., {et~al.} 2017, \mnras,
	464, 4204, \dodoi{10.1093/mnras/stw2586}
	
	\bibitem[{{Crighton} {et~al.}(2015){Crighton}, {Murphy}, {Prochaska},
		{Worseck}, {Rafelski}, {Becker}, {Ellison}, {Fumagalli}, {Lopez}, {Meiksin},
		\& {O'Meara}}]{Crighton2015}
	{Crighton}, N. H.~M., {Murphy}, M.~T., {Prochaska}, J.~X., {et~al.} 2015,
	\mnras, 452, 217, \dodoi{10.1093/mnras/stv1182}
	
	\bibitem[{{Cunnington} {et~al.}(2023){Cunnington}, {Li}, {Santos}, {Wang},
		{Carucci}, {Irfan}, {Pourtsidou}, {Spinelli}, {Wolz}, {Soares}, {Blake},
		{Bull}, {Engelbrecht}, {Fonseca}, {Grainge}, \& {Ma}}]{Cunnington2023}
	{Cunnington}, S., {Li}, Y., {Santos}, M.~G., {et~al.} 2023, \mnras, 518, 6262,
	\dodoi{10.1093/mnras/stac3060}
	
	\bibitem[{{Darvish} {et~al.}(2018){Darvish}, {Scoville}, {Martin}, {Mobasher},
		{Diaz-Santos}, \& {Shen}}]{Darvish2018}
	{Darvish}, B., {Scoville}, N.~Z., {Martin}, C., {et~al.} 2018, \apj, 860, 111,
	\dodoi{10.3847/1538-4357/aac836}
	
	\bibitem[{{Dav{\'e}} {et~al.}(2020){Dav{\'e}}, {Crain}, {Stevens}, {Narayanan},
		{Saintonge}, {Catinella}, \& {Cortese}}]{Dave2020}
	{Dav{\'e}}, R., {Crain}, R.~A., {Stevens}, A. R.~H., {et~al.} 2020, \mnras,
	497, 146, \dodoi{10.1093/mnras/staa1894}
	
	\bibitem[{{Dawson} {et~al.}(2016){Dawson}, {Kneib}, {Percival}, {Alam},
		{Albareti}, {Anderson}, {Armengaud}, {Aubourg}, {Bailey}, {Bautista},
		{Berlind}, {Bershady}, {Beutler}, {Bizyaev}, {Blanton}, {Blomqvist},
		{Bolton}, {Bovy}, {Brandt}, {Brinkmann}, {Brownstein}, {Burtin}, {Busca},
		{Cai}, {Chuang}, {Clerc}, {Comparat}, {Cope}, {Croft}, {Cruz-Gonzalez}, {da
			Costa}, {Cousinou}, {Darling}, {de la Macorra}, {de la Torre}, {Delubac}, {du
			Mas des Bourboux}, {Dwelly}, {Ealet}, {Eisenstein}, {Eracleous}, {Escoffier},
		{Fan}, {Finoguenov}, {Font-Ribera}, {Frinchaboy}, {Gaulme}, {Georgakakis},
		{Green}, {Guo}, {Guy}, {Ho}, {Holder}, {Huehnerhoff}, {Hutchinson}, {Jing},
		{Jullo}, {Kamble}, {Kinemuchi}, {Kirkby}, {Kitaura}, {Klaene}, {Laher},
		{Lang}, {Laurent}, {Le Goff}, {Li}, {Liang}, {Lima}, {Lin}, {Lin}, {Lin},
		{Long}, {Lundgren}, {MacDonald}, {Geimba Maia}, {Malanushenko},
		{Malanushenko}, {Mariappan}, {McBride}, {McGreer}, {M{\'e}nard}, {Merloni},
		{Meza}, {Montero-Dorta}, {Muna}, {Myers}, {Nandra}, {Naugle}, {Newman},
		{Noterdaeme}, {Nugent}, {Ogando}, {Olmstead}, {Oravetz}, {Oravetz},
		{Padmanabhan}, {Palanque-Delabrouille}, {Pan}, {Parejko}, {P{\^a}ris},
		{Peacock}, {Petitjean}, {Pieri}, {Pisani}, {Prada}, {Prakash}, {Raichoor},
		{Reid}, {Rich}, {Ridl}, {Rodriguez-Torres}, {Carnero Rosell}, {Ross},
		{Rossi}, {Ruan}, {Salvato}, {Sayres}, {Schneider}, {Schlegel}, {Seljak},
		{Seo}, {Sesar}, {Shandera}, {Shu}, {Slosar}, {Sobreira}, {Streblyanska},
		{Suzuki}, {Taylor}, {Tao}, {Tinker}, {Tojeiro}, {Vargas-Maga{\~n}a}, {Wang},
		{Weaver}, {Weinberg}, {White}, {Wood-Vasey}, {Yeche}, {Zhai}, {Zhao}, {Zhao},
		{Zheng}, {Ben Zhu}, \& {Zou}}]{Dawson2016}
	{Dawson}, K.~S., {Kneib}, J.-P., {Percival}, W.~J., {et~al.} 2016, \aj, 151,
	44, \dodoi{10.3847/0004-6256/151/2/44}
	
	\bibitem[{{Decarli} {et~al.}(2016){Decarli}, {Walter}, {Aravena}, {Carilli},
		{Bouwens}, {da Cunha}, {Daddi}, {Ivison}, {Popping}, {Riechers}, {Smail},
		{Swinbank}, {Weiss}, {Anguita}, {Assef}, {Bauer}, {Bell}, {Bertoldi},
		{Chapman}, {Colina}, {Cortes}, {Cox}, {Dickinson}, {Elbaz},
		{G{\'o}nzalez-L{\'o}pez}, {Ibar}, {Infante}, {Hodge}, {Karim}, {Le Fevre},
		{Magnelli}, {Neri}, {Oesch}, {Ota}, {Rix}, {Sargent}, {Sheth}, {van der Wel},
		{van der Werf}, \& {Wagg}}]{Decarli2016}
	{Decarli}, R., {Walter}, F., {Aravena}, M., {et~al.} 2016, \apj, 833, 69,
	\dodoi{10.3847/1538-4357/833/1/69}
	
	\bibitem[{{Decarli} {et~al.}(2019){Decarli}, {Walter},
		{G{\'o}nzalez-L{\'o}pez}, {Aravena}, {Boogaard}, {Carilli}, {Cox}, {Daddi},
		{Popping}, {Riechers}, {Uzgil}, {Weiss}, {Assef}, {Bacon}, {Bauer},
		{Bertoldi}, {Bouwens}, {Contini}, {Cortes}, {da Cunha}, {D{\'\i}az-Santos},
		{Elbaz}, {Inami}, {Hodge}, {Ivison}, {Le F{\`e}vre}, {Magnelli}, {Novak},
		{Oesch}, {Rix}, {Sargent}, {Smail}, {Swinbank}, {Somerville}, {van der Werf},
		{Wagg}, \& {Wisotzki}}]{Decarli2019}
	{Decarli}, R., {Walter}, F., {G{\'o}nzalez-L{\'o}pez}, J., {et~al.} 2019, \apj,
	882, 138, \dodoi{10.3847/1538-4357/ab30fe}
	
	\bibitem[{{Delhaize} {et~al.}(2013){Delhaize}, {Meyer}, {Staveley-Smith}, \&
		{Boyle}}]{Delhaize2013}
	{Delhaize}, J., {Meyer}, M.~J., {Staveley-Smith}, L., \& {Boyle}, B.~J. 2013,
	\mnras, 433, 1398, \dodoi{10.1093/mnras/stt810}
	
	\bibitem[{{Dev} {et~al.}(2023){Dev}, {Driver}, {Meyer}, {Roychowdhury}, {Rhee},
		{Stevens}, {Lagos}, {Bland-Hawthorn}, {Catinella}, {Hopkins}, {Loveday},
		{Obreschkow}, {Phillipps}, \& {Robotham}}]{Dev2023}
	{Dev}, A., {Driver}, S.~P., {Meyer}, M., {et~al.} 2023, \mnras, 523, 2693,
	\dodoi{10.1093/mnras/stad1575}
	
	\bibitem[{{Diemer} {et~al.}(2019){Diemer}, {Stevens}, {Lagos}, {Calette},
		{Tacchella}, {Hernquist}, {Marinacci}, {Nelson}, {Pillepich},
		{Rodriguez-Gomez}, {Villaescusa-Navarro}, \& {Vogelsberger}}]{Diemer2019}
	{Diemer}, B., {Stevens}, A. R.~H., {Lagos}, C. d.~P., {et~al.} 2019, \mnras,
	487, 1529, \dodoi{10.1093/mnras/stz1323}
	
	\bibitem[{{Drinkwater} {et~al.}(2010){Drinkwater}, {Jurek}, {Blake}, {Woods},
		{Pimbblet}, {Glazebrook}, {Sharp}, {Pracy}, {Brough}, {Colless}, {Couch},
		{Croom}, {Davis}, {Forbes}, {Forster}, {Gilbank}, {Gladders}, {Jelliffe},
		{Jones}, {Li}, {Madore}, {Martin}, {Poole}, {Small}, {Wisnioski}, {Wyder}, \&
		{Yee}}]{Drinkwater2010}
	{Drinkwater}, M.~J., {Jurek}, R.~J., {Blake}, C., {et~al.} 2010, \mnras, 401,
	1429, \dodoi{10.1111/j.1365-2966.2009.15754.x}
	
	\bibitem[{{Dutta} \& {Khandai}(2021)}]{Dutta2021}
	{Dutta}, S., \& {Khandai}, N. 2021, \mnras, 500, L37,
	\dodoi{10.1093/mnrasl/slaa178}
	
	\bibitem[{{Feroz} {et~al.}(2009){Feroz}, {Hobson}, \& {Bridges}}]{Feroz2009}
	{Feroz}, F., {Hobson}, M.~P., \& {Bridges}, M. 2009, \mnras, 398, 1601,
	\dodoi{10.1111/j.1365-2966.2009.14548.x}
	
	\bibitem[{{Fletcher} {et~al.}(2021){Fletcher}, {Saintonge}, {Soares}, \&
		{Pontzen}}]{Fletcher2021}
	{Fletcher}, T.~J., {Saintonge}, A., {Soares}, P.~S., \& {Pontzen}, A. 2021,
	\mnras, 501, 411, \dodoi{10.1093/mnras/staa3025}
	
	\bibitem[{{Fu} {et~al.}(2013){Fu}, {Kauffmann}, {Huang}, {Yates}, {Moran},
		{Heckman}, {Dav{\'e}}, {Guo}, \& {Henriques}}]{Fu2013}
	{Fu}, J., {Kauffmann}, G., {Huang}, M.-l., {et~al.} 2013, \mnras, 434, 1531,
	\dodoi{10.1093/mnras/stt1117}
	
	\bibitem[{{Giovanelli} {et~al.}(2005){Giovanelli}, {Haynes}, {Kent},
		{Perillat}, {Saintonge}, {Brosch}, {Catinella}, {Hoffman}, {Stierwalt},
		{Spekkens}, {Lerner}, {Masters}, {Momjian}, {Rosenberg}, {Springob},
		{Boselli}, {Charmandaris}, {Darling}, {Davies}, {Garcia Lambas}, {Gavazzi},
		{Giovanardi}, {Hardy}, {Hunt}, {Iovino}, {Karachentsev}, {Karachentseva},
		{Koopmann}, {Marinoni}, {Minchin}, {Muller}, {Putman}, {Pantoja}, {Salzer},
		{Scodeggio}, {Skillman}, {Solanes}, {Valotto}, {van Driel}, \& {van
			Zee}}]{Giovanelli2005}
	{Giovanelli}, R., {Haynes}, M.~P., {Kent}, B.~R., {et~al.} 2005, \aj, 130,
	2598, \dodoi{10.1086/497431}
	
	\bibitem[{{Grogin} {et~al.}(2011){Grogin}, {Kocevski}, {Faber}, {Ferguson},
		{Koekemoer}, {Riess}, {Acquaviva}, {Alexander}, {Almaini}, {Ashby}, {Barden},
		{Bell}, {Bournaud}, {Brown}, {Caputi}, {Casertano}, {Cassata}, {Castellano},
		{Challis}, {Chary}, {Cheung}, {Cirasuolo}, {Conselice}, {Roshan Cooray},
		{Croton}, {Daddi}, {Dahlen}, {Dav{\'e}}, {de Mello}, {Dekel}, {Dickinson},
		{Dolch}, {Donley}, {Dunlop}, {Dutton}, {Elbaz}, {Fazio}, {Filippenko},
		{Finkelstein}, {Fontana}, {Gardner}, {Garnavich}, {Gawiser}, {Giavalisco},
		{Grazian}, {Guo}, {Hathi}, {H{\"a}ussler}, {Hopkins}, {Huang}, {Huang},
		{Jha}, {Kartaltepe}, {Kirshner}, {Koo}, {Lai}, {Lee}, {Li}, {Lotz}, {Lucas},
		{Madau}, {McCarthy}, {McGrath}, {McIntosh}, {McLure}, {Mobasher},
		{Moustakas}, {Mozena}, {Nandra}, {Newman}, {Niemi}, {Noeske}, {Papovich},
		{Pentericci}, {Pope}, {Primack}, {Rajan}, {Ravindranath}, {Reddy}, {Renzini},
		{Rix}, {Robaina}, {Rodney}, {Rosario}, {Rosati}, {Salimbeni}, {Scarlata},
		{Siana}, {Simard}, {Smidt}, {Somerville}, {Spinrad}, {Straughn}, {Strolger},
		{Telford}, {Teplitz}, {Trump}, {van der Wel}, {Villforth}, {Wechsler},
		{Weiner}, {Wiklind}, {Wild}, {Wilson}, {Wuyts}, {Yan}, \& {Yun}}]{Grogin2011}
	{Grogin}, N.~A., {Kocevski}, D.~D., {Faber}, S.~M., {et~al.} 2011, \apjs, 197,
	35, \dodoi{10.1088/0067-0049/197/2/35}
	
	\bibitem[{{Guo} {et~al.}(2020){Guo}, {Jones}, {Haynes}, \& {Fu}}]{Guo2020}
	{Guo}, H., {Jones}, M.~G., {Haynes}, M.~P., \& {Fu}, J. 2020, \apj, 894, 92,
	\dodoi{10.3847/1538-4357/ab886f}
	
	\bibitem[{{Guo} {et~al.}(2022){Guo}, {Jones}, \& {Wang}}]{Guo2022}
	{Guo}, H., {Jones}, M.~G., \& {Wang}, J. 2022, \apjl, 933, L12,
	\dodoi{10.3847/2041-8213/ac794f}
	
	\bibitem[{{Guo} {et~al.}(2021){Guo}, {Jones}, {Wang}, \& {Lin}}]{Guo2021}
	{Guo}, H., {Jones}, M.~G., {Wang}, J., \& {Lin}, L. 2021, \apj, 918, 53,
	\dodoi{10.3847/1538-4357/ac062e}
	
	\bibitem[{{Guo} {et~al.}(2017){Guo}, {Li}, {Zheng}, {Mo}, {Jing}, {Zu}, {Lim},
		\& {Xu}}]{Guo2017}
	{Guo}, H., {Li}, C., {Zheng}, Z., {et~al.} 2017, \apj, 846, 61,
	\dodoi{10.3847/1538-4357/aa85e7}
	
	\bibitem[{{Haynes} {et~al.}(2011){Haynes}, {Giovanelli}, {Martin}, {Hess},
		{Saintonge}, {Adams}, {Hallenbeck}, {Hoffman}, {Huang}, {Kent}, {Koopmann},
		{Papastergis}, {Stierwalt}, {Balonek}, {Craig}, {Higdon}, {Kornreich},
		{Miller}, {O'Donoghue}, {Olowin}, {Rosenberg}, {Spekkens}, {Troischt}, \&
		{Wilcots}}]{Haynes2011}
	{Haynes}, M.~P., {Giovanelli}, R., {Martin}, A.~M., {et~al.} 2011, \aj, 142,
	170, \dodoi{10.1088/0004-6256/142/5/170}
	
	\bibitem[{{Haynes} {et~al.}(2018){Haynes}, {Giovanelli}, {Kent}, {Adams},
		{Balonek}, {Craig}, {Fertig}, {Finn}, {Giovanardi}, {Hallenbeck}, {Hess},
		{Hoffman}, {Huang}, {Jones}, {Koopmann}, {Kornreich}, {Leisman}, {Miller},
		{Moorman}, {O'Connor}, {O'Donoghue}, {Papastergis}, {Troischt}, {Stark}, \&
		{Xiao}}]{Haynes2018}
	{Haynes}, M.~P., {Giovanelli}, R., {Kent}, B.~R., {et~al.} 2018, \apj, 861, 49,
	\dodoi{10.3847/1538-4357/aac956}
	
	\bibitem[{{Heintz} {et~al.}(2021){Heintz}, {Watson}, {Oesch}, {Narayanan}, \&
		{Madden}}]{Heintz2021}
	{Heintz}, K.~E., {Watson}, D., {Oesch}, P.~A., {Narayanan}, D., \& {Madden},
	S.~C. 2021, \apj, 922, 147, \dodoi{10.3847/1538-4357/ac2231}
	
	\bibitem[{{Heintz} {et~al.}(2022){Heintz}, {Oesch}, {Aravena}, {Bouwens},
		{Dayal}, {Ferrara}, {Fudamoto}, {Graziani}, {Inami}, {Sommovigo}, {Smit},
		{Stefanon}, {Topping}, {Pallottini}, \& {van der Werf}}]{Heintz2022}
	{Heintz}, K.~E., {Oesch}, P.~A., {Aravena}, M., {et~al.} 2022, \apjl, 934, L27,
	\dodoi{10.3847/2041-8213/ac8057}
	
	\bibitem[{{Janowiecki} {et~al.}(2020){Janowiecki}, {Catinella}, {Cortese},
		{Saintonge}, \& {Wang}}]{Janowiecki2020}
	{Janowiecki}, S., {Catinella}, B., {Cortese}, L., {Saintonge}, A., \& {Wang},
	J. 2020, \mnras, 493, 1982, \dodoi{10.1093/mnras/staa178}
	
	\bibitem[{{Jones} {et~al.}(2018){Jones}, {Haynes}, {Giovanelli}, \&
		{Moorman}}]{Jones2018}
	{Jones}, M.~G., {Haynes}, M.~P., {Giovanelli}, R., \& {Moorman}, C. 2018,
	\mnras, 477, 2, \dodoi{10.1093/mnras/sty521}
	
	\bibitem[{{Jones} {et~al.}(2020){Jones}, {Hess}, {Adams}, \&
		{Verdes-Montenegro}}]{Jones2020}
	{Jones}, M.~G., {Hess}, K.~M., {Adams}, E. A.~K., \& {Verdes-Montenegro}, L.
	2020, \mnras, 494, 2090, \dodoi{10.1093/mnras/staa810}
	
	\bibitem[{{Kanekar} {et~al.}(2016){Kanekar}, {Sethi}, \&
		{Dwarakanath}}]{Kanekar2016}
	{Kanekar}, N., {Sethi}, S., \& {Dwarakanath}, K.~S. 2016, \apjl, 818, L28,
	\dodoi{10.3847/2041-8205/818/2/L28}
	
	\bibitem[{{Keres} {et~al.}(2003){Keres}, {Yun}, \& {Young}}]{Keres2003}
	{Keres}, D., {Yun}, M.~S., \& {Young}, J.~S. 2003, \apj, 582, 659,
	\dodoi{10.1086/344820}
	
	\bibitem[{{Kim} {et~al.}(2017){Kim}, {Wyithe}, {Baugh}, {Lagos}, {Power}, \&
		{Park}}]{Kim2017}
	{Kim}, H.-S., {Wyithe}, J. S.~B., {Baugh}, C.~M., {et~al.} 2017, \mnras, 465,
	111, \dodoi{10.1093/mnras/stw2779}
	
	\bibitem[{{Klypin} {et~al.}(2016){Klypin}, {Yepes}, {Gottl{\"o}ber}, {Prada},
		\& {He{\ss}}}]{Klypin2016}
	{Klypin}, A., {Yepes}, G., {Gottl{\"o}ber}, S., {Prada}, F., \& {He{\ss}}, S.
	2016, \mnras, 457, 4340, \dodoi{10.1093/mnras/stw248}
	
	\bibitem[{{Koekemoer} {et~al.}(2011){Koekemoer}, {Faber}, {Ferguson}, {Grogin},
		{Kocevski}, {Koo}, {Lai}, {Lotz}, {Lucas}, {McGrath}, {Ogaz}, {Rajan},
		{Riess}, {Rodney}, {Strolger}, {Casertano}, {Castellano}, {Dahlen},
		{Dickinson}, {Dolch}, {Fontana}, {Giavalisco}, {Grazian}, {Guo}, {Hathi},
		{Huang}, {van der Wel}, {Yan}, {Acquaviva}, {Alexander}, {Almaini}, {Ashby},
		{Barden}, {Bell}, {Bournaud}, {Brown}, {Caputi}, {Cassata}, {Challis},
		{Chary}, {Cheung}, {Cirasuolo}, {Conselice}, {Roshan Cooray}, {Croton},
		{Daddi}, {Dav{\'e}}, {de Mello}, {de Ravel}, {Dekel}, {Donley}, {Dunlop},
		{Dutton}, {Elbaz}, {Fazio}, {Filippenko}, {Finkelstein}, {Frazer}, {Gardner},
		{Garnavich}, {Gawiser}, {Gruetzbauch}, {Hartley}, {H{\"a}ussler},
		{Herrington}, {Hopkins}, {Huang}, {Jha}, {Johnson}, {Kartaltepe},
		{Khostovan}, {Kirshner}, {Lani}, {Lee}, {Li}, {Madau}, {McCarthy},
		{McIntosh}, {McLure}, {McPartland}, {Mobasher}, {Moreira}, {Mortlock},
		{Moustakas}, {Mozena}, {Nandra}, {Newman}, {Nielsen}, {Niemi}, {Noeske},
		{Papovich}, {Pentericci}, {Pope}, {Primack}, {Ravindranath}, {Reddy},
		{Renzini}, {Rix}, {Robaina}, {Rosario}, {Rosati}, {Salimbeni}, {Scarlata},
		{Siana}, {Simard}, {Smidt}, {Snyder}, {Somerville}, {Spinrad}, {Straughn},
		{Telford}, {Teplitz}, {Trump}, {Vargas}, {Villforth}, {Wagner}, {Wandro},
		{Wechsler}, {Weiner}, {Wiklind}, {Wild}, {Wilson}, {Wuyts}, \&
		{Yun}}]{Koekemoer2011}
	{Koekemoer}, A.~M., {Faber}, S.~M., {Ferguson}, H.~C., {et~al.} 2011, \apjs,
	197, 36, \dodoi{10.1088/0067-0049/197/2/36}
	
	\bibitem[{{Lagos} {et~al.}(2011){Lagos}, {Baugh}, {Lacey}, {Benson}, {Kim}, \&
		{Power}}]{Lagos2011}
	{Lagos}, C. D.~P., {Baugh}, C.~M., {Lacey}, C.~G., {et~al.} 2011, \mnras, 418,
	1649, \dodoi{10.1111/j.1365-2966.2011.19583.x}
	
	\bibitem[{{Lagos} {et~al.}(2015){Lagos}, {Crain}, {Schaye}, {Furlong}, {Frenk},
		{Bower}, {Schaller}, {Theuns}, {Trayford}, {Bah{\'e}}, \& {Dalla
			Vecchia}}]{Lagos2015}
	{Lagos}, C. d.~P., {Crain}, R.~A., {Schaye}, J., {et~al.} 2015, \mnras, 452,
	3815, \dodoi{10.1093/mnras/stv1488}
	
	\bibitem[{{Lah} {et~al.}(2007){Lah}, {Chengalur}, {Briggs}, {Colless}, {de
			Propris}, {Pracy}, {de Blok}, {Fujita}, {Ajiki}, {Shioya}, {Nagao},
		{Murayama}, {Taniguchi}, {Yagi}, \& {Okamura}}]{Lah2007}
	{Lah}, P., {Chengalur}, J.~N., {Briggs}, F.~H., {et~al.} 2007, \mnras, 376,
	1357, \dodoi{10.1111/j.1365-2966.2007.11540.x}
	
	\bibitem[{{Landy} \& {Szalay}(1993)}]{Landy1993}
	{Landy}, S.~D., \& {Szalay}, A.~S. 1993, \apj, 412, 64, \dodoi{10.1086/172900}
	
	\bibitem[{{Lenki{\'c}} {et~al.}(2020){Lenki{\'c}}, {Bolatto}, {F{\"o}rster
			Schreiber}, {Tacconi}, {Neri}, {Combes}, {Walter}, {Garc{\'\i}a-Burillo},
		{Genzel}, {Lutz}, \& {Cooper}}]{Lenkic2020}
	{Lenki{\'c}}, L., {Bolatto}, A.~D., {F{\"o}rster Schreiber}, N.~M., {et~al.}
	2020, \aj, 159, 190, \dodoi{10.3847/1538-3881/ab7458}
	
	\bibitem[{{Leroy} {et~al.}(2009){Leroy}, {Walter}, {Bigiel}, {Usero}, {Weiss},
		{Brinks}, {de Blok}, {Kennicutt}, {Schuster}, {Kramer}, {Wiesemeyer}, \&
		{Roussel}}]{Leroy2009}
	{Leroy}, A.~K., {Walter}, F., {Bigiel}, F., {et~al.} 2009, \aj, 137, 4670,
	\dodoi{10.1088/0004-6256/137/6/4670}
	
	\bibitem[{{Li} {et~al.}(2012){Li}, {Kauffmann}, {Fu}, {Wang}, {Catinella},
		{Fabello}, {Schiminovich}, \& {Zhang}}]{Li2012}
	{Li}, C., {Kauffmann}, G., {Fu}, J., {et~al.} 2012, \mnras, 424, 1471,
	\dodoi{10.1111/j.1365-2966.2012.21337.x}
	
	\bibitem[{{Li} {et~al.}(2022{\natexlab{a}}){Li}, {Li}, {Mo}, {Xiao}, \&
		{Wang}}]{Li2022a}
	{Li}, X., {Li}, C., {Mo}, H.~J., {Xiao}, T., \& {Wang}, J. 2022{\natexlab{a}},
	\apj, 941, 48, \dodoi{10.3847/1538-4357/ac9ccb}
	
	\bibitem[{{Li} {et~al.}(2022{\natexlab{b}}){Li}, {Guo}, \& {Mao}}]{Li2022}
	{Li}, Z., {Guo}, H., \& {Mao}, Y. 2022{\natexlab{b}}, arXiv e-prints,
	arXiv:2207.10414.
	\newblock \doarXiv{2207.10414}
	
	\bibitem[{{Lilly} {et~al.}(2013){Lilly}, {Carollo}, {Pipino}, {Renzini}, \&
		{Peng}}]{Lilly2013}
	{Lilly}, S.~J., {Carollo}, C.~M., {Pipino}, A., {Renzini}, A., \& {Peng}, Y.
	2013, \apj, 772, 119, \dodoi{10.1088/0004-637X/772/2/119}
	
	\bibitem[{{Lim} {et~al.}(2017){Lim}, {Mo}, {Lu}, {Wang}, \& {Yang}}]{Lim2017}
	{Lim}, S.~H., {Mo}, H.~J., {Lu}, Y., {Wang}, H., \& {Yang}, X. 2017, \mnras,
	470, 2982, \dodoi{10.1093/mnras/stx1462}
	
	\bibitem[{{Liu} {et~al.}(2019){Liu}, {Schinnerer}, {Groves}, {Magnelli},
		{Lang}, {Leslie}, {Jim{\'e}nez-Andrade}, {Riechers}, {Popping}, {Magdis},
		{Daddi}, {Sargent}, {Gao}, {Fudamoto}, {Oesch}, \& {Bertoldi}}]{Liu2019}
	{Liu}, D., {Schinnerer}, E., {Groves}, B., {et~al.} 2019, \apj, 887, 235,
	\dodoi{10.3847/1538-4357/ab578d}
	
	\bibitem[{{Ma} {et~al.}(2022){Ma}, {Liu}, {Guo}, {Cui}, {Jones}, {Wang},
		{Zhang}, \& {Dav{\'e}}}]{Ma2022}
	{Ma}, W., {Liu}, K., {Guo}, H., {et~al.} 2022, \apj, 941, 205,
	\dodoi{10.3847/1538-4357/aca326}
	
	\bibitem[{{Magnelli} {et~al.}(2020){Magnelli}, {Boogaard}, {Decarli},
		{G{\'o}nzalez-L{\'o}pez}, {Novak}, {Popping}, {Smail}, {Walter}, {Aravena},
		{Assef}, {Bauer}, {Bertoldi}, {Carilli}, {Cortes}, {Cunha}, {Daddi},
		{D{\'\i}az-Santos}, {Inami}, {Ivison}, {F{\`e}vre}, {Oesch}, {Riechers},
		{Rix}, {Sargent}, {Werf}, {Wagg}, \& {Weiss}}]{Magnelli2020}
	{Magnelli}, B., {Boogaard}, L., {Decarli}, R., {et~al.} 2020, \apj, 892, 66,
	\dodoi{10.3847/1538-4357/ab7897}
	
	\bibitem[{{Martin} {et~al.}(2010){Martin}, {Papastergis}, {Giovanelli},
		{Haynes}, {Springob}, \& {Stierwalt}}]{Martin2010}
	{Martin}, A.~M., {Papastergis}, E., {Giovanelli}, R., {et~al.} 2010, \apj, 723,
	1359, \dodoi{10.1088/0004-637X/723/2/1359}
	
	\bibitem[{{Masui} {et~al.}(2013){Masui}, {Switzer}, {Banavar}, {Bandura},
		{Blake}, {Calin}, {Chang}, {Chen}, {Li}, {Liao}, {Natarajan}, {Pen},
		{Peterson}, {Shaw}, \& {Voytek}}]{Masui2013}
	{Masui}, K.~W., {Switzer}, E.~R., {Banavar}, N., {et~al.} 2013, \apjl, 763,
	L20, \dodoi{10.1088/2041-8205/763/1/L20}
	
	\bibitem[{{Meyer} {et~al.}(2004){Meyer}, {Zwaan}, {Webster}, {Staveley-Smith},
		{Ryan-Weber}, {Drinkwater}, {Barnes}, {Howlett}, {Kilborn}, {Stevens},
		{Waugh}, {Pierce}, {Bhathal}, {de Blok}, {Disney}, {Ekers}, {Freeman},
		{Garcia}, {Gibson}, {Harnett}, {Henning}, {Jerjen}, {Kesteven}, {Knezek},
		{Koribalski}, {Mader}, {Marquarding}, {Minchin}, {O'Brien}, {Oosterloo},
		{Price}, {Putman}, {Ryder}, {Sadler}, {Stewart}, {Stootman}, \&
		{Wright}}]{Meyer2004}
	{Meyer}, M.~J., {Zwaan}, M.~A., {Webster}, R.~L., {et~al.} 2004, \mnras, 350,
	1195, \dodoi{10.1111/j.1365-2966.2004.07710.x}
	
	\bibitem[{{Neeleman} {et~al.}(2016){Neeleman}, {Prochaska}, {Ribaudo},
		{Lehner}, {Howk}, {Rafelski}, \& {Kanekar}}]{Neeleman2016}
	{Neeleman}, M., {Prochaska}, J.~X., {Ribaudo}, J., {et~al.} 2016, \apj, 818,
	113, \dodoi{10.3847/0004-637X/818/2/113}
	
	\bibitem[{{Nelson} {et~al.}(2019){Nelson}, {Springel}, {Pillepich},
		{Rodriguez-Gomez}, {Torrey}, {Genel}, {Vogelsberger}, {Pakmor}, {Marinacci},
		{Weinberger}, {Kelley}, {Lovell}, {Diemer}, \& {Hernquist}}]{Nelson2019}
	{Nelson}, D., {Springel}, V., {Pillepich}, A., {et~al.} 2019, Computational
	Astrophysics and Cosmology, 6, 2, \dodoi{10.1186/s40668-019-0028-x}
	
	\bibitem[{{Noble} {et~al.}(2017){Noble}, {McDonald}, {Muzzin}, {Nantais},
		{Rudnick}, {van Kampen}, {Webb}, {Wilson}, {Yee}, {Boone}, {Cooper},
		{DeGroot}, {Delahaye}, {Demarco}, {Foltz}, {Hayden}, {Lidman},
		{Manilla-Robles}, \& {Perlmutter}}]{Noble2017}
	{Noble}, A.~G., {McDonald}, M., {Muzzin}, A., {et~al.} 2017, \apjl, 842, L21,
	\dodoi{10.3847/2041-8213/aa77f3}
	
	\bibitem[{{Noterdaeme} {et~al.}(2009){Noterdaeme}, {Petitjean}, {Ledoux}, \&
		{Srianand}}]{Noterdaeme2009}
	{Noterdaeme}, P., {Petitjean}, P., {Ledoux}, C., \& {Srianand}, R. 2009, \aap,
	505, 1087, \dodoi{10.1051/0004-6361/200912768}
	
	\bibitem[{{Noterdaeme} {et~al.}(2012){Noterdaeme}, {Petitjean}, {Carithers},
		{P{\^a}ris}, {Font-Ribera}, {Bailey}, {Aubourg}, {Bizyaev}, {Ebelke},
		{Finley}, {Ge}, {Malanushenko}, {Malanushenko}, {Miralda-Escud{\'e}},
		{Myers}, {Oravetz}, {Pan}, {Pieri}, {Ross}, {Schneider}, {Simmons}, \&
		{York}}]{Noterdaeme2012}
	{Noterdaeme}, P., {Petitjean}, P., {Carithers}, W.~C., {et~al.} 2012, \aap,
	547, L1, \dodoi{10.1051/0004-6361/201220259}
	
	\bibitem[{{Obuljen} {et~al.}(2019){Obuljen}, {Alonso}, {Villaescusa-Navarro},
		{Yoon}, \& {Jones}}]{Obuljen2019}
	{Obuljen}, A., {Alonso}, D., {Villaescusa-Navarro}, F., {Yoon}, I., \& {Jones},
	M. 2019, \mnras, 486, 5124, \dodoi{10.1093/mnras/stz1118}
	
	\bibitem[{{Oman}(2022)}]{Oman2022}
	{Oman}, K.~A. 2022, \mnras, 509, 3268, \dodoi{10.1093/mnras/stab3164}
	
	\bibitem[{{O'Meara} {et~al.}(2007){O'Meara}, {Prochaska}, {Burles}, {Prochter},
		{Bernstein}, \& {Burgess}}]{OMeara2007}
	{O'Meara}, J.~M., {Prochaska}, J.~X., {Burles}, S., {et~al.} 2007, \apj, 656,
	666, \dodoi{10.1086/510711}
	
	\bibitem[{{Padmanabhan} \& {Loeb}(2020)}]{Padmanabhan2020}
	{Padmanabhan}, H., \& {Loeb}, A. 2020, \mnras, 496, 1124,
	\dodoi{10.1093/mnras/staa1565}
	
	\bibitem[{{Paul} {et~al.}(2018){Paul}, {Choudhury}, \& {Paranjape}}]{Paul2018}
	{Paul}, N., {Choudhury}, T.~R., \& {Paranjape}, A. 2018, \mnras, 479, 1627,
	\dodoi{10.1093/mnras/sty1539}
	
	\bibitem[{{P{\'e}roux} \& {Howk}(2020)}]{Peroux2020}
	{P{\'e}roux}, C., \& {Howk}, J.~C. 2020, \araa, 58, 363,
	\dodoi{10.1146/annurev-astro-021820-120014}
	
	\bibitem[{{P{\'e}roux} {et~al.}(2003){P{\'e}roux}, {McMahon},
		{Storrie-Lombardi}, \& {Irwin}}]{Peroux2003}
	{P{\'e}roux}, C., {McMahon}, R.~G., {Storrie-Lombardi}, L.~J., \& {Irwin},
	M.~J. 2003, \mnras, 346, 1103, \dodoi{10.1111/j.1365-2966.2003.07129.x}
	
	\bibitem[{{Planck Collaboration} {et~al.}(2016){Planck Collaboration}, {Ade},
		{Aghanim}, {Arnaud}, {Ashdown}, {Aumont}, {Baccigalupi}, {Banday},
		{Barreiro}, {Bartlett}, {Bartolo}, {Battaner}, {Battye}, {Benabed},
		{Beno{\^\i}t}, {Benoit-L{\'e}vy}, {Bernard}, {Bersanelli}, {Bielewicz},
		{Bock}, {Bonaldi}, {Bonavera}, {Bond}, {Borrill}, {Bouchet}, {Boulanger},
		{Bucher}, {Burigana}, {Butler}, {Calabrese}, {Cardoso}, {Catalano},
		{Challinor}, {Chamballu}, {Chary}, {Chiang}, {Chluba}, {Christensen},
		{Church}, {Clements}, {Colombi}, {Colombo}, {Combet}, {Coulais}, {Crill},
		{Curto}, {Cuttaia}, {Danese}, {Davies}, {Davis}, {de Bernardis}, {de Rosa},
		{de Zotti}, {Delabrouille}, {D{\'e}sert}, {Di Valentino}, {Dickinson},
		{Diego}, {Dolag}, {Dole}, {Donzelli}, {Dor{\'e}}, {Douspis}, {Ducout},
		{Dunkley}, {Dupac}, {Efstathiou}, {Elsner}, {En{\ss}lin}, {Eriksen},
		{Farhang}, {Fergusson}, {Finelli}, {Forni}, {Frailis}, {Fraisse},
		{Franceschi}, {Frejsel}, {Galeotta}, {Galli}, {Ganga}, {Gauthier}, {Gerbino},
		{Ghosh}, {Giard}, {Giraud-H{\'e}raud}, {Giusarma}, {Gjerl{\o}w},
		{Gonz{\'a}lez-Nuevo}, {G{\'o}rski}, {Gratton}, {Gregorio}, {Gruppuso},
		{Gudmundsson}, {Hamann}, {Hansen}, {Hanson}, {Harrison}, {Helou},
		{Henrot-Versill{\'e}}, {Hern{\'a}ndez-Monteagudo}, {Herranz}, {Hildebrandt},
		{Hivon}, {Hobson}, {Holmes}, {Hornstrup}, {Hovest}, {Huang}, {Huffenberger},
		{Hurier}, {Jaffe}, {Jaffe}, {Jones}, {Juvela}, {Keih{\"a}nen}, {Keskitalo},
		{Kisner}, {Kneissl}, {Knoche}, {Knox}, {Kunz}, {Kurki-Suonio}, {Lagache},
		{L{\"a}hteenm{\"a}ki}, {Lamarre}, {Lasenby}, {Lattanzi}, {Lawrence}, {Leahy},
		{Leonardi}, {Lesgourgues}, {Levrier}, {Lewis}, {Liguori}, {Lilje},
		{Linden-V{\o}rnle}, {L{\'o}pez-Caniego}, {Lubin}, {Mac{\'\i}as-P{\'e}rez},
		{Maggio}, {Maino}, {Mandolesi}, {Mangilli}, {Marchini}, {Maris}, {Martin},
		{Martinelli}, {Mart{\'\i}nez-Gonz{\'a}lez}, {Masi}, {Matarrese}, {McGehee},
		{Meinhold}, {Melchiorri}, {Melin}, {Mendes}, {Mennella}, {Migliaccio},
		{Millea}, {Mitra}, {Miville-Desch{\^e}nes}, {Moneti}, {Montier}, {Morgante},
		{Mortlock}, {Moss}, {Munshi}, {Murphy}, {Naselsky}, {Nati}, {Natoli},
		{Netterfield}, {N{\o}rgaard-Nielsen}, {Noviello}, {Novikov}, {Novikov},
		{Oxborrow}, {Paci}, {Pagano}, {Pajot}, {Paladini}, {Paoletti}, {Partridge},
		{Pasian}, {Patanchon}, {Pearson}, {Perdereau}, {Perotto}, {Perrotta},
		{Pettorino}, {Piacentini}, {Piat}, {Pierpaoli}, {Pietrobon}, {Plaszczynski},
		{Pointecouteau}, {Polenta}, {Popa}, {Pratt}, {Pr{\'e}zeau}, {Prunet},
		{Puget}, {Rachen}, {Reach}, {Rebolo}, {Reinecke}, {Remazeilles}, {Renault},
		{Renzi}, {Ristorcelli}, {Rocha}, {Rosset}, {Rossetti}, {Roudier},
		{Rouill{\'e} d'Orfeuil}, {Rowan-Robinson}, {Rubi{\~n}o-Mart{\'\i}n},
		{Rusholme}, {Said}, {Salvatelli}, {Salvati}, {Sandri}, {Santos},
		{Savelainen}, {Savini}, {Scott}, {Seiffert}, {Serra}, {Shellard}, {Spencer},
		{Spinelli}, {Stolyarov}, {Stompor}, {Sudiwala}, {Sunyaev}, {Sutton},
		{Suur-Uski}, {Sygnet}, {Tauber}, {Terenzi}, {Toffolatti}, {Tomasi},
		{Tristram}, {Trombetti}, {Tucci}, {Tuovinen}, {T{\"u}rler}, {Umana},
		{Valenziano}, {Valiviita}, {Van Tent}, {Vielva}, {Villa}, {Wade}, {Wandelt},
		{Wehus}, {White}, {White}, {Wilkinson}, {Yvon}, {Zacchei}, \&
		{Zonca}}]{Planck2016}
	{Planck Collaboration}, {Ade}, P.~A.~R., {Aghanim}, N., {et~al.} 2016, \aap,
	594, A13, \dodoi{10.1051/0004-6361/201525830}
	
	\bibitem[{{Ponomareva} {et~al.}(2023){Ponomareva}, {Jarvis}, {Pan}, {Maddox},
		{Jones}, {Frank}, {Rajohnson}, {Mulaudzi}, {Meyer}, {Adams}, {Baes}, {Hess},
		{Kurapati}, {Prandoni}, {Sinigaglia}, {Spekkens}, {Tudorache}, {Heywood},
		{Collier}, \& {Sekhar}}]{Ponomareva2023}
	{Ponomareva}, A.~A., {Jarvis}, M.~J., {Pan}, H., {et~al.} 2023, \mnras, 522,
	5308, \dodoi{10.1093/mnras/stad1249}
	
	\bibitem[{{Popping} {et~al.}(2014){Popping}, {Somerville}, \&
		{Trager}}]{Popping2014}
	{Popping}, G., {Somerville}, R.~S., \& {Trager}, S.~C. 2014, \mnras, 442, 2398,
	\dodoi{10.1093/mnras/stu991}
	
	\bibitem[{{Popping} {et~al.}(2015){Popping}, {Caputi}, {Trager}, {Somerville},
		{Dekel}, {Kassin}, {Kocevski}, {Koekemoer}, {Faber}, {Ferguson}, {Galametz},
		{Grogin}, {Guo}, {Lu}, {Wel}, \& {Weiner}}]{Popping2015}
	{Popping}, G., {Caputi}, K.~I., {Trager}, S.~C., {et~al.} 2015, \mnras, 454,
	2258, \dodoi{10.1093/mnras/stv2136}
	
	\bibitem[{{Prochaska} {et~al.}(2005){Prochaska}, {Herbert-Fort}, \&
		{Wolfe}}]{Prochaska2005}
	{Prochaska}, J.~X., {Herbert-Fort}, S., \& {Wolfe}, A.~M. 2005, \apj, 635, 123,
	\dodoi{10.1086/497287}
	
	\bibitem[{{Prochaska} \& {Wolfe}(2009)}]{Prochaska2009}
	{Prochaska}, J.~X., \& {Wolfe}, A.~M. 2009, \apj, 696, 1543,
	\dodoi{10.1088/0004-637X/696/2/1543}
	
	\bibitem[{{Rhee} {et~al.}(2018){Rhee}, {Lah}, {Briggs}, {Chengalur}, {Colless},
		{Willner}, {Ashby}, \& {Le F{\`e}vre}}]{Rhee2018}
	{Rhee}, J., {Lah}, P., {Briggs}, F.~H., {et~al.} 2018, \mnras, 473, 1879,
	\dodoi{10.1093/mnras/stx2461}
	
	\bibitem[{{Rhee} {et~al.}(2013){Rhee}, {Zwaan}, {Briggs}, {Chengalur}, {Lah},
		{Oosterloo}, \& {van der Hulst}}]{Rhee2013}
	{Rhee}, J., {Zwaan}, M.~A., {Briggs}, F.~H., {et~al.} 2013, \mnras, 435, 2693,
	\dodoi{10.1093/mnras/stt1481}
	
	\bibitem[{{Rhee} {et~al.}(2023){Rhee}, {Meyer}, {Popping}, {Bellstedt},
		{Driver}, {Robotham}, {Whiting}, {Baldry}, {Brough}, {Brown}, {Bunton},
		{Dodson}, {Holwerda}, {Hopkins}, {Koribalski}, {Lee-Waddell},
		{L{\'o}pez-S{\'a}nchez}, {Loveday}, {Mahony}, {Roychowdhury}, {Rozgonyi}, \&
		{Staveley-Smith}}]{Rhee2023}
	{Rhee}, J., {Meyer}, M., {Popping}, A., {et~al.} 2023, \mnras, 518, 4646,
	\dodoi{10.1093/mnras/stac3065}
	
	\bibitem[{{Riechers} {et~al.}(2019){Riechers}, {Pavesi}, {Sharon}, {Hodge},
		{Decarli}, {Walter}, {Carilli}, {Aravena}, {da Cunha}, {Daddi}, {Dickinson},
		{Smail}, {Capak}, {Ivison}, {Sargent}, {Scoville}, \& {Wagg}}]{Riechers2019}
	{Riechers}, D.~A., {Pavesi}, R., {Sharon}, C.~E., {et~al.} 2019, \apj, 872, 7,
	\dodoi{10.3847/1538-4357/aafc27}
	
	\bibitem[{{Saintonge} \& {Catinella}(2022)}]{Saintonge2022}
	{Saintonge}, A., \& {Catinella}, B. 2022, \araa, 60, 319,
	\dodoi{10.1146/annurev-astro-021022-043545}
	
	\bibitem[{{Saintonge} {et~al.}(2016){Saintonge}, {Catinella}, {Cortese},
		{Genzel}, {Giovanelli}, {Haynes}, {Janowiecki}, {Kramer}, {Lutz},
		{Schiminovich}, {Tacconi}, {Wuyts}, \& {Accurso}}]{Saintonge2016}
	{Saintonge}, A., {Catinella}, B., {Cortese}, L., {et~al.} 2016, \mnras, 462,
	1749, \dodoi{10.1093/mnras/stw1715}
	
	\bibitem[{{Saintonge} {et~al.}(2017){Saintonge}, {Catinella}, {Tacconi},
		{Kauffmann}, {Genzel}, {Cortese}, {Dav{\'e}}, {Fletcher},
		{Graci{\'a}-Carpio}, {Kramer}, {Heckman}, {Janowiecki}, {Lutz}, {Rosario},
		{Schiminovich}, {Schuster}, {Wang}, {Wuyts}, {Borthakur}, {Lamperti}, \&
		{Roberts-Borsani}}]{Saintonge2017}
	{Saintonge}, A., {Catinella}, B., {Tacconi}, L.~J., {et~al.} 2017, \apjs, 233,
	22, \dodoi{10.3847/1538-4365/aa97e0}
	
	\bibitem[{{Scoville} {et~al.}(2017){Scoville}, {Lee}, {Vanden Bout},
		{Diaz-Santos}, {Sanders}, {Darvish}, {Bongiorno}, {Casey}, {Murchikova},
		{Koda}, {Capak}, {Vlahakis}, {Ilbert}, {Sheth}, {Morokuma-Matsui}, {Ivison},
		{Aussel}, {Laigle}, {McCracken}, {Armus}, {Pope}, {Toft}, \&
		{Masters}}]{Scoville2017}
	{Scoville}, N., {Lee}, N., {Vanden Bout}, P., {et~al.} 2017, \apj, 837, 150,
	\dodoi{10.3847/1538-4357/aa61a0}
	
	\bibitem[{{Sinigaglia} {et~al.}(2022){Sinigaglia}, {Rodighiero}, {Elson},
		{Vaccari}, {Maddox}, {Frank}, {Jarvis}, {Oosterloo}, {Dav{\'e}}, {Salvato},
		{Baes}, {Bellstedt}, {Bisigello}, {Collier}, {Cook}, {Davies}, {Delhaize},
		{Driver}, {Foster}, {Kurapati}, {Lagos}, {Lidman}, {Mancera Pi{\~n}a},
		{Meyer}, {Mogotsi}, {Pan}, {Ponomareva}, {Prandoni}, {Rajohnson}, {Robotham},
		{Santos}, {Sekhar}, {Spekkens}, {Thorne}, {van der Hulst}, \&
		{Wong}}]{Sinigaglia2022}
	{Sinigaglia}, F., {Rodighiero}, G., {Elson}, E., {et~al.} 2022, \apjl, 935,
	L13, \dodoi{10.3847/2041-8213/ac85ae}
	
	\bibitem[{{Spinelli} {et~al.}(2020){Spinelli}, {Zoldan}, {De Lucia}, {Xie}, \&
		{Viel}}]{Spinelli2020}
	{Spinelli}, M., {Zoldan}, A., {De Lucia}, G., {Xie}, L., \& {Viel}, M. 2020,
	\mnras, 493, 5434, \dodoi{10.1093/mnras/staa604}
	
	\bibitem[{{Stevens} {et~al.}(2019){Stevens}, {Diemer}, {Lagos}, {Nelson},
		{Pillepich}, {Brown}, {Catinella}, {Hernquist}, {Weinberger}, {Vogelsberger},
		\& {Marinacci}}]{Stevens2019}
	{Stevens}, A. R.~H., {Diemer}, B., {Lagos}, C. d.~P., {et~al.} 2019, \mnras,
	483, 5334, \dodoi{10.1093/mnras/sty3451}
	
	\bibitem[{{Stiskalek} {et~al.}(2021){Stiskalek}, {Desmond}, {Holvey}, \&
		{Jones}}]{Stiskalek2021}
	{Stiskalek}, R., {Desmond}, H., {Holvey}, T., \& {Jones}, M.~G. 2021, \mnras,
	506, 3205, \dodoi{10.1093/mnras/stab1845}
	
	\bibitem[{{Switzer} {et~al.}(2013){Switzer}, {Masui}, {Bandura}, {Calin},
		{Chang}, {Chen}, {Li}, {Liao}, {Natarajan}, {Pen}, {Peterson}, {Shaw}, \&
		{Voytek}}]{Switzer2013}
	{Switzer}, E.~R., {Masui}, K.~W., {Bandura}, K., {et~al.} 2013, \mnras, 434,
	L46, \dodoi{10.1093/mnrasl/slt074}
	
	\bibitem[{{Tacconi} {et~al.}(2020){Tacconi}, {Genzel}, \&
		{Sternberg}}]{Tacconi2020}
	{Tacconi}, L.~J., {Genzel}, R., \& {Sternberg}, A. 2020, \araa, 58, 157,
	\dodoi{10.1146/annurev-astro-082812-141034}
	
	\bibitem[{{Tacconi} {et~al.}(2018){Tacconi}, {Genzel}, {Saintonge}, {Combes},
		{Garc{\'\i}a-Burillo}, {Neri}, {Bolatto}, {Contini}, {F{\"o}rster Schreiber},
		{Lilly}, {Lutz}, {Wuyts}, {Accurso}, {Boissier}, {Boone}, {Bouch{\'e}},
		{Bournaud}, {Burkert}, {Carollo}, {Cooper}, {Cox}, {Feruglio}, {Freundlich},
		{Herrera-Camus}, {Juneau}, {Lippa}, {Naab}, {Renzini}, {Salome}, {Sternberg},
		{Tadaki}, {{\"U}bler}, {Walter}, {Weiner}, \& {Weiss}}]{Tacconi2018}
	{Tacconi}, L.~J., {Genzel}, R., {Saintonge}, A., {et~al.} 2018, \apj, 853, 179,
	\dodoi{10.3847/1538-4357/aaa4b4}
	
	\bibitem[{{Tadaki} {et~al.}(2019){Tadaki}, {Kodama}, {Hayashi}, {Shimakawa},
		{Koyama}, {Lee}, {Tanaka}, {Hatsukade}, {Iono}, {Kohno}, {Matsuda}, {Suzuki},
		{Tamura}, {Toshikawa}, \& {Umehata}}]{Tadaki2019}
	{Tadaki}, K.-i., {Kodama}, T., {Hayashi}, M., {et~al.} 2019, \pasj, 71, 40,
	\dodoi{10.1093/pasj/psz005}
	
	\bibitem[{{Tumlinson} {et~al.}(2017){Tumlinson}, {Peeples}, \&
		{Werk}}]{Tumlinson2017}
	{Tumlinson}, J., {Peeples}, M.~S., \& {Werk}, J.~K. 2017, \araa, 55, 389,
	\dodoi{10.1146/annurev-astro-091916-055240}
	
	\bibitem[{{Villaescusa-Navarro} {et~al.}(2018){Villaescusa-Navarro}, {Genel},
		{Castorina}, {Obuljen}, {Spergel}, {Hernquist}, {Nelson}, {Carucci},
		{Pillepich}, {Marinacci}, {Diemer}, {Vogelsberger}, {Weinberger}, \&
		{Pakmor}}]{Villaescusa2018}
	{Villaescusa-Navarro}, F., {Genel}, S., {Castorina}, E., {et~al.} 2018, \apj,
	866, 135, \dodoi{10.3847/1538-4357/aadba0}
	
	\bibitem[{{Walter} {et~al.}(2008){Walter}, {Brinks}, {de Blok}, {Bigiel},
		{Kennicutt}, {Thornley}, \& {Leroy}}]{Walter2008}
	{Walter}, F., {Brinks}, E., {de Blok}, W.~J.~G., {et~al.} 2008, \aj, 136, 2563,
	\dodoi{10.1088/0004-6256/136/6/2563}
	
	\bibitem[{{Walter} {et~al.}(2020){Walter}, {Carilli}, {Neeleman}, {Decarli},
		{Popping}, {Somerville}, {Aravena}, {Bertoldi}, {Boogaard}, {Cox}, {da
			Cunha}, {Magnelli}, {Obreschkow}, {Riechers}, {Rix}, {Smail}, {Weiss},
		{Assef}, {Bauer}, {Bouwens}, {Contini}, {Cortes}, {Daddi}, {Diaz-Santos},
		{Gonz{\'a}lez-L{\'o}pez}, {Hennawi}, {Hodge}, {Inami}, {Ivison}, {Oesch},
		{Sargent}, {van der Werf}, {Wagg}, \& {Yung}}]{Walter2020}
	{Walter}, F., {Carilli}, C., {Neeleman}, M., {et~al.} 2020, \apj, 902, 111,
	\dodoi{10.3847/1538-4357/abb82e}
	
	\bibitem[{{Wang} {et~al.}(2020){Wang}, {Catinella}, {Saintonge}, {Pan},
		{Serra}, \& {Shao}}]{Wang2020}
	{Wang}, J., {Catinella}, B., {Saintonge}, A., {et~al.} 2020, \apj, 890, 63,
	\dodoi{10.3847/1538-4357/ab68dd}
	
	\bibitem[{{Wang} {et~al.}(2016){Wang}, {Koribalski}, {Serra}, {van der Hulst},
		{Roychowdhury}, {Kamphuis}, \& {Chengalur}}]{Wang2016}
	{Wang}, J., {Koribalski}, B.~S., {Serra}, P., {et~al.} 2016, \mnras, 460, 2143,
	\dodoi{10.1093/mnras/stw1099}
	
	\bibitem[{{Wang} {et~al.}(2021{\natexlab{a}}){Wang}, {Staveley-Smith},
		{Westmeier}, {Catinella}, {Shao}, {Reynolds}, {For}, {Lee}, {Liang}, {Wang},
		{Elagali}, {D{\'e}nes}, {Kleiner}, {Koribalski}, {Lee-Waddell}, {Oh}, {Rhee},
		{Serra}, {Spekkens}, {Wong}, {Bekki}, {Bigiel}, {Courtois}, {Hess},
		{Holwerda}, {McQuinn}, {Pandey-Pommier}, {van der Hulst}, \&
		{Verdes-Montenegro}}]{Wang2021}
	{Wang}, J., {Staveley-Smith}, L., {Westmeier}, T., {et~al.} 2021{\natexlab{a}},
	\apj, 915, 70, \dodoi{10.3847/1538-4357/abfc52}
	
	\bibitem[{{Wang} {et~al.}(2022){Wang}, {Magnelli}, {Schinnerer}, {Liu},
		{Modak}, {Jim{\'e}nez-Andrade}, {Karoumpis}, {Kokorev}, \&
		{Bertoldi}}]{Wang2022}
	{Wang}, T.-M., {Magnelli}, B., {Schinnerer}, E., {et~al.} 2022, \aap, 660,
	A142, \dodoi{10.1051/0004-6361/202142299}
	
	\bibitem[{{Wang} {et~al.}(2021{\natexlab{b}}){Wang}, {Chen}, {Mao}, {Mo},
		{Wang}, {Guo}, {Li}, {Fu}, {Jing}, {Wang}, {Yang}, \& {Zheng}}]{WangZY2021}
	{Wang}, Z., {Chen}, Y., {Mao}, Y., {et~al.} 2021{\natexlab{b}}, \apj, 907, 4,
	\dodoi{10.3847/1538-4357/abcb8a}
	
	\bibitem[{{Wechsler} {et~al.}(2006){Wechsler}, {Zentner}, {Bullock},
		{Kravtsov}, \& {Allgood}}]{Wechsler2006}
	{Wechsler}, R.~H., {Zentner}, A.~R., {Bullock}, J.~S., {Kravtsov}, A.~V., \&
	{Allgood}, B. 2006, \apj, 652, 71, \dodoi{10.1086/507120}
	
	\bibitem[{{Wolfe} {et~al.}(2005){Wolfe}, {Gawiser}, \& {Prochaska}}]{Wolfe2005}
	{Wolfe}, A.~M., {Gawiser}, E., \& {Prochaska}, J.~X. 2005, \araa, 43, 861,
	\dodoi{10.1146/annurev.astro.42.053102.133950}
	
	\bibitem[{{Wolz} {et~al.}(2022){Wolz}, {Pourtsidou}, {Masui}, {Chang},
		{Bautista}, {M{\"u}ller}, {Avila}, {Bacon}, {Percival}, {Cunnington},
		{Anderson}, {Chen}, {Kneib}, {Li}, {Liao}, {Pen}, {Peterson}, {Rossi},
		{Schneider}, {Yadav}, \& {Zhao}}]{Wolz2022}
	{Wolz}, L., {Pourtsidou}, A., {Masui}, K.~W., {et~al.} 2022, \mnras, 510, 3495,
	\dodoi{10.1093/mnras/stab3621}
	
	\bibitem[{{Xi} {et~al.}(2021){Xi}, {Staveley-Smith}, {For}, {Freudling},
		{Zwaan}, {Hoppmann}, {Liang}, \& {Peng}}]{Xi2021}
	{Xi}, H., {Staveley-Smith}, L., {For}, B.-Q., {et~al.} 2021, \mnras, 501, 4550,
	\dodoi{10.1093/mnras/staa3931}
	
	\bibitem[{{Xie} {et~al.}(2017){Xie}, {De Lucia}, {Hirschmann}, {Fontanot}, \&
		{Zoldan}}]{Xie2017}
	{Xie}, L., {De Lucia}, G., {Hirschmann}, M., {Fontanot}, F., \& {Zoldan}, A.
	2017, \mnras, 469, 968, \dodoi{10.1093/mnras/stx889}
	
	\bibitem[{{York} {et~al.}(2000){York}, {Adelman}, {Anderson}, {Anderson},
		{Annis}, {Bahcall}, {Bakken}, {Barkhouser}, {Bastian}, {Berman}, {Boroski},
		{Bracker}, {Briegel}, {Briggs}, {Brinkmann}, {Brunner}, {Burles}, {Carey},
		{Carr}, {Castander}, {Chen}, {Colestock}, {Connolly}, {Crocker}, {Csabai},
		{Czarapata}, {Davis}, {Doi}, {Dombeck}, {Eisenstein}, {Ellman}, {Elms},
		{Evans}, {Fan}, {Federwitz}, {Fiscelli}, {Friedman}, {Frieman}, {Fukugita},
		{Gillespie}, {Gunn}, {Gurbani}, {de Haas}, {Haldeman}, {Harris}, {Hayes},
		{Heckman}, {Hennessy}, {Hindsley}, {Holm}, {Holmgren}, {Huang}, {Hull},
		{Husby}, {Ichikawa}, {Ichikawa}, {Ivezi{\'c}}, {Kent}, {Kim}, {Kinney},
		{Klaene}, {Kleinman}, {Kleinman}, {Knapp}, {Korienek}, {Kron}, {Kunszt},
		{Lamb}, {Lee}, {Leger}, {Limmongkol}, {Lindenmeyer}, {Long}, {Loomis},
		{Loveday}, {Lucinio}, {Lupton}, {MacKinnon}, {Mannery}, {Mantsch}, {Margon},
		{McGehee}, {McKay}, {Meiksin}, {Merelli}, {Monet}, {Munn}, {Narayanan},
		{Nash}, {Neilsen}, {Neswold}, {Newberg}, {Nichol}, {Nicinski}, {Nonino},
		{Okada}, {Okamura}, {Ostriker}, {Owen}, {Pauls}, {Peoples}, {Peterson},
		{Petravick}, {Pier}, {Pope}, {Pordes}, {Prosapio}, {Rechenmacher}, {Quinn},
		{Richards}, {Richmond}, {Rivetta}, {Rockosi}, {Ruthmansdorfer}, {Sandford},
		{Schlegel}, {Schneider}, {Sekiguchi}, {Sergey}, {Shimasaku}, {Siegmund},
		{Smee}, {Smith}, {Snedden}, {Stone}, {Stoughton}, {Strauss}, {Stubbs},
		{SubbaRao}, {Szalay}, {Szapudi}, {Szokoly}, {Thakar}, {Tremonti}, {Tucker},
		{Uomoto}, {Vanden Berk}, {Vogeley}, {Waddell}, {Wang}, {Watanabe},
		{Weinberg}, {Yanny}, {Yasuda}, \& {SDSS Collaboration}}]{York2000}
	{York}, D.~G., {Adelman}, J., {Anderson}, John~E., J., {et~al.} 2000, \aj, 120,
	1579, \dodoi{10.1086/301513}
	
	\bibitem[{{Young} {et~al.}(1995){Young}, {Xie}, {Tacconi}, {Knezek}, {Viscuso},
		{Tacconi-Garman}, {Scoville}, {Schneider}, {Schloerb}, {Lord}, {Lesser},
		{Kenney}, {Huang}, {Devereux}, {Claussen}, {Case}, {Carpenter}, {Berry}, \&
		{Allen}}]{Young1995}
	{Young}, J.~S., {Xie}, S., {Tacconi}, L., {et~al.} 1995, \apjs, 98, 219,
	\dodoi{10.1086/192159}
	
	\bibitem[{{Zafar} {et~al.}(2013){Zafar}, {P{\'e}roux}, {Popping}, {Milliard},
		{Deharveng}, \& {Frank}}]{Zafar2013}
	{Zafar}, T., {P{\'e}roux}, C., {Popping}, A., {et~al.} 2013, \aap, 556, A141,
	\dodoi{10.1051/0004-6361/201321154}
	
	\bibitem[{{Zhang} {et~al.}(2013){Zhang}, {Li}, {Kauffmann}, \&
		{Xiao}}]{Zhang2013}
	{Zhang}, W., {Li}, C., {Kauffmann}, G., \& {Xiao}, T. 2013, \mnras, 429, 2191,
	\dodoi{10.1093/mnras/sts490}
	
	\bibitem[{{Zhao} {et~al.}(2009){Zhao}, {Jing}, {Mo}, \&
		{B{\"o}rner}}]{Zhao2009}
	{Zhao}, D.~H., {Jing}, Y.~P., {Mo}, H.~J., \& {B{\"o}rner}, G. 2009, \apj, 707,
	354, \dodoi{10.1088/0004-637X/707/1/354}
	
	\bibitem[{{Zwaan} {et~al.}(2005){Zwaan}, {Meyer}, {Staveley-Smith}, \&
		{Webster}}]{Zwaan2005}
	{Zwaan}, M.~A., {Meyer}, M.~J., {Staveley-Smith}, L., \& {Webster}, R.~L. 2005,
	\mnras, 359, L30, \dodoi{10.1111/j.1745-3933.2005.00029.x}
	
\end{thebibliography}

\end{document}